\documentclass[runningheads]{llncs}
\usepackage{booktabs}   

\usepackage[paperheight=235mm,paperwidth=155mm,textwidth=12.2cm,textheight=19.3cm]{geometry}
\usepackage{bbding}
\usepackage{graphicx}

\usepackage[colorinlistoftodos,prependcaption,backgroundcolor=orange!50]{todonotes}

\usepackage{float}
\usepackage{amssymb,pifont,bbding,amsmath}
\usepackage{pgf}
\usepackage{tikz}
\usetikzlibrary{calc,positioning,backgrounds,decorations.pathmorphing,shapes,calc}
\usepackage{enumitem}
\usepackage{mathcomp}
\usepackage{graphicx}

\usepackage{wrapfig}
\usepackage{stmaryrd}
\usepackage[algoruled,vlined,linesnumbered,titlenotnumbered,noend]{algorithm2e}
\usepackage{adjustbox}
\usepackage{bold-extra}
\usepackage{xspace}

\usepackage{color,colortbl}
\usepackage{mathpartir}
\usepackage{listings}

\usepackage{thmtools}
\usepackage{thm-restate}
\usepackage{tabularx,environ}
\usepackage{mdframed}

\usepackage{multirow}
\usepackage{empheq}
\usepackage[most]{tcolorbox}
\usepackage{lineno}

\usepackage{epstopdf}

\newtcbox{\mymath}[1][]{%
    nobeforeafter, math upper, tcbox raise base,
    enhanced, 
    colframe=blue!30!black,
    boxrule=0.5pt,
    #1}
\newtcbox{\mymathh}[1][]{%
   nobeforeafter, math upper, 
    colframe=white,
    boxrule=0pt,
    #1}

\newtheorem{lemm}{Lemma}[section]

\definecolor{reasonred}{RGB}{169, 92, 104}
\definecolor{rulered}{RGB}{227, 36, 28}
\definecolor{varblue}{RGB}{36, 28, 227}

\SetKwBlock{KwVar}{Var}{}{}

\DeclareSymbolFont{matha}{OML}{txmi}{m}{it}
\DeclareMathSymbol{\varv}{\mathord}{matha}{118}

\makeatletter

\newcolumntype{\expand}{}
\long\@namedef{NC@rewrite@\string\expand}{\expandafter\NC@find}

\NewEnviron{myproblem}[2][]{%
  \def\problem@arg{#1}%
  \def\problem@framed{framed}%
  \def\problem@lined{lined}%
  \def\problem@doublelined{doublelined}%
  \ifx\problem@arg\@empty%
    \def\problem@hline{}%
  \else%
    \ifx\problem@arg\problem@doublelined%
      \def\problem@hline{\hline\hline}%
    \else%
      \def\problem@hline{\hline}%
    \fi%
  \fi%
  \ifx\problem@arg\problem@framed%
    \def\problem@tablelayout{|>{\bfseries}lX|c}%
    \def\problem@title{\multicolumn{2}{|l|}{%
        \raisebox{-\fboxsep}{\text{\large #2}}%
      }}%
  \else
    \def\problem@tablelayout{>{\bfseries}lXc}%
    \def\problem@title{\multicolumn{2}{l}{%
        \raisebox{-\fboxsep}{\text{\large #2}}%
      }}%
  \fi%
  \bigskip\par\noindent%
  \renewcommand{\arraystretch}{1}%
  \begin{tabularx}{\textwidth}{\expand\problem@tablelayout}%
    \problem@hline%
    \problem@title\\[2\fboxsep]%
    \BODY\\\problem@hline%
  \end{tabularx}%
  \medskip\par%
}
\makeatother

\lstdefinestyle{splgrammar}{
  mathescape,
  belowcaptionskip=1\baselineskip,
  breaklines=true,
  xleftmargin=\parindent,
  language=C,
  showstringspaces=false,
  basicstyle=\footnotesize\ttfamily,
  keywords={if,then,term,:=,CAS},
  keywords=[2]{prog,stmt,instr,},
  keywords=[3]{reg,var,lbl,expr},
  keywordstyle=\footnotesize\bfseries\color{mygreen},
  keywordstyle=[2]\footnotesize\bfseries\color{brown},
  keywordstyle=[3]\bfseries\color{black},
  commentstyle=\itshape\color{mymauve},
  identifierstyle=\color{black},
  stringstyle=\color{orange},
  tabsize=2
}

\lstdefinestyle{customlang}{
  belowcaptionskip=1\baselineskip,
  breaklines=true,
  xleftmargin=\parindent,
  language=C,
  showstringspaces=false,
  basicstyle=\small\ttfamily,
  keywords={CAS,fence,while,for,goto,term},
  keywordstyle=\bfseries\color{mygreen},
  commentstyle=\itshape\color{mymauve},
  identifierstyle=\color{blue},
  stringstyle=\color{orange},
  tabsize=2
}

\lstdefinestyle{litmusresults}{
  belowcaptionskip=1\baselineskip,
  breaklines=true,
  xleftmargin=\parindent,
  language=C,
  showstringspaces=false,
  basicstyle=\scriptsize\ttfamily,
  morekeywords={Histogram, Test},
  keywordstyle=\bfseries\color{mygreen},
  commentstyle=\itshape\color{mymauve},
  identifierstyle=\color{black},
  stringstyle=\color{orange},
  tabsize=2
}

\newcommand\rulename[1]{\textcolor{rulered}{\texttt{\textbf{#1}}}}

\newcommand\bb{b}

\newcommand\ii{i}
\newcommand\jj{j}
\newcommand\kk{k}
\newcommand\mm{m}
\newcommand\nn{n}
\newcommand\xx{x}

\newcommand\nat{\mathbb{N}}
\newcommand\pnat{\nat^{>0}}
\newcommand\reals{\mathbb{R}}
\newcommand\preals{\reals^{>0}}
\newcommand\bool{\mathbb{B}}
\newcommand\true{{\tt true}}
\newcommand\false{{\tt false}}

\newcommand\intrvlof[2]{[#1,#2]}
\newcommand\tuple[1]{\left\langle#1\right\rangle}

\newcommand\set[1]{\left\{#1\right\}}
\newcommand\setcomp[2]{\left\{#1~|~\;{#2}\right\}}

\newcommand\aset{A}

\newcommand\sizeof[1]{\left| {#1} \right|}
\newcommand\lengthof[1]{\left| {#1} \right|}
\newcommand\ordering\sqsubseteq

\newcommand\alphabet\Sigma
\newcommand\aelem{a}
\newcommand\belem{b}
\newcommand\wordsover[1]{{#1}^*}
\newcommand\kwordsover[2]{{#2}^{#1}}

\newcommand\infwordsover[1]{{#1}^\omega}
\newcommand\prminfwordsover[2]{{#1}_{#2}^\omega}

\newcommand\emptyword\epsilon
\newcommand\word{w}
\newcommand\vord{v}

\newcommand\wordof[1]{\word[#1]}

\newcommand\app{\boldsymbol{\cdot}} 
\newcommand\subword\preceq
\newcommand\shuffle\otimes
\newcommand\headof[1]{{\tt head}\left({#1}\right)}
\newcommand\tailof[1]{{\tt tail}\left({#1}\right)}

\newcommand\fun{f}

\newcommand\undf\bot

\newcommand\funtype[3]{{#1}\!:{#2}\rightarrow{#3}}

\newcommand\updatefun[3]{{#1}\left[#2\leftarrow#3\right]}
\newcommand\relcompose\circ

\newcommand\prog{{\mathcal P}}
\newcommand\pprog{\prog'}
\newcommand\progtuple{\tuple{\procset,\weight}}
\newcommand\procset{{\sf Procs}}
\newcommand\proc{p}
\newcommand\pproc{\proc'}
\newcommand\weight{{\tt Sched}}
\newcommand\weightof[1]{\weight\left(#1\right)}
\newcommand\rweight{{\tt Rweight}}
\newcommand\rweightof[2]{\rweight\left(#1\right)\left(#2\right)}
\newcommand\instr{\mathsf{i}}
\newcommand\instrset{\mathsf{Instr}}
\newcommand\instrsetof[1]{\instrset_{#1}}
\newcommand\lbl{\mathsf{l}}

\newcommand\lblset{\mathsf{Lbl}}
\newcommand\lblsetof[1]{\lblset_{#1}}
\newcommand\rmlbl\ominus

\newcommand\stmt{\mathsf{s}}
\newcommand\mkinstr[2]{{#1}:{#2}}
\newcommand\regset{{\sf Regs}}
\newcommand\regsetof[1]{\regset_{#1}}
\newcommand\valset{{\mathcal V}}
\newcommand\val{\varv}
\newcommand\zero{0}
\newcommand\expr{e}
\newcommand\stmtof[1]{{\tt stmt}\left(#1\right)}

\newcommand\terminated{\text{\lstinline[style=customlang]{term}}}
\newcommand\termlblof[1]{\lbl^{\terminated}_{#1}}
\newcommand\goto{\text{\lstinline[style=customlang]{goto}}}
\newcommand\disabled{{\sf disab}}
\newcommand\disabledof[1]{\disabled\left(#1\right)}
\newcommand\enabled{{\sf enab}}
\newcommand\enabledof[1]{\enabled\left(#1\right)}

\newcommand\nextof[1]{{\tt next}\left(#1\right)}
\newcommand\areg{{\tt a}}
\newcommand\breg{{\tt b}}
\newcommand\varset{{\mathcal X}}
\newcommand\xvar{\mathsf{x}}

\newcommand\assigned{:=}
\newcommand\cas{\text{\lstinline[style=customlang]{CAS}}}
\newcommand\casof[3]{\cas\left(#1,#2,#3\right)}

\newcommand\conftuple{\tuple{\labeling,\regstate,\bufferstate,\memstate}}

\newcommand\pconftuple{\tuple{\plabeling,\pregstate,\pbufferstate,\pmemstate}}
\newcommand\ppconftuple{\tuple{\pplabeling,\ppregstate,\ppbufferstate,\ppmemstate}}

\newcommand\labeling\lambda
\newcommand\labelingof[1]{\labeling\left(#1\right)}
\newcommand\plabeling{\labeling'}

\newcommand\pplabeling{\labeling''}

\newcommand\regstate{{\mathcal R}}
\newcommand\regstateof[1]{\regstate(#1)}
\newcommand\pregstate{\regstate'}

\newcommand\ppregstate{\regstate''}

\newcommand\bufferstate{{\mathcal B}}

\newcommand\bufferstateof[1]{\bufferstate\left(#1\right)}
\newcommand\pbufferstate{\bufferstate'}
\newcommand\pbufferstateof[1]{\pbufferstate\left(#1\right)}
\newcommand\ppbufferstate{\bufferstate''}

\newcommand\memstate{{\mathcal M}}
\newcommand\memstateof[1]{\memstate\left(#1\right)}
\newcommand\pmemstate{\memstate'}

\newcommand\ppmemstate{\memstate''}

\newcommand\fetchval{{\tt FetchVal}}

\newcommand\fetchvalof[3]{\fetchval\left(#1\right)\left(#2\right)\left(#3\right)}
\newcommand\probval\theta
\newcommand\fiprobval\phi
\newcommand\costval\psi
\newcommand\precision\varepsilon

\newcommand\movesto[1]{\rightarrow_{\mbox{\tiny ${#1}$}}}
\newcommand\starmovesto[1]{\stackrel*\rightarrow_{\mbox{\tiny ${#1}$}}}
\newcommand\plainmovesto[1]{\stackrel{#1}\rightarrow}
\newcommand\progmovesto[2]{\xrightarrow{#2}_{#1}}
\newcommand\procmovesto[1]{\xrightarrow{#1}_{{\tt proc}}}
\newcommand\updatemovesto[1]{\xrightarrow{#1}_{{\tt update}}}

\newcommand\procseq\alpha

\newcommand\figureprogmovesto[1]{%
  \mathrel{\begin{tikzpicture}[baseline={($(current bounding box.south)+(0pt,5pt)$)}]
      \node[inner sep=.44ex, align=center,font=\tiny,minimum width=10pt] (tmp) at (0pt,0pt) 
           {};
           \path[draw,->,line width=1pt] 
           ($(tmp.south west)+(-1pt,0pt)$)-- ($(tmp.south east)+(1pt,0pt)$);
           \node[inner sep=.44ex, align=center,font=\fontsize{6}{6},anchor= north west] at ($(tmp.south east)+(-1pt,2pt)$) 
                {{\normalfont\textsc{\textbf{$\prog$}}}};
        \end{tikzpicture}}}

\newcommand\confset{\Gamma}
\newcommand\confsetof[1]{\Gamma_{#1}}

\newcommand\lblconfsetof[2]{\Gamma^{#2}_{#1}}
\newcommand\numconfsetof[2]{\confset_{#1}^{#2}}
\newcommand\plainconfsetof[1]{\confset^{\tt plain}_{#1}}
\newcommand\smallconfsetof[1]{\confset^{\tt small}_{#1}}
\newcommand\largeconfsetof[1]{\confset^{\tt large}_{#1}}
\newcommand\bplainconfsetof[1]{\confset^{\tt Bplain}_{#1}}

\newcommand\conf{\gamma}

\newcommand\confs{G}
\newcommand\fconfs{F}
\newcommand\initconf{\conf_{\it init}}
\newcommand\targetstates{{\tt Target}}

\renewcommand\ts{{\mathcal T}}

\newcommand\tstuple{\tuple{\confset,\plainmovesto{}}}
\newcommand\pconf{\conf'}
\newcommand\ppconf{\conf''}
\newcommand\mchain{{\mathcal C}}

\newcommand\mchainof[1]{\llbracket{#1}\rrbracket^{\tt MC}}
\newcommand\mchaintuple{\tuple{\confset,\pmtrx}}
\newcommand\mchaintupleof[1]{\tuple{\confsetof{#1},\prmpmtrx{#1}}}
\newcommand\pmtrx{{\normalfont\texttt{M}}}
\newcommand\pmtrxof[1]{\pmtrx\left(#1\right)}
\newcommand\procpmtrx{{\tt M}_{\tt proc}}
\newcommand\procpmtrxof[1]{\procpmtrx\left(#1\right)}
\newcommand\updatepmtrx{{\tt M}_{\tt update}}
\newcommand\updatepmtrxof[1]{\updatepmtrx\left(#1\right)}
\newcommand\progpmtrx[1]{{\tt M}_{#1}}
\newcommand\progpmtrxof[2]{\progpmtrx{#1}\left(#2\right)}

\newcommand\prmpmtrx[1]{\pmtrx_{#1}}
\newcommand\prmpmtrxof[2]{\prmpmtrx{#1}\left(#2\right)}

\newcommand\tsdenotationof[1]{\llbracket{#1}\rrbracket^{\tt TS}}
\newcommand\mcdenotationof[1]{\llbracket{#1}\rrbracket^{\tt MC}}
\newcommand\prob{{\it Prob}}
\newcommand\probof[2]{\prob_{#1}\left(#2\right)}
\newcommand\measure[1]{{\it Prob}_{#1}}
\newcommand\measureof[2]{\measure{#1}\left(#2\right)}
\newcommand\mctots[1]{{#1}^\downarrow}

\newcommand\run\rho
\newcommand\runset{{\tt Runs}}
\newcommand\runsetof[1]{\runset\left(#1\right)}
\newcommand\frunset{{\tt FRuns}}
\newcommand\frunsetof[2]{\frunset\left(#1\right)\left(#2\right)}
\newcommand\drunset{{\tt DRuns}}
\newcommand\drunsetof[3]{\drunset\left(#1\right)\left(#2\right)\left(#3\right)}
\newcommand\srunset{{\tt SRuns}}
\newcommand\srunsetof[2]{\srunset\left(#1\right)\left(#2\right)}

\newcommand\prun{\run'}
\newcommand\pth\pi

\def\Om{\Omega}

\newcommand{\ef}{\exists\Diamond}
\newcommand{\eventually}{\Diamond}
\newcommand{\neventually}[1]{\eventually^{\!\!#1}}
\newcommand{\nxt}{\bigcirc}
\newcommand{\nnxt}[1]{\nxt^{\!#1}}
\newcommand{\always}{\Box}

\newcommand{\ctlstar}{\mathit{CTL}^*}
\newcommand\modelswrt[1]{\models_{#1}}

\newcommand{\qualreach}{{\sc Qual\_Reach}\xspace}
\newcommand{\qualrepreach}{{\sc Qual\_Rep\_Reach}\xspace}
\newcommand{\neverqualreach}{{\sc Never\_Qual\_Reach}\xspace}
\newcommand{\neverqualrepreach}{{\sc Never\_Qual\_Rep\_Reach}\xspace}
\newcommand{\quantreach}{{\sc Quant\_Reach}\xspace}
\newcommand{\quantrepreach}{{\sc Quant\_Rep\_Reach}\xspace}
\newcommand{\expaveragecost}{{\sc Exp\_Ave\_Cost}\xspace}

\newcommand\flagvar{\textcolor{varblue}{\tt flag}}
\newcommand\posflagvar{\textcolor{varblue}{\tt posflag}}

\newcommand\bplainvar{\textcolor{varblue}{\tt BPlain}}
\newcommand\waitingvar{\textcolor{varblue}{\tt waiting}}

\newcommand\pwaitingvar{\textcolor{varblue}{\waitingvar'}}
\newcommand{\posapprxvar}{\textcolor{varblue}{\tt PosApprx}}
\newcommand{\posapprxvarof}[1]{\textcolor{varblue}{\posapprxvar^{(#1)}}}
\newcommand{\negapprxvar}{\textcolor{varblue}{\tt NegApprx}}
\newcommand{\negapprxvarof}[1]{\textcolor{varblue}{\negapprxvar^{(#1)}}}

\newcommand\costapprxvar{\textcolor{varblue}{\tt CostApprx}}
\newcommand\costapprxvarof[1]{\textcolor{varblue}{\costapprxvar^{(#1)}}}
\newcommand\probapprxvar{\textcolor{varblue}{\tt ProbApprx}}
\newcommand\probapprxvarof[1]{\textcolor{varblue}{\probapprxvar^{(#1)}}}
\newcommand\costerrorvar{\textcolor{varblue}{\tt CostError}}
\newcommand\costerrorvarof[1]{\textcolor{varblue}{\costerrorvar^{(#1)}}}
\newcommand\proberrorvar{\textcolor{varblue}{\tt ProbError}}
\newcommand\proberrorvarof[1]{\textcolor{varblue}{\proberrorvar^{(#1)}}}
\newcommand\kvar{\textcolor{varblue}{k}}
\newcommand\nvar{\textcolor{varblue}{n}}

\newcommand\cost{{\tt Cost}}
\newcommand\costof[1]{\cost\left(#1\right)}
\newcommand\runcostof[2]{\cost\left(#1\right)\left(#2\right)}

\newcommand\ccostof[2]{\cost\left(#1,#2\right)}
\newcommand\maxcost{\tt MaxCost}
\newcommand\maxcostof[1]{\maxcost\left(#1\right)}

\newcommand\rcost[3]{X_{#1,#2,#3}}
\newcommand\rcostof[4]{\rcost{#1}{#2}{#3}\left(#4\right)}
\newcommand\cecostof[3]{E\left(\rcost{#1}{#2}{#3}~\mid~{#1}\models\exists\eventually{#2}\right)}

\newcommand\ecostof[3]{E\left(\rcost{#1}{#2}{#3}\right)}
\DeclareSymbolFont{matha}{OML}{txmi}{m}{it}
\DeclareMathSymbol{\varv}{\mathord}{matha}{118}

\newcommand\eagerness{{\mathcal E}}
\newcommand\eagernessof[1]{\eagerness_{#1}}
\newcommand\eagernessthreshold{\eta}
\newcommand\eagernessthresholdof[1]{\eagernessthreshold_{#1}}

\newcommand\seagerness{{\mathcal E}^{\tt S}}
\newcommand\seagernessof[1]{\seagerness_{#1}}

\newcommand\deagerness{{\mathcal E}^{\tt D}}
\newcommand\deagernessof[1]{\deagerness_{#1}}
\newcommand\deagernessthreshold{\eta^{\tt D}}
\newcommand\deagernessthresholdof[1]{\deagernessthreshold_{#1}}

\newcommand\sdeagerness{{\mathcal E}^{\tt SD}}
\newcommand\sdeagernessof[1]{\sdeagerness_{#1}}
\newcommand\sdeagernessthreshold{\eta^{\tt SD}}
\newcommand\sdeagernessthresholdof[1]{\sdeagernessthreshold_{#1}}

\newcommand\neagernessof[2]{\eagerness^{#1}_{#2}}
\newcommand\keagernessthresholdof[2]{\eagernessthreshold_{#1}^{#2}}

\newcommand\gravity{{\mathcal G}}

\newcommand\gravityof[1]{\gravity_{#1}}

\newcommand\reason[2]{\;\textcolor{reasonred}{{#1}\;\left\{#2\right\}}}
\newcommand\textreason[2]{\;\textcolor{reasonred}{{#1}\;\left\{\mbox{#2}\right\}}}

\newcommand\visit{{\tt Visit}}
\newcommand\visitof[3]{\visit_{#1}\left(#2,#3\right)}

\newcommand\partition\oplus

\newcommand\pthapp\odot

\newcommand\before{{\tt Before}}
\newcommand\onebefore{\,\before\,}
\newcommand\kbefore[1]{\,\before^{#1}\,}
\newcommand\border\nu
\newcommand\aaset{{\mathcal A}}

\newcommand\qq{q}
\newcommand\pp{p}
\newcommand\qqstar{\hat{q}}
\newcommand\ppstar{\hat{p}}
\newcommand\gmchain{\mchain^G}
\newcommand\gmchainof[1]{\mchain_{#1}^G}
\newcommand\gpmtrx{\pmtrx^G}
\newcommand\gpmtrxof[1]{\pmtrx^G_{#1}}
\newcommand\gmchaintupleof[1]{\tuple{\nat,\gpmtrxof{#1}}}

\newcommand\yy{y}
\newcommand\yfun\yy

\newcommand\yyfun\yy
\newcommand\yfunof[2]{\yyfun\left(#1,#2\right)}

\newcommand{\levelfun}{level}

\definecolor{mygreen}{rgb}{0,0.6,0}
\definecolor{mygray}{rgb}{0.5,0.5,0.5}
\definecolor{mymauve}{rgb}{0.58,0,0.82}

\title{Probabilistic Total Store Ordering}

\author{Parosh Aziz Abdulla\inst{1} \and Mohamed Faouzi Atig\inst{1} \and Raj Aryan Agarwal\inst{2} \and \\ Adwait Godbole\inst{3} \and Krishna S.\inst{2}}

\institute{Uppsala University, Sweden \and IIT Bombay, India \and University of California Berkeley, USA}

\authorrunning{P. A. Abdulla et al.}

\begin{document}

\maketitle

\begin{abstract}
We present {\it Probabilistic Total Store Ordering (PTSO)}
-- a probabilistic extension 
of the classical TSO semantics.
For a given (finite-state) program, the operational semantics of
PTSO induces an infinite-state Markov chain.
We resolve the inherent non-determinism due to 
process schedulings and memory updates
according to given probability distributions.
We provide a comprehensive set of results showing the decidability of 
several properties for PTSO, namely
(i)
{\it Almost-Sure (Repeated) Reachability}:
whether a run, starting from a given initial configuration,
almost surely visits (resp.\ almost surely repeatedly visits) a given 
set of target configurations.
(ii)
{\it Almost-Never (Repeated) Reachability}:
whether a run from the initial configuration,
almost never visits (resp.\ almost never repeatedly visits) the target.
(iii)
{\it Approximate Quantitative (Repeated) Reachability}:
to approximate, up to an arbitrary degree of precision,
the measure of runs that start from the initial configuration
and (repeatedly) visit the target.
(iv) 
{\it Expected Average Cost:}
to approximate, up to an arbitrary degree of precision,
the expected average cost of a run from the initial configuration to the target.
We derive our results through a nontrivial combination 
of results from the classical theory of (infinite-state) Markov chains, the
theories of {\it decisive} and {\it eager} Markov chains,
specific techniques from combinatorics,
as well as,
decidability and complexity results for the classical
(non-probabilistic) TSO semantics.
As far as we know, this is the first work that considers probabilistic  verification of programs running on weak memory models. \end{abstract}

\section{Introduction}
The classical Sequential Consistency (SC) 
semantics \cite{Lamport:multiprocess:executes} has been
a fundamental assumption in concurrent programming.
SC guarantees that process operations are atomic.
A write
operation, performed  by a given process, 
is immediately visible to all the other processes.
However, 
designers of modern computer systems,
in their quest of increased system efficiency, often sacrifice
the SC guarantee.
Instead,
 the processes communicate asynchronously,
allowing a {\it delay} in the 
propagation of write operations.
Due to the propagation delay,
written values can become available
to processes at different time points, 
and in an order that may be different from the order in 
which they are generated.
This asynchronous behavior gives rise to new semantics,
collectively referred to as {\it weak memory models}
\cite{DBLP:journals/computer/AdveG96}.
In the presence of weak memory models,
programs exhibit new, and often unexpected, 
behaviors, 
bringing about complex challenges in the design and analysis 
of concurrent systems.
Even text-book programs may behave erroneously.
The classical Dekker mutual exclusion protocol is a case in point.
The ubiquity of weak memory models has led to an extensive research effort
for the testing and verification of concurrent programs running 
under such semantics.

Existing works on the verification of programs running on weak memory models,
consider safety properties such as state reachability, 
assertion violation, and robustness.
While safety properties are fundamental, we need also
to prove liveness properties, i.e.,
to show that the program indeed makes progress.
This is, of course, true already in the case of SC.
A program, such as a mutual exclusion protocol, needs to
guarantee that each process will eventually
reach its critical section.
The satisfiability of liveness properties
is often  dependent on
the type of {\it fairness conditions} on process executions
that are provided by the underlying platform
\cite{DBLP:books/daglib/0067978,DBLP:books/daglib/0077033}.
The reason is the presence of {\it concurrency non-determinism},
i.e., the inherent non-determinism in program
behavior due to the different possible ways in which
the scheduler can interleave the processes.
The scheduler may always neglect a given 
process, which means that the
process will never make progress (e.g., never reaches its critical section). 
Therefore, we need 
the scheduler to follow a fair selection policy  
that allows each process
to advance in its execution.
The situation is even more complicated in the
case of weak memory models, since we also need to deal with a second
source of non-determinism, besides concurrency non-determinism,
namely  {\it (data) propagation non-determinism}.
Since write operations are propagated asynchronously,
there is in general no way to predict {\it if},  {\it when}, and in which {\it order}, 
write operations become visible to the processes.

In this paper we present a framework for the verification
of liveness properties for concurrent programs running 
under the classical Total Store Ordering (TSO) semantics
\cite{SSONM2010}.
The TSO model
 puts an unbounded {\it store (write) buffer}
between each process and the main memory.
The buffer carries pending write operations that have been performed
by the process.
These operations are propagated from the buffer to 
the shared memory in a FIFO manner.
When a process performs a write operation, it appends the operation
as a message to its buffer.
When a process reads a variable, it searches its buffer
for a pending write operation on that variable.
If such operations exist then it reads from the most recent
one.
If no such operation exists, it fetches the value
of the variable from the main memory.
The TSO propagation mechanism is a typical example of how propagation
non-determinism arises: the write operations are propagated to 
the shared memory {\it non-deterministically},
and a process sees the other processes' write operations
only when the latter are available in the memory.
Therefore, having a scheduler that fairly selects the processes
is not sufficient.
We also need to ensure that the write operations propagate to the
processes sufficiently often.

Traditional fairness conditions such as strong or weak
fairness 
\cite{DBLP:books/daglib/0067978,DBLP:books/daglib/0077033,KWIATKOWSKA1989371}
cannot capture propagation policies adequately since
they irrationally allow {\it slow propagation}, i.e.,
they allow write operations to propagate at a lower rate than the
rate by which they are issued.
For instance, strong fairness guarantees that messages are transferred 
infinitely
often from the buffers to the memory.
Still, it does not constrain the relative frequency of
 write and update operations, and hence it  does not
prevent the buffer contents from growing unboundedly.
In such a scenario, more and more un-propagated messages 
may be clustered inside the buffers,
and a given process may, from some point on, 
be confined only to read its own writes, since it will
not see the memory updates by the other processes.
Accordingly, verifying liveness properties subject to strong 
fairness  may wrongly deem the system to be incorrect:
even if a process is selected infinitely often by the scheduler
and write operations are propagated infinitely often to the memory,
a given process may incorrectly be judged not to make progress due to slow
propagation.

While slow propagation can arise theoretically under the above
mentioned fairness conditions,  
it is almost never observed in practice.
%
Existing  platforms implement different policies,
such as invalidation or write-back policies, to
flush the buffers at regular intervals \cite{aros-micro16,ElverN14}.
This prevents the buffer sizes from growing beyond
certain sizes, and implicitly ensure
propagation fairness.
In fact, this is true to the degree 
that non-SC behaviors are (relatively)
rarely observed on TSO platforms
\cite{DBLP:conf/tacas/AlglaveMSS11,DBLP:conf/IEEEpact/LinNG10}.

%

%
%
In this paper, we perform
verification of liveness properties
for concurrent programs under TSO using
 {\it probabilistic fairness} \cite{DBLP:journals/entcs/Alfaro99}.
As far as we know, this is the first work that considers probabilistic  verification of programs running on weak memory models. In our model, both process scheduling and message propagation
are carried out  according
to given probability distributions.
We assign a weight (a natural number) to each process.
We resolve concurrency non-determinism probabilistically by letting
the scheduler select the next process to execute 
with a probability that reflects the weight of the process 
compared to the weights of the other processes that are
enabled in the same configuration.
After each process step, we allow an update step,
in which the buffers transfer parts of
their contents to the memory.
We make the probability distribution equal among all possible 
update operations in the given 
configuration\footnote{Our framework allows 
several other types of probability distributions 
(see Sec.~\ref{conclusions:section}.)}.
As we will see later in the paper, 
defining the model in this way implies that
we assign low probabilities to program runs
that unboundedly increase the number of messages
inside the buffers.
Accordingly, our model
is more faithful to real program behavior compared to models 
induced by  non-probabilistic fairness conditions.

We perform  a comprehensive analysis of the decidability 
of verifying
liveness properties for concurrent programs running under 
the TSO semantics, subject to probabilistic fairness.
In fact, verifying programs running on the TSO memory model,
even with respect to safety properties, poses
a difficult challenge.
The  unboundedness of the buffers implies that
the state space of the system is infinite, even in the case
where the input program is finite-state
\cite{DBLP:conf/popl/AtigBBM10,DBLP:journals/lmcs/AbdullaABN18}.
Similarly, the operational semantics of our model gives rise to Markov chains
with infinite state spaces.
Furthermore, in general, liveness properties give rise
to more difficult problems than safety properties, since the former
are interpreted over infinite program executions while the latter
are interpreted over finite executions.
Our results rely on nontrivial combinations of results from the
classical theory of (infinite-state) Markov chains 
\cite{Feller:book,Kulkarni:book}, the
theories of {\it decisive} and {\it eager} Markov chains
\cite{AbdullaHM07,DBLP:conf/atva/AbdullaHMS06},
specific techniques from combinatorics  \cite{Binomial:2001},
as well as,
decidability and complexity results for the classical
(non-probabilistic) TSO semantics 
\cite{DBLP:conf/esop/AtigBBM12,DBLP:journals/lmcs/AbdullaABN18}.
Concretely, 
we show the  decidability of the following problems, each of which is defined
by giving an initial configuration $\initconf$ and
a set $\targetstates$ of process target  states.

\smallskip
\noindent{\bf Qualitative Analysis} (Sec.~\ref{qual:reachability:section}).
In qualitative reasoning, 
we are interested in knowing whether the given property
is satisfied with probability $1$ (almost surely satisfied), or
with probability $0$ (almost never satisfied).
We show that the satisfiability of these properties can be reduced to 
similar problems on the underlying (non-probabilistic) transition systems
for classical TSO.
%
The actual probabilities appearing in the induced Markov chains 
then are inconsequential 
and only their non-zeroness matters.
This is useful whenever the probabilities have not been measured
exactly, or the portion of the system giving rise to probabilistic 
behavior has not been designed yet.
We consider the following different flavors of qualitative analysis:
%
{\it Almost-Sure (Repeated) Reachability}\footnote{While repeated reachability
is a liveness property, plain reachability in the non-probabilistic case is a safety property. However, in the presence of probabilities, plain reachability
measures the probability of convergence towards a target state, and hence it can be considered a form of liveness property. In any case, this is a matter of definition and has no bearing on the rest of the paper.}:
whether a run of the system from $\initconf$ will almost surely visit (resp. repeatedly visit) $\targetstates$;
{\it Almost-Never (Repeated) Reachability}:
whether a run of the system from $\initconf$ will almost never
visit (resp. repeatedly visit) $\targetstates$.
%
Furthermore, we show that all these problems 
have non-primitive-recursive complexities.

\smallskip
\noindent{\bf Quantitative Analysis} (Sec.~\ref{quan:reachability:section}).
The task is to estimate to an arbitrary degree of precision 
the probability by which a run from 
$\initconf$ (repeatedly)
visits $\targetstates$, rather than only checking
whether the probability is equal to one or zero.
%
%

\smallskip
\noindent{\bf Expected Average Cost} (Sec.~\ref{cost:section}).
We study the expected cost for runs that
start from $\initconf$ until they reach $\targetstates$.
To that end, we extend our model by providing a cost function
that assigns a fixed cost to each instruction
in the language.
Calculating expected costs of runs has many potential applications.
For instance, one might be interested in the {\it mean-time}
of reaching a target, i.e., the average number of steps 
before reaching the target \cite{PMC:book}.
In the context of weak memory models, in general, and TSO in particular,
one can perform a more refined analysis by also taking into account
the fact that specific instructions, e.g., memory fences, 
have higher costs \cite{benchmarkingWMM}.
Incorporating instruction costs in the model
makes average cost analysis reflect more faithfully
the efficiency of the program 
compared to an instruction count based metric.
There have been several approaches towards
optimizing fence implementations in hardware
\cite{fenceScoping,weeFence,cfences} which exploit the fact that non-SC
behaviours are rare even in unfenced code.
A quantitative analysis of the prevalence of behaviours and
cost of executing instructions
can help determine the efficacy of such implementations.

%
\smallskip 
\noindent {\bf {The  supplementary material}} contains detailed proofs of all the lemmas and theorems.


\section{Preliminaries}
\label{prels:section}
In this section, we introduce notation,
 recall basics of
transition systems, Temporal logic and Markov chains.

\paragraph{Basic Notation}
%
%
%

The size of a set $\aset$ is denoted by $\sizeof\aset$.
We use $\wordsover\aset$ and 
$\infwordsover\aset$
to denote the set of finite resp.\ infinite words over (a possibly infinite set) $\aset$, and
let $\emptyword$ be the empty word.
For $\word\in\wordsover\aset$,
$\lengthof\word$ denotes the length of $\word$
($\lengthof\word=\infty$ if $\word$ is infinite).
For
$\ii:1\leq\ii\leq\lengthof\word$, we use $\wordof\ii$ to denote
the $\ii^{th}$ element of $\word$.
We define $\headof\word:=\wordof{1}$ and $\tailof\word:=\wordof2\cdots\wordof{\lengthof\word}$.
We  
use $\aelem\in\word$ to denote that $\wordof\ii=\aelem$ for some
$\ii:1\leq\ii\leq\lengthof\word$.
For words $\word_1\in\wordsover\aset$ and $\word_2\in(\wordsover\aset\cup\infwordsover\aset)$,
we use $\word_1\app\word_2$ to denote their concatenation.
For $\kk\in\nat$, we define 
$\kwordsover\kk\aset:=\setcomp{\word\in\wordsover\aset}{\lengthof\word=\kk}$,
i.e., it is the set of words over $\aset$ of length $\kk$.

%
%
%

\paragraph{Transition Systems}
A {\it transition system} is a pair 
$\tstuple$ where
$\confset$ is a (potentially) infinite set of {\it configurations}, and
$\plainmovesto{}\subseteq\confset\times\confset$ 
is the {\it transition relation}.
We write $\conf\plainmovesto{}{}\pconf$ to denote that 
$\tuple{\conf,\pconf}\in\plainmovesto{}$, and use
$\plainmovesto*$ to be the reflexive transitive closure of  $\movesto{}{}$.
For $\kk\in\nat$, we write
$\conf\plainmovesto\kk\pconf$ to denote that there is a sequence
$\conf_0\plainmovesto{}\conf_1\plainmovesto{}\cdots\plainmovesto{}\conf_\kk$
where $\conf_0=\conf$ and $\conf_\kk=\pconf$, i.e., there is
a sequence of $\kk$ transition steps leading from $\conf$
to $\pconf$.
For $\sim\in\set{<,\leq,=}$, we
write $\conf\plainmovesto{\sim\kk}\pconf$ to denote that
$\conf\plainmovesto{\mm}\pconf$ for some 
$\mm:0\leq\mm\sim\kk$.
%

\paragraph{Temporal Logic}
\label{runs:subsection}
A {\it run} $\run$ of transition system $\ts=\tstuple$ is an infinite word
$\conf_0\conf_1\ldots$ of configurations such that 
$\conf_i\plainmovesto{}\conf_{i+1}$
for $i\geq 0$. 
%
%
We use $\run[i]$ to denote $\conf_i$.
We say that $\run$ is a
$\conf$-run if $\run[0]=\conf$.
%
We use $\runsetof\conf$ to denote the set of $\conf$-runs.
A {\it path} $\pth$ is a finite prefix of a run, and a $\conf$-path is a finite
prefix of a $\conf$-run. 
%
%
%
We use the standard notation $\conf\modelswrt\ts\phi$ 
to represent that $\conf$ satisfies the $\ctlstar$ \textit{state} 
formula $\phi$ and $\run \modelswrt\ts \phi$ to 
 mean that $\run$ satisfies the \textit{path}\footnote{We term infinite sequences as runs and finite sequences as paths. However, traditionally, $\ctlstar$ refers to properties of infinite-sequences (our runs) as \emph{path}-formulae.} formula $\phi$.
 We refer the reader to \cite{CGP:book} for details of CTL. 


For $\conf\in\confset$ and $\confs\subseteq\confset$, we say that $\confs$ is 
{\it reachable} from $\conf$, denoted $\conf\modelswrt\ts\ef\confs$,
if there is a $\conf$-run $\rho$ such that $\run[\ii]\in\confs$ for some $\ii$.
For $\kk\in\nat$, $\conf\in\confset$, and
$\confs\subseteq\confset$,
$\rho\modelswrt\ts\neventually{\kk}\confs$ says that 
$\rho$ reaches $\confs$ first at the $\kk^{\it th}$ step.
For $\sim~\in\set{<,\leq,=,\geq,>}$,
$\rho\modelswrt\ts\neventually{\sim\kk}\confs$ says that
$\rho\modelswrt\ts\neventually{\mm}\confs$ holds
for some $\mm:0\leq\mm\sim\kk$.
The statement $\rho\modelswrt\ts\nnxt{\kk}\confs$ says that $\rho$
visits $\confs$ at the $\kk^{\it th}$ step
(but possibly earlier).
%
%

\paragraph{Markov Chains}

A Markov chain $\mchain$ is a pair
$\mchaintuple$ where $\confset$ is a (potentially infinite) set
of {\it configurations}, and 
$\funtype{\pmtrx}{\confset\times\confset}{\intrvlof01}$ is 
a transition probability matrix over $\confset$,
called the {\it probability matrix} of $\mchain$, 
i.e. $\pmtrx$ satisfies: $\forall\aelem\in\aset. \sum_{\belem\in\aset}\pmtrxof{\aelem,\belem}=1$.
%
A Markov chain $\mchain=\tuple{\confset,\pmtrx}$ 
induces an {\it underlying transition system},
denoted $\mctots\mchain$.
We define $\mctots\mchain:=\tuple{\confset,\plainmovesto{}}$,
where 
$\plainmovesto{}:=
\setcomp{\tuple{\conf,\pconf}}{\pmtrxof{\conf,\pconf}>0}$.
The underlying transition system has the same configuration set,
with transitions between configurations that 
have non-zero transition probability under $\mchain$.
This allows us to lift the temporal logic concepts defined above to Markov chains.

\paragraph{Probability Measures}
\label{measure:subsection}
Consider a Markov chain $\mchain=\mchaintuple$.
%
The {\it probability} of taking path $\pth$
is the product of single step probabilities 
along $\pth$:
%
$$\probof\mchain\pth:=\prod_{i=0,...,\lengthof\pth-1}\pmtrxof{\pth[\ii],\pth[\ii+1]}$$
For a configuration $\conf$, we adopt the usual probability space
on $\conf$-runs with the $\sigma$-algebra over cylindrical sets
starting from $\conf$ (see \cite{KSK:book,PMC:book} for details).
%
For path formula $\phi$, we define $\measure{\mchain}\left(\conf\models\phi\right)= 
\measureof{\mchain}{\{\rho \in \runsetof\conf ~|~ \run\modelswrt\mchain\phi\}}$ (which is measurable
by \cite{Vardi:probabilistic}),
e.g. given a set $\fconfs\subseteq\confs$, 
$\measure{\mchain}\left(\conf\models\eventually\fconfs\right)$
is the measure of $\conf$-runs which reach $\fconfs$.
%
%
If $\measure{\mchain}\left(\conf\models\phi\right)=1$ the we say that
{\em almost all $\conf$-runs of $\mchain$ satisfy $\phi$}.
Following the literature, we say that $\conf\modelswrt\mchain\phi$ holds
{\em almost surely} ({\em almost certainly}), or that $\phi$ holds almost surely from $\conf$.

\section{Concurrent Programs}
\label{section:concprograms}
A (concurrent) program consists of a set of {\it processes} that run in parallel
and communicate through a set of {\it shared variables}.
The operation of the program is controlled by a 
central {\it scheduler} that selects the processes to execute one after the
other.
%
%
%
We assume a finite set $\procset$ of processes that share
a set $\varset$ of variables.
Fig.~\ref{planguage:fig} gives the grammar for a 
small but general assembly-like language that we use for defining
the syntax of concurrent programs.
A program instance, $\prog$ is described by 
a set of shared variables, \lstinline[style=splgrammar]{var$^*$},
followed by the codes of the processes, \lstinline[style=splgrammar]{(proc reg$^*$ instr$^*$)$^*$}.
Each process $\proc\in\procset$ has a finite 
set
$\regsetof\proc$ of (local) {\it registers}.
We assume that the sets of  registers
of the different processes are disjoint,
and define $\regsetof\prog:=\cup_{\proc\in\procset}\regsetof\proc$.
\begin{wrapfigure}[9]{r}{0.58\textwidth}
\vspace{-0.2cm}
\centering
\begin{lstlisting}[style=splgrammar]
  prog  ::= var$^*$(proc reg$^*$ instr$^*$)$^*$				
  instr ::= lbl : stmt
  stmt  ::= | var:=reg 
            | reg:=var 
            | reg:=expr 
            | reg:=CAS(var,reg,reg)  
            | if reg then lbl 
            | term
\end{lstlisting}
\vspace{-0.4cm}
\caption{A simple programming language.}
\label{planguage:fig}
\end{wrapfigure}
Each process
declares its set of  
registers, \lstinline[style=splgrammar]{reg$^*$}, followed by a sequence of instructions.
We assume that the data domain of $\varset$ and $\regsetof\prog$
is a finite set $\valset$, with
a special element $\zero\in\valset$.

\paragraph{Instructions}
An instruction $\instr$ is of the form $\mkinstr\lbl\stmt$ 
 where $\lbl$ is a unique (across processes)
label and $\stmt$ is a statement.
Labels represent program counters of processes and indicate
the instruction that the process executes the next time it is scheduled.
A {\it read/write} statement either
writes the value of a register to a shared variable,
reads the value of a shared variable into a register,
or updates the value of a register by evaluating an
expression.
We assume a set \lstinline[style=splgrammar]{expr} of expressions 
over constants and registers,
but not referring to the shared variables.
The $\cas$ statement is the standard {\it compare-and-swap} operation,
and \lstinline[style=splgrammar]{if}-statements have their usual interpretations. 
Iterative constructs such as \lstinline[style=customlang]{while} and \lstinline[style=customlang]{for},
as well as \lstinline[style=customlang]{goto}-statements,
can be encoded with branching \lstinline[style=splgrammar]{if}-statements as usual.
The \lstinline[style=customlang]{fence} statement, that flushes the contents of the 
buffer of the process,
can be simulated using the $\cas$ statement.
The statement $\terminated$ will cause the process
to terminate its execution.
%
%
Sometimes, we will refer to an instruction by its statement,
e.g. the instruction 
\lstinline[style=customlang]{r:=x}, (where \lstinline[style=customlang]{r} is a register and \lstinline[style=customlang]{x} is a shared variable) a {\it read} instruction, similarly for a {\it write} instruction, etc.
Semantics of these instructions 
are explained through a set of inference rules
in Sec.~\ref{semantics:section}.

\paragraph{Labels} We define $\lblsetof\proc$ to be the set of labels that occur
in the code of the process $\proc$, and
define
$\lblsetof\prog:=\cup_{\proc\in\procset}\lblsetof\proc$.
We assume that  $\terminated$ has the label $\termlblof\proc$.
We define $\instrsetof\proc$ to be the set of instructions
occurring in $\proc$, and
define $\instrsetof\prog:=\cup_{\proc\in\procset}\instrsetof\proc$.
For instruction $\instr$  of the form $\mkinstr\lbl\stmt$
we define $\labelingof\instr:=\lbl$ and
$\stmtof\instr:=\stmt$.
Abusing notation, we also define $\stmtof\lbl:=\stmt$.
For a process $\proc\in\procset$ 
instruction $\instr\in\instrsetof\proc$, with 
$\stmtof\instr\neq\terminated$,
we define $\nextof\instr$ to be the (unique)
instruction next to $\instr$ in the code of $\proc$.
%
%
For an instruction 
$\mkinstr{\lbl_1}{({\tt if} \;\areg\; {\tt then} \; \lbl_2)}$,
we assume, without loss of generality\footnote{We make the restriction
for technical convenience. The case where $\lbl_1=\lbl_2$ do not introduce conceptual
difficulties. However, it simplifies the presentation by eliminating some corner cases when we define probability measures (Sec.~\ref{basics:section}) 
and when we introduce our cost model (Sec.~\ref{cost:section}).}, 
that
$\lbl_1\neq\lbl_2$.

\paragraph{Scheduler}
%
%
%
%
The scheduler selects the process from $\procset$ to run next. 
The operational model for classical TSO \cite{OSS2009} 
uses a non-deterministic scheduler.
We adopt a scheduler that 
selects the next process probabilistically.
The scheduler policy is defined by a function $\weight$:
$\weightof{\proc}\in\nat$ denotes the scheduling weight assigned to
to the process $\proc$.
If $\proc$ is \textit{enabled} (i.e. the process can execute the 
next instruction, formally defined in Sec.~\ref{semantics:section}) 
then $\proc$ is scheduled
at the next step with a probability that is proportional to $\weightof\proc$.

\section{Operational Semantics}
\label{semantics:section}

The operational model for classical TSO \cite{OSS2009} describes the semantics as a transition system. We also take an operational approach. However, we differ in a fundamental aspect: classical TSO models choice between transitions as \textit{non-deterministic choice}. We on the other hand, model this as \textit{probabilistic} choice, to get a system called as Probabilistic TSO (PTSO for short). Adding probabilities induces a Markov chain, which governs the behaviours of PTSO. 

%
%
A program is described by a pair: the set of processes, $\procset$ and
the scheduler policy $\weight$.
In this section, we fix such a program $\prog=\progtuple$.
%
We develop the operational semantics of $\prog$ under PTSO as an infinite-state Markov chain
$\mcdenotationof\prog:=\mchaintupleof{\prog}$.
%
%
%
We begin by defining the set of configurations $\confsetof\prog$
(Sec.~\ref{confs:semantics:section}).
Then we describe the behavior of $\prog$
under classical TSO using a transition system
$\tsdenotationof\prog$
(Sec.~\ref{TSO:semantics:section});
Finally, we extend the transition system to a
Markov chain
$\mcdenotationof\prog$ by giving probability distributions that 
define govern process scheduling, and
memory updates.

\subsection{Configurations}
\label{confs:semantics:section}
The central feature of TSO is the \textit{store buffer}: a FIFO buffer in which pending write operations are queued as messages. 
The semantics equips each process $\proc\in\procset$ with an unbounded buffer,
here called the {\it $\proc$-buffer}, that carries pending write operations
issued by $\proc$, but that have yet not reached the shared memory.
%
%

A configuration, $\conftuple$, describes four attributes: 
a labeling state ($\labeling$),
a register state ($\regstate$),
a buffer state ($\bufferstate$), and
a memory state ($\memstate$).
We use $\confsetof\prog$ to denote the set of configurations of 
$\prog$. 

%
A {\it labeling state} is a function
$\funtype\labeling\procset{\lblsetof\prog}$
that defines, for
$\proc\in\procset$, the label
$\labelingof\proc\in\lblsetof\proc$ of the next instruction 
to be executed by $\proc$.

A {\it register state} is a function
$\funtype{\regstate}{\regsetof\prog}\valset$ that
maps each register
$\areg\in\regsetof\prog$, to its current value $\regstateof\areg\in\valset$.
For an expression $\expr$,
we use $\regstateof\expr$ to denote the evaluation of $\expr$
against the register state $\regstate$.

%
A {\it single-buffer state} $\word$ is a word
in $\wordsover{(\varset\times\valset)}$,
describing the content of
the $\proc$-buffer for some process $\proc\in\procset$.
The buffer contains a sequence of pending {\it write messages}, i.e.
pairs of form $\tuple{\xvar,\val}$ representing a write to $\xvar$, 
with value $\val$.
%
%
A {\it buffer state} is a function 
$\funtype\bufferstate\procset{\wordsover{(\varset\times\valset)}}$
that defines, for each process $\proc\in\procset$, a single-buffer state
describing the content of the $\proc$-buffer.

A {\it memory state} is a function
$\funtype\memstate\varset\valset$ that
assigns to each variable $\xvar\in\varset$ its current
value $\memstateof\xvar\in\valset$ in the shared memory.

\makeatletter
  \newenvironment{lines}{\endgroup}{\begingroup\def\@currenvir{lines}}
\makeatother

\tikzset{background rectangle/.style={fill=none
}}
\begin{figure}[h]
\centering
\small
\resizebox{\textwidth}{!}{
\begin{tikzpicture}[codeblock/.style={line width=0.5pt, inner xsep=0pt, inner ysep=5pt}  , show background rectangle]
\node[codeblock] (init) at (current bounding box.north west) {
$
\def\arraystretch{1.5}
\begin{array}{c}
  \rowcolor{green!10!white}
  \begin{array}{ccc}
    \begin{array}{c}
      \rulename{write} \\
      \inferrule{
          \stmtof{\labelingof\proc}= (\xvar\assigned\areg) \\\\
          \pbufferstate=\updatefun\bufferstate\proc{\tuple{\xvar,\regstateof\areg}\app\bufferstateof\proc} \\\\
          \plabeling=\updatefun\labeling\proc{\nextof{\labelingof\proc}} 
      }{
        \tuple{\labeling,\regstate,\bufferstate,\memstate} \procmovesto{\proc}
        \tuple{\plabeling,\regstate,\pbufferstate,\memstate}
      }
    \end{array}
    &
    \begin{array}{c}
      \rulename{read} \\
      \inferrule{
        \stmtof{\labelingof\proc}= (\areg\assigned\xvar) \\\\
        \fetchvalof\xvar{\bufferstateof\proc}{\memstateof\xvar}=\val \\
        \pregstate=\updatefun\regstate\areg\val \\\\
        \plabeling=\updatefun\labeling\proc{\nextof{\labelingof\proc}}
      }{
        \tuple{\labeling,\regstate,\bufferstate,\memstate} \procmovesto{\proc}
        \tuple{\plabeling,\pregstate,\bufferstate,\memstate}
      }  
    \end{array}
    &  
    \begin{array}{c}
    \rulename{expr} \\
    \inferrule{
      \stmtof{\labelingof\proc}= (\areg\assigned\expr) \\\\
      \pregstate=\updatefun\regstate\areg{\regstateof\expr} \\\\
      \plabeling=\updatefun\labeling\proc{\nextof{\labelingof\proc}}
    }{
      \tuple{\labeling,\regstate,\bufferstate,\memstate} \procmovesto{\proc}
      \tuple{\plabeling,\pregstate,\bufferstate,\memstate}
    }
    \end{array}
  \end{array} \\ [1cm]
  \rowcolor{green!10!white}
  \begin{array}{cc}
  \begin{array}{c}
      \rulename{CAS-true} \\
      \inferrule{
        \stmtof{\labelingof\proc}= (\breg\assigned\casof\xvar{\areg_1}{\areg_2}) \\
        \memstateof{\xvar} = \regstateof{\areg_1} \\\\
        \pregstate=\updatefun\regstate\breg\true \\
        \bufferstateof\proc=\emptyword \\
        \pmemstate=\updatefun\memstate\xvar{\regstateof{\areg_2}} \\\\
        \plabeling=\updatefun\labeling\proc{\nextof{\labelingof\proc}}
      }{
        \tuple{\labeling,\regstate,\bufferstate,\memstate} \procmovesto{\proc}
        \tuple{\plabeling,\pregstate,\bufferstate,\pmemstate}
      }
    \end{array}
    & 
    \begin{array}{c}
      \rulename{CAS-false} \\
      \inferrule{
        \stmtof{\labelingof\proc}= (\breg\assigned\casof\xvar{\areg_1}{\areg_2}) \\
        \memstateof{\xvar} \neq \regstateof{\areg_1} \\\\
        \pregstate=\updatefun\regstate\breg\false \\
        \bufferstateof\proc=\emptyword \\\\
        \plabeling=\updatefun\labeling\proc{\nextof{\labelingof\proc}}
      }{
        \tuple{\labeling,\regstate,\bufferstate,\memstate} \procmovesto{\proc}
        \tuple{\plabeling,\pregstate,\bufferstate,\memstate}
      }  
    \end{array}
  \end{array} \\ [1cm]
  \rowcolor{green!10!white}
  \begin{array}{cccc}
    \begin{array}{c}
      \rulename{if-true} \\
      \inferrule{
        \stmtof{\labelingof\proc}=({\tt if} \;\areg\; {\tt then} \; \lbl) \\\\
        \regstateof\areg=\true \\\\
        \plabeling=\updatefun\labeling\proc{\lbl}
      }{
        \tuple{\labeling,\regstate,\bufferstate,\memstate} \procmovesto{\proc}
        \tuple{\plabeling,\regstate,\bufferstate,\memstate}
      }
    \end{array} &
    \begin{array}{c}
      \rulename{if-false} \\
      \inferrule{
        \stmtof{\labelingof\proc}=({\tt if} \;\areg\; {\tt then} \; \lbl) \\\\
        \regstateof\areg=\false \\\\
        \updatefun\labeling\proc{\nextof{\labelingof\proc}}
      }{
        \tuple{\labeling,\regstate,\bufferstate,\memstate} \procmovesto{\proc}
        \tuple{\plabeling,\regstate,\bufferstate,\memstate}
      }
    \end{array} &
    \begin{array}{c}
      \rulename{proc}\\
      \inferrule{
        \tuple{\labeling,\regstate,\bufferstate,\memstate} \procmovesto{\proc}
        \tuple{\plabeling,\pregstate,\pbufferstate,\pmemstate}
      }{
        \tuple{\labeling,\regstate,\bufferstate,\memstate} \procmovesto{}
        \tuple{\plabeling,\pregstate,\pbufferstate,\pmemstate}
      }
    \end{array} &
    \begin{array}{c}
      \rulename{disabled} \\
      \inferrule{
        \conf \text{ is disabled}
      }{
        \conf \procmovesto{} \conf
      } 
    \end{array}
  \end{array} \\ [1cm]
  \rowcolor{orange!15}
  \begin{array}{ccc}
    \begin{array}{c}
      \rulename{empty-update} \\
      \inferrule{~}{
        \tuple{\labeling,\regstate,\bufferstate,\memstate} \updatemovesto{\epsilon}
        \tuple{\labeling,\regstate,\bufferstate,\memstate}
      }
    \end{array} &
    \begin{array}{c}
      \rulename{single-update}\\
      \inferrule{
        \pbufferstateof\proc=\word\app\tuple{\xvar,\val} \\
        \ppbufferstate=\updatefun\pbufferstate\proc\word \\\\
        \ppmemstate=\updatefun\pmemstate\xvar\val \\\\
        \tuple{\labeling,\regstate,\bufferstate,\memstate} \updatemovesto{\procseq}
        \tuple{\labeling,\regstate,\pbufferstate,\pmemstate}
      }{
        \tuple{\labeling,\regstate,\bufferstate,\memstate} \updatemovesto{\proc\app\procseq}
        \tuple{\labeling,\regstate,\ppbufferstate,\ppmemstate}
      }
    \end{array}  &
    \begin{array}{c}
      \rulename{update} \\
      \inferrule{
        \tuple{\labeling,\regstate,\bufferstate,\memstate} \updatemovesto{\procseq}
        \tuple{\labeling,\regstate,\pbufferstate,\pmemstate}
      }{
        \tuple{\labeling,\regstate,\bufferstate,\memstate} \updatemovesto{}
        \tuple{\labeling,\regstate,\pbufferstate,\pmemstate}
      }
    \end{array}
  \end{array} \\ [1cm]
  \begin{array}{c}
    \rulename{Full-TSO} \\
    \inferrule{
      \tuple{\labeling,\regstate,\bufferstate,\memstate} \procmovesto{}
      \tuple{\plabeling,\pregstate,\pbufferstate,\pmemstate} \\
      \tuple{\plabeling,\pregstate,\pbufferstate,\pmemstate} \updatemovesto{}
      \tuple{\pplabeling,\ppregstate,\ppbufferstate,\ppmemstate}
    }{
      \tuple{\labeling,\regstate,\bufferstate,\memstate} 
      \progmovesto\prog{}
      \tuple{\pplabeling,\ppregstate,\ppbufferstate,\ppmemstate}
    }
  \end{array}
\end{array}$
};
\end{tikzpicture}
}
\vspace*{-5mm}
\caption{The classical TSO semantics: process transitions (green), update transitions (orange) and overall transition (\rulename{Full-TSO})}.
\label{TSO:semantics:fig}
\end{figure}
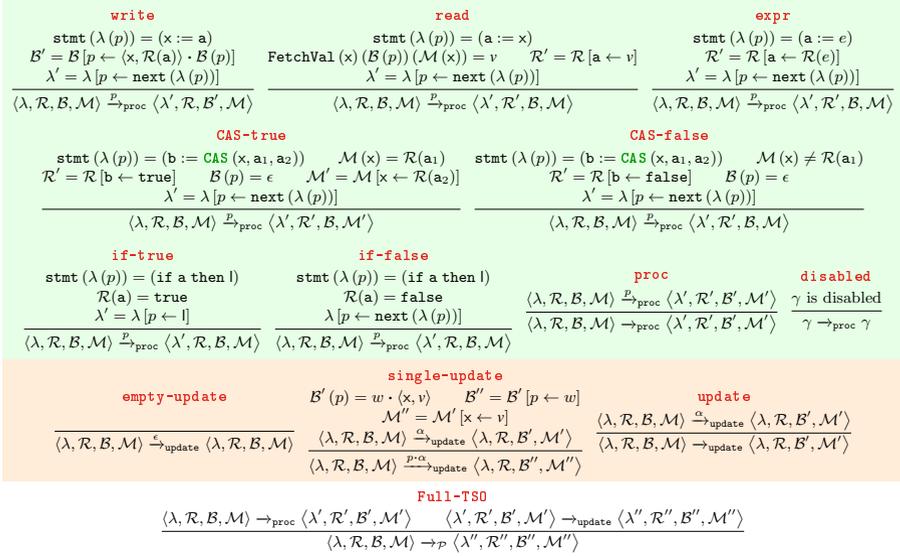

%

Consider a configuration $\conf=\conftuple$.
We say that $\conf$ is {\it plain} if
$\bufferstateof\proc=\emptyword$ for all $\proc\in\procset$,
i.e., all the buffers in $\conf$ are empty.
We use $\plainconfsetof\prog$  
to denote the set of plain configurations of $\prog$.
Notice that $\plainconfsetof\prog\subseteq\confsetof\prog$ and that
$\plainconfsetof\prog$ is finite.
For a label $\lbl\in\lblsetof\prog$, we write $\lbl\in\conf$
if $\labelingof\proc=\lbl$ for some $\proc\in\procset$.
We define 
$\lblconfsetof\prog\lbl:=\setcomp{\conf\in\confsetof\prog}{\lbl\in\conf}$,
i.e., configurations in which $\lbl$ occurs.
For a configuration $\conf=\conftuple$
we define the size of $\conf$ by 
$\sizeof\conf:=\sum_{\proc\in\procset}\sizeof{\bufferstateof\proc}$,
i.e., it is the total number of messages in the buffers in $\conf$.
For
$\sim\in\set{<,\leq,=,\geq,>}$,
we define
$\numconfsetof\prog{\sim\ell}:=
\setcomp{\conf\in\confsetof\prog}{\sizeof\conf{\sim\ell}}$.
, i.e. configurations where
the total number of messages, $\mm$, relates to $\ell$ by $\mm\sim\ell$.
\subsection{The Classical TSO Semantics}
\label{TSO:semantics:section}
We recall the classical semantics of TSO, using a transition system 
$\tsdenotationof\prog=\tuple{\confsetof\prog,\progmovesto\prog{}}$.
We define the transition relation 
$\progmovesto\prog{}$ 
through the set of inference rules in Fig.~\ref{TSO:semantics:fig}.
The relation $\progmovesto\prog{}$ is the composition of two
relations:
the relation 
$\procmovesto{}$ describes the processes' execution steps,
and the relation $\updatemovesto{}$
describes memory updates, where pending writes are propagated to the memory.
\subsubsection{Process Transitions}
We define the process transition relation
$\procmovesto{}:=
\cup_{\proc\in\procset}\procmovesto\proc$ 
as a union of relations each corresponding to one process
(the rule \rulename{proc}).
The inference rules defining $\procmovesto\proc$, 
for a process $\proc\in\procset$
are depicted in Fig.\ref{TSO:semantics:fig}.
Each rule corresponds to one step performed by $\proc$.
After executing an instruction,
$\proc$ will move on to the next instruction in its code.
It executes the latter instruction when again selected by the scheduler.

A \rulename{write} instruction ($\xvar\assigned\areg$) assigns
the value of the local register $\areg$ to 
the shared variable $\xvar$.
The process appends a write
message consisting of $\xvar$ together 
with the value $\regstateof\areg$ of $\areg$, to the head of the $\proc$-buffer.
A \rulename{read} instruction, ($\areg\assigned\xvar$), 
assigns the value of the shared variable $\xvar$
to the local register $\areg$.
The value of $\xvar$ is either fetched from the $\proc$-buffer
({\it read-own-write}), or from the shared
memory ({\it read-from-memory}).
We capture both cases in one inference rule,
using the function $\fetchval$ defined as follows.
Let $\word$ be the contents of the $\proc$-buffer.
We write $\xvar\in\word$ if $\tuple{\xvar,\val}\in\word$ for some
$\val\in\valset$, and write $\xvar\not\in\word$ otherwise.
%
%
%
We define 
(i) $\fetchvalof\xvar\word\memstate:=\val$ 
if $\xvar\in\word$ and 
$\word=\word_1\app\tuple{\xvar,\val}\app\word_2$
with $\xvar\not\in\word_1$; 
and (ii) define $\fetchvalof\xvar\word\memstate:=\memstateof{\xvar}$ if 
$\xvar\not\in\word$.
In case (i), the value of $\xvar$ is taken from the
latest $\xvar$-message from the $\proc$-buffer.
In case (ii), no $\xvar$-messages exist
in the $\proc$-buffer, and the value is 
read from the shared memory.
%
%

%
The instruction $\breg\assigned\casof\xvar{\areg_1}{\areg_2}$ checks 
whether the $\proc$-buffer is empty and
the value of the shared variable $\xvar$ is equal to
the value of the register $\areg_1$.
If {\it yes}, we assign atomically the value of the register $\areg_2$
to $\xvar$, and assign the value $\true$ to $\breg$
 (the rule \rulename{CAS-true}).
If the value of $\xvar$ is different from
the value of $\areg_1$ then we do not change the value of $\xvar$,
but assign the value $\false$ to $\breg$
(the rule \rulename{CAS-false}).
If the $\proc$-buffer is not empty then $\proc$ is disabled
in the current configuration.
We define the set of disabled processes at configuration $\conf$:
\begin{equation*}
\resizebox*{\textwidth}{!}{
	$\small\disabledof\conf:=
	\setcomp{\proc~}
	{
	(\stmtof{\proc}=\terminated)
	\lor
	((\stmtof{\proc}=(b\assigned\casof\xvar{\areg_1}{\areg_2}))\land
	(\bufferstateof\proc\neq\emptyword))}$
}
\end{equation*}


In other words, it is the set of processes that are disabled in 
$\conf$ either because they have terminated or because
they are about to perform a $\cas$ operation and their buffers
are not empty.
We say that $\proc$ is {\it disabled} in $\conf$ if
$\proc\in\disabledof\conf$, and
that $\conf$ is {\it disabled} if 
all the processes are disabled in $\conf$.
If a process (resp. configuration) is not disabled then it is enabled.
%
%
If $\conf$ is disabled, we
make a dummy transition that does not change $\conf$
(the rule \rulename{disabled})\footnote{The latter transition is not strictly needed, but it is included
for technical convenience.}.
Notice that if $\conf\procmovesto\proc\pconf$ then there is 
unique process $\proc\in\procset$ 
such that $\conf\procmovesto\proc\pconf$.

\subsubsection{Update Transitions}
\label{update:transitions:sub:section}
Between two process transitions, the system
may perform a (possibly empty) sequence of update steps.
The rule \rulename{empty-update} describes an empty update step.
Each \rulename{single-update} step pops one write message at the end of 
the $\proc$-buffer for some process $p$ and uses it to update the memory.
The \rulename{update} rule captures the effect of a sequence of such
\rulename{single-update} steps.
We define the update transition relation
$\updatemovesto{} :=
\cup_{\procseq\in\wordsover\procset}\updatemovesto\procseq$ 
as a union of relations each corresponding to a given sequence 
of update steps.
The word $\procseq$ gives the sequence of processes that perform the 
updates.
%
%
%
%
%
The net effect is that 
the system 
(i) pops a sequence of ( possibly empty) suffixes
from the buffer of each process,
(ii) shuffles these into one sequence, and
(iii) uses the resulting sequence to update the memory.
Notice that each selection of possible suffixes in step (i)
may result in several different 
sequences due to multiple interleavings in step (ii).
Observe that  $\progmovesto\prog{}$ is deadlock-free, i.e., for each
configuration $\conf\in\confset$, there is at least one configuration
$\pconf\in\confset$ such that
$\conf\progmovesto\prog{}\pconf$.

\subsection{Adding Probabilities: PTSO}
\label{mc:semantics:section}
We define the Markov Chain $\mcdenotationof\prog=\mchaintupleof{\prog}$.
The set $\confsetof\prog$ of configurations is defined as above.
The probability matrix $\progpmtrx\prog$ is defined as the composition
of two probability distributions:
(i)
the {\it process} probability distribution $\procpmtrx$
(ii) 
the {\it update} probability distribution $\updatepmtrx$
which add probabilities to
the process transition relation $\procmovesto{}$, and
the update transition relation $\updatemovesto{}$
respectively.

\subsubsection{The Process Probability Distribution: the Scheduler}
At each program step ($\progmovesto\prog{}$), a process is
selected for execution 
according to a probability given by the scheduler.
In a configuration
$\conf$, the scheduler selects an enabled  process 
$\proc\in\enabledof\conf$
with a probability that reflects the relative weight of $\proc$ compared
to those of the other enabled processes, $\rweightof\conf\proc$:
%
\begin{equation}
	\rweightof\conf\proc = 
	\begin{cases*}
	    0 & if  $\proc\in\disabledof\conf$  \\
	    \frac{\weightof{\proc}}{\sum_{\pproc\in\enabledof\conf}\weightof{\pproc}} & if $\proc\in\enabledof\conf$
	 \end{cases*}
\end{equation}
%
This gives the probability that $\proc$ 
to execute in the next step from $\conf$. 
For configurations $\conf$  and $\pconf$,
with $\conf\procmovesto\proc\pconf$,
we define 
$\procpmtrxof{\conf,\pconf}:=\rweightof\conf\proc$.
In other words, we move from $\conf$ to $\pconf$ with a probability that is given
by the relative weight of $\proc$ in $\conf$.
We define $\progpmtrxof\prog{\conf,\pconf}:=0$ if 
$\conf \not{\procmovesto{}}\pconf$.
To account for the case where all the processes are disabled in $\conf$,
we define $\procpmtrxof{\conf,\conf}:=1$
if $\conf$ is disabled.

\paragraph{Faithfulness}

Our model uses a scheduling policy that assigns a fixed scheduling weight, 
$\weightof\proc$, to each process $\proc$ in the system. 
This is a case of
{\it memoryless} scheduling, i.e., the probability distribution 
over processes does not depend on the execution history. 
However, we can relax this constraint to allow for any scheduling policy
that satisfies the {\it faithfulness condition}:
\[
\forall\proc \in\procset~~ \rweightof\conf\proc = 0 \iff \proc \in \disabledof\conf
\]
In words, at each step, each enabled process should be 
scheduled with non-zero probability.
A scheduler that assigns scheduling weights such that the above condition
holds is said to be a faithful scheduler.


%

\paragraph{Schedulers with memory} The above criterion
allows for schedulers that are more refined as compared to the memoryless scheduler.
As an example, on implementations of TSO, processes are often scheduled for 
multiple consecutive steps since unnecessary context switching wastes processor resources.
To reflect this detail, we can consider a scheduler that assigns a higher probability 
to the previously scheduled process, $p_\textit{prv}$. 
For some choice of constant weights, $\weight$, 
we can define a new choice of weights $\weight'$ where $\lambda > 1$ is some parameter.
\begin{equation*}
	\weight'(\proc) = \weightof\proc ~~\text{ if }~~ \proc \neq p_\textit{prv} ~~\text{ and } ~~\lambda\cdot\weightof\proc ~~\text{otherwise}
\end{equation*}
In this case, $p_{prv}$ is re-scheduled with a weight which is larger by a factor of $\lambda$. 
A larger $\lambda$ implies a stronger tendency to re-schedule a process.
This scheduling policy still satisfies faithfulness. 
One can extend this by formulating more intricate policies, e.g. 
ones that account for $k$ previous steps.

To better illustrate the concerns and challenges 
of verification, we continue to adopt the simple (memoryless) scheduler proposed earlier. 
However, we emphasize that our results extend to faithful schedulers.

\noindent{\bf{The Update Probability Distribution: the Memory update policy}}
Between the process steps, pending messages from the store buffers are propagated 
to the shared memory (the update transition).
The details of this write propagation are implementation-specific, with policies tuned towards
system performance.
%
Classical TSO models this update propagation non-deterministically.
We, on the other hand, consider a probabilistic update policy.
In a similar manner to the scheduling probabilities, the
update probability distribution defines the probability by which
a configuration $\conf$ reaches another configuration $\pconf$
through an update step ($\updatemovesto{}$).
Recall that an update step consists of a sequence of (single) update operations.
The number of possible update sequences from $\conf$ is finite
since the sizes of each buffer is finite.
In our model, we assume that the update distribution is the uniform distrbution
over all possible update sequences. 
%
We note that starting from $\conf$, different update sequences can lead to 
the same configuration $\pconf$.
The reason is that
different shufflings of the selected suffixes 
(see Sec.~\ref{update:transitions:sub:section})
may lead to the same memory state.
To reflect this, for configurations
$\conf$ and $\pconf$, we define
$
\updatepmtrxof{\conf,\pconf}:=
\frac
{\sizeof{\setcomp{\procseq}{\conf\updatemovesto\procseq\pconf}}}
{\sizeof{\setcomp{\procseq}{\exists\ppconf.\;\conf\updatemovesto\procseq\ppconf}}}$, 
i.e. the fraction of update sequences that lead to the configuration $\pconf$.

\paragraph{Left-Biasedness}
Though we adopt a specific update distribution, we provide a generic condition on 
that update policy that is sufficient for our results to hold. 
We call this the left-biasedness property. 
Here we provide an intuitive description of left-biasedness and defer the
formal definition to Sec.~\ref{cost:section}.

Intuitively, left-biasedness requires that for sufficiently large configurations, 
the probability that 
the configuration size reduces in a single $\progmovesto\prog{}$ step is 
strictly greater than $p$ for some $p > \frac{1}{2}$.
Left-biasedness allows a wide class of more refined 
scheduling policies, e.g., where no message propagation is performed
when the number of messages is smaller than a certain value, or where only the messages
inside the buffers of some (probabilistically selected) processes are propagated.

Though our results apply more generally to models characterized by
faithfulness (scheduler policy), and left-biasedness (update policy), 
we continue to adopt the fixed-weight (memoryless) scheduler and
uniform update policy for reasons described above.

\noindent{\bf{The Full Probability Distribution}}. 
We combine the process and update
probability distributions, 
to derive the probability matrix $\prmpmtrx\prog$,
and thus obtain the Markov chain $\mcdenotationof\prog$.
Consider configurations $\conf$ and $\pconf$ where
$\conf\progmovesto\prog{}\pconf$.
Let $\ppconf$ be the unique configuration such that 
$\conf\procmovesto{}\ppconf\updatemovesto{}\pconf$.
Then, we define
$\progpmtrxof\prog{\conf,\pconf}:=\procpmtrxof{\conf,\ppconf}\cdot\updatepmtrxof{\ppconf,\pconf}$.

\begin{restatable}{lem}{lemTSOProbDist}
\label{tso:prob:dist:lemma}
$\progpmtrx\prog$ is a prob. distribution on $\confsetof\prog$; hence, $\mcdenotationof\prog$ is a Markov chain.
\end{restatable}

\section{PTSO: Concepts and Properties}
\label{basics:section}
Now, we intuit some  concepts underlying Probabilistic TSO
and its properties.

\smallskip 
\noindent{\bf{PTSO Refines Classical TSO}}. 
%
After introducing $\tsdenotationof\prog$ and $\mcdenotationof\prog$ in  Sec. \ref{semantics:section}, we s.t.
they are closely related; 
$\tsdenotationof\prog$ is the underlying transition system of $\mcdenotationof\prog$.

\begin{restatable}{lem}{lemMChainTS}
\label{program:mchain:ts}
$\mctots{\left(\mcdenotationof\prog\right)}=
\tsdenotationof\prog$ for any program $\prog$.
\end{restatable}
In particular, this means that the PTSO system $\mcdenotationof\prog$ is a refinement of $\tsdenotationof\prog$:
a behaviour is observed in $\tsdenotationof\prog$ iff it is seen in $\mcdenotationof\prog$ with non-zero probability.
Whenever the context is clear, we write $\prog$
instead of $\tsdenotationof\prog$, $\mcdenotationof\prog$.

%

\smallskip 
\noindent{\bf{Label Reachability}}. 
We formulate our verification problems in terms
of reachability to instruction labels.
%
To simplify the notation,
we identify a label $\lbl\in\lblsetof\prog$ with the set
$\lblconfsetof\prog\lbl$ of configurations
in which $\lbl$ occurs.
We say that ``$\lbl$ is reachable'' rather than 
``$\lblconfsetof\prog\lbl$ is reachable'', and
write $\eventually\lbl$ instead of 
$\eventually\setcomp{\conf\in\confsetof\prog}{\lbl\in\conf}$. 
In \cite{DBLP:journals/lmcs/AbdullaABN18,DBLP:conf/popl/AtigBBM10} 
the authors show that label reachability from a plain configuration
is decidable. The following lemma,
generalizes this to the case where the source configuration need not be plain and destination can be
a particular plain configuration.

\begin{restatable}{lem}{lemTSOConfReachability}
\label{tso:conf:reachability:lemma}
For a program $\prog$, a configuration $\conf\in\confsetof\prog$, and 
a plain configuration $\pconf\in\plainconfsetof\prog$,   
it is decidable whether $\conf\progmovesto\prog*\pconf$.
%
\end{restatable}

%
Extending this, we have Lemma~\ref{tso:lbl:reachability:lemma}: 
we can query whether $\conf\progmovesto\prog*\pconf$ for each 
$\pconf\in\lblconfsetof\prog\lbl \cup \plainconfsetof\prog$. 
Decidability of Lemma~\ref{tso:lbl:reachability:lemma} 
follows since $\plainconfsetof\prog$ is finite and the 
subroutine is decidable by Lemma \ref{tso:conf:reachability:lemma}.


\begin{restatable}{lem}{lemTSOLblReachability}
\label{tso:lbl:reachability:lemma}
For a program $\prog$, a configuration $\conf\in\confsetof\prog$,
and a label $\lbl\in\lblsetof\prog$, it is decidable
whether $\conf\progmovesto\prog*\lbl$.
\end{restatable}

\subsection{Left-Orientedness and Attractors}
\label{plain:attractor:subsection}
%
We show that the set of plain configurations $\plainconfsetof\prog$ set has an {\it attractor} property in the sense
of \cite{AbdullaHM07}. 
In our setting, this means that any run of $\mcdenotationof\prog$
almost surely visits $\plainconfsetof\prog$ infinitely often.

\paragraph{Small and large configurations}
To arrive at this result, we consider a generalization of plain configurations,
called {\it small} configurations,
denoted $\smallconfsetof\prog$.
$\smallconfsetof\prog$ consists of
configurations with a small number of messages inside
their buffers.
Concretely, a configuration $\conf$ is small if $\sizeof\conf\leq4$, i.e., the total number
of messages inside the buffers does not exceed $4$.
\footnote{This value is an artifact of the probabilistic policies we have adopted in Sec. \ref{semantics:section}}
We define the set of {\it large configurations} by
$\largeconfsetof\prog:=\confsetof\prog-\smallconfsetof\prog =\numconfsetof\prog{\geq5}$.
We show that the Markov chain $\mcdenotationof\prog$ is {\it left-oriented}
in the sense of \cite{10.5555/1126294.1126298}.
That is, for any large configuration 
$\conf\in\largeconfsetof\prog$, 
the expected change in configuration size for a single $\progmovesto\prog{}$ step
is negative.

%

\newcommand\procL{\texttt{procL}}
\newcommand\procR{\texttt{procR}}

\paragraph{An illustrative example}
\begin{wrapfigure}[3]{r}{0.35\textwidth}
\centering
\vspace{-0.1cm}
\begin{tabular}{l || l}
  \begin{lstlisting}[style=customlang]
 0: x = 1
 1: goto 0
  \end{lstlisting}      
  &
  \begin{lstlisting}[style=customlang]
 2: x = 2
 3: a = x
 4: goto 2
  \end{lstlisting}
\end{tabular}
\end{wrapfigure}
\setlength{\columnsep}{5pt}%
We explain the 
update probability distribution
through the code snippet on the right.
%
To begin with let us only consider the process on the left ($\procL$).
It executes an infinite loop, writing 
\lstinline[style=customlang]{1} to variable \lstinline[style=customlang]{x}.
\newcommand{\bufsize}[2]{\texttt{#1}@\texttt{#2}}
Let us consider the evolution of the buffer-sizes of $\procL$,
i.e. the number of \lstinline[style=customlang]{(x,1)}
messages in the $\procL$-buffer.
%
Assume that on reaching label \texttt{0}, 
$\procL$ has $6$ messages in its buffer.
%
%
%
%
%
%
The $\progmovesto\prog{}$ step consists of a process transition, $\procmovesto{}$
followed by an update transition, $\updatemovesto{}$.
%
%
In the $\procmovesto{}$ step, the write increases the size
of the buffer by one, thus obtaining a buffer of size 7.
Following this the $\updatemovesto{}$ step may push any number of messages
to the memory.
Since the update policy chooses uniformly amongst possible update sequences, 
the resulting configuration has one amongst $\{0, \dots, 7\}$ messages
in the $\procL$-buffer, each occurring with an equal probability of $1/8$.
%
%
The next $\procmovesto{}$ step (a \lstinline[style=customlang]{goto}),
does not change the buffer size, but the 
$\updatemovesto{}$ step can still propagate messages.
The reasoning for the next steps follows similarly.

\paragraph{Comparison with other notions of fairness}
At each $\procmovesto{}$ step atmost one message is
added to the process buffers (when the process performs a write), however in the
following $\updatemovesto{}$ can still remove large number of messages.
Hence, from sufficient large configuration sizes, the system
has a tendency to move towards configurations with smaller buffer sizes.
Formally, we prove the following lemma, using the left-orientedness property 
mentioned earlier.
\begin{restatable}{lem}{lemFiniteAttractor}
  \label{finite:attractor:lemma}
  $\measureof\prog{\conf\models\always\eventually\plainconfsetof\prog}=1$ for
  all configurations $\conf\in\confsetof\prog$.
\end{restatable}
For the above example, PTSO
guarantees that the process on the right ($\procR$) eventually reads value
\lstinline[style=customlang]{1} into register \lstinline[style=customlang]{a}.
This follows since in a plain configuration, the buffer of $\procR$ is empty and hence 
it can read the value from the memory - this happens almost surely.
We highlight that other notions of fairness such as strong fairness in process scheduling
(discussed in \cite{DBLP:journals/corr/abs-2107-09930})
as well memory fairness \cite{Lahav2020MakingWM}, cannot provide this guarantee.
In particular, memory fairness from \cite{Lahav2020MakingWM}, would consider the execution 
which exactly alternates writes of both processes but $\procR$ reads before its own 
write is pushed memory to be fair and hence permissible.
\begin{center}
  \lstinline[style=customlang]{x = 1} ~~~ \lstinline[style=customlang]{x = 2} ~~~ \lstinline[style=customlang]{a = x}  
  ~{\color{gray}\footnotesize \texttt{// 2}  } ~~~ \lstinline[style=customlang]{x = 1} ~~~ \lstinline[style=customlang]{x = 2} ~~~ \lstinline[style=customlang]{a = x}  
  ~{\color{gray}\footnotesize \texttt{// 2}  } ~~~ \lstinline[style=customlang]{x = 1} ~~~ $\cdots$
\end{center}

%
%
%
%


\paragraph{B-Plain Configurations}
We can refine our analysis of the attraction property
enjoyed by the set $\plainconfsetof\prog$ of plain configurations.
We consider a subset of $\plainconfsetof\prog$ which we call
the set of {\it bottom plain configurations},
(or  {\it B-plain} configurations, for short),
denoted $\bplainconfsetof\prog$.
Intuitively,
a B-plain configuration is a member of 
a bottom strongly connected component in the graph 
of plain configurations.
Formally, a configuration $\conf\in\confsetof\prog$
is said to be {\it B-plain} if 
(i) $\conf\in\plainconfsetof\prog$,
and
(ii)
for any $\pconf\in\plainconfsetof\prog$, 
if $\conf\progmovesto\prog*\pconf$
then  $\conf'\progmovesto\prog*\conf$. 
%
Since any run of the system almost surely visits the set of 
$\plainconfsetof\prog$ infinitely often, it will also almost surely visit a
B-plain configuration infinitely often.
%
\begin{restatable}{lem}{lemBPlain}
$\measureof\prog{\conf\models\always\eventually\bplainconfsetof\prog}=1$ for
all configurations $\conf\in\confsetof\prog$.
\label{bplain:finite:attractor:lemma}
\end{restatable}

%

%

\section{Qualitative (Repeated) Reachability}
\label{qual:reachability:section}
\begin{figure}[H]
	\centering
\begin{mdframed}[backgroundcolor=black!5,linecolor=black]
	\begin{tabular}{l}	
	\textbf{Given:} a program $\prog$, a configuration $\initconf \in \confsetof\prog$, a label $\lbl \in \lblsetof\prog$ \\[0.2cm]
	\begin{tabular}{l l}
	\qualreach:& \textbf{Determine} whether $\measureof\prog{\initconf\models\eventually\lbl} = 1$  \\[0.2cm]
	\qualrepreach:& \textbf{Determine} whether $\measureof\prog{\initconf\models\always\eventually\lbl} = 1$
	\end{tabular}
	\end{tabular}
\end{mdframed}
	\vspace{-0.2cm}
\end{figure}
In this section, we perform {\it qualitative reachability} analysis
for PTSO.
%
%
Given a program $\prog$, configuration $\initconf$, 
and label $\lbl$, we check 
whether a $\initconf$-run almost surely reaches $\lbl$.
%
%
%
We also consider {\it  qualitative repeated reachability},
where, we ask whether a $\initconf$-run
repeatedly visits $\lbl$ (visits $\lbl$ infinitely often) w.p. 1.
We also consider almost-never variants of the problems, where
we check whether the probabilities are $0$ rather than $1$.
We prove that these problems are decidable, and have 
non-primitive-recursive complexities.
\subsection{Almost-Sure Reachability}
\label{qual:plain:reachability:section}
The qualitative reachability problem, \qualreach,
is defined above.
%
%
The algorithm in Figure~\ref{qual:reach:algorithm:fig} solves \qualreach
by analyzing the transition system $\tsdenotationof\prog$,
the underlying transition system of PTSO.
%
%
%
If $\lbl$ occurs in $\initconf$ then the property
trivially holds, and hence we answer positively.
Otherwise, the algorithm considers a new program $\pprog$
obtained by replacing the statement labeled $\lbl$, 
by a new statement
that makes $\pprog$ terminate immediately if $\lbl$ is reached.
%
%
Let $\proc\in\procset$ be the unique process such that
$\lbl\in\lblsetof\proc$.
We define 
$\prog\rmlbl\lbl:=
\tuple{\procset-\set\proc\cup\set{\pproc},\weight}$
where $\pproc$ is a fresh process derived from
$\proc$ by replacing $\stmtof\lbl$ by
$\goto~\termlblof{new}$ for a fresh label $\goto~\termlblof{new}\not\in\lblsetof\prog$ and adding a \lstinline[style=customlang]{term} at label $\termlblof{new}$. The remaining instructions of $\pproc$ are identical to $\proc$.
%
%


\newcommand{\algolinenum}[1]{{\small \tt #1}}

\begin{wrapfigure}[11]{r}{0.5\textwidth}
\resizebox{0.5\textwidth}{!}{
\begin{minipage}{0.6\textwidth}
\begin{algorithm}[H]
  \SetAlgoRefName{}
  \KwIn{
    $\prog$: program;
    $\initconf\in\confsetof\prog$: configuration;
    $\lbl\in\lblset_\prog$: label.
  }
\lIf {$\lbl\in\initconf$}{\KwRet{$\true$}}\label{init:line:qual:reach:fig}
$\pprog\assigned\prog\rmlbl\lbl$\;
   \For{{\bf each} $\conf\in\plainconfsetof{\prog}$}{\label{qual:reach:fig:for:line}
     \lIf{$\initconf\progmovesto\pprog*\conf$ {\bf and} $\neg\left(\conf\progmovesto\pprog*\lbl\right)$}{\label{qual:reach:fig:if:line}
       \KwRet{$\false$}
       }
   } 
   \KwRet{$\true$}
  \caption{\qualreach}
\end{algorithm}
\end{minipage}}
\vspace{-0.2cm}
\caption{Almost-sure reachability algorithm.}
\label{qual:reach:algorithm:fig}
\end{wrapfigure}

The loop on line \algolinenum{3} cycles through the (finite) set of plain configurations.
%
%
%
For each plain configuration $\conf$ from the \textit{original} program $\prog$, we check:
(i) Whether $\conf$ is reachable from the initial configuration $\initconf$ 
in $\pprog$.
By the construction of $\pprog$, this is equivalent to checking whether $\conf$ is reachable from $\initconf$
in $\prog$ without observing label $\lbl$.
(ii) Whether it can reach the label $\lbl$.
If the answer to (i) is {\it yes}, {\it and} the answer to (ii) is {\it no},
then we have found a finite path $\pth$ in $\prog$ that starting
from $\initconf$, without visiting
$\lbl$, reaches configuration $\conf$ 
from which $\lbl$ is not reachable.
This implies that $\measureof\prog{\initconf\models\eventually\lbl} <1$.
%
If none of the plain configurations satisfy the condition, 
then each plain configuration $\conf$ reachable from $\initconf$
has a path to $\lbl$. Now by the attractor lemma, any 
run will almost surely visit $\plainconfsetof\prog$
infinitely often and by the fairness property of Markov chains,
it almost surely visits $\lbl$.

%
%
%
%

\subsection{Almost-Sure Repeated Reachability}
\label{qual:plain:repeated:reachability:section}

For almost-sure repeated reachability we 
are interested in determining whether the $\initconf$-runs 
visit $\lbl$ infinitely often with probability 1.
The algorithm for this is similar to the case for almost-sure reachability:
we check whether $\exists$ a plain configuration $\conf$ that satisfies 
$\initconf\progmovesto\prog*\conf \land \neg\left(\conf\progmovesto\prog*\lbl\right)$, 
in which case we return  false. The difference is that we do not need to 
transform the program as in the case of almost-sure reachability.
Details are in the supplementary material.

\subsection{Almost-Never (Repeated) Reachability}
\smallskip
\begin{figure}[h]
	\centering
\begin{mdframed}[backgroundcolor=black!5,linecolor=black]
	\begin{tabular}{l}
	\textbf{Given:} a program $\prog$, a configuration $\initconf \in \confsetof\prog$, a label $\lbl \in \lblsetof\prog$ \\[0.2cm]
	\begin{tabular}{l l}
	\neverqualreach:&\textbf{Determine} whether $\measureof\prog{\initconf\models\eventually\lbl} = 0$ \\[0.2cm]
	\neverqualrepreach:& \textbf{Determine} whether $\measureof\prog{\initconf\models\always\eventually\lbl} = 0$
	\end{tabular}
	\end{tabular}
\end{mdframed}
	\vspace{-0.2cm}
\end{figure}
The almost-never variants 
of the (repeated) reachability problems, \neverqualreach resp.\ \neverqualrepreach, ask
%
whether the probabilities equal to $0$ rather than $1$.
%
%
The solution to \neverqualreach is straightforward, since 
$\measureof\prog{\initconf\models\eventually \lbl}=0$
iff $\neg(\initconf\progmovesto\prog*\lbl)$.
%
%
On the other hand, the \neverqualrepreach problem requires a 
search over B-plain configurations $\conf$ satisfying 
$\initconf \progmovesto\prog* \conf \progmovesto\prog* \lbl$. Due to
space constraints, we defer the algorithm and proofs to the appendix.

\subsection{Decidability and Complexity}
The algorithms can be effectively implemented
since
(i) $\plainconfsetof\prog$ is finite; and
(ii)
the conditions of the for-loops and if-statements 
can be checked  effectively, as implied
by Lemma~\ref{tso:lbl:reachability:lemma}.
This gives Theorem \ref{qual:decidability:theorem}.
Theorem \ref{qual:complexity:theorem} is proved 
through reductions from the reachability
problem under the classical (non-probabilistic) TSO semantics
\cite{DBLP:conf/esop/AtigBBM12}.
The non-primitive-recursive lower bounds follow from the corresponding
result for reachability of classical TSO.

\begin{restatable}{thm}{thmQualReachDec}
\label{qual:decidability:theorem}
\qualreach, \qualrepreach, \neverqualreach,
\neverqualrepreach are all decidable.
\end{restatable}
%
\begin{restatable}{thm}{thmQualReachNPR}
\label{qual:complexity:theorem}
\qualreach, \qualrepreach, \neverqualreach,
\neverqualrepreach all have non-primitive-recursive complexities.
\end{restatable}

\section{Quantitative (Repeated)  Reachability}
\label{quan:reachability:section}

In this section we discuss {\it quantitative} reachability problems
for PTSO.
In contrast to qualitative analysis from Sec. \ref{qual:reachability:section},
the task here is to {\it compute} the actual probability.
We are not able to compute the probabilities exactly, 
but we can approximate the probability with an arbitrary
degree of precision.

\subsection{Approximate Quantitative Reachability}
\label{quan:plain:reachability:section}

\begin{figure}[H]
	\centering
\begin{mdframed}[backgroundcolor=black!5,linecolor=black]
	\resizebox{\textwidth}{!}{
	\begin{tabular}{l}	
	\textbf{Given:} program $\prog$, configuration $\initconf \in \confsetof\prog$, label $\lbl \in \lblsetof\prog$, precision value $\epsilon \in \mathbb{R}^{+}$ \\[0.2cm]
	\begin{tabular}{l l}
	\quantreach:& ~\textbf{Determine} $\probval$ s.t. $\measureof\prog{\initconf\models\eventually\lbl} \in [\probval, \probval + \precision]$ \\[0.2cm]
	\quantrepreach:& ~\textbf{Determine} $\probval$ s.t. $\measureof\prog{\initconf\models\always\eventually\lbl} \in [\probval, \probval + \precision]$
	\end{tabular}
	\end{tabular}
	}
\end{mdframed}
	\vspace{-0.2cm}
\end{figure}

%
In the approximate quantitative reachability problem, \quantreach,
given a precision parameter $\precision$, 
we are interested in determining an approximation $\probval$ satisfying
$\probval \leq \measureof\prog{\initconf\models\eventually\lbl}  \leq \probval + \precision$.
%
%

The algorithm in Fig.~\ref{quant:reach:alg:fig} 
solves the problem by successively improving the approximation at each iteration 
until we are within $\precision$-precision of the exact value. 
%
The algorithm maintains two variables:
$\posapprxvar$ (positive approximation)
is an under-approximation
of the probability with which $\lbl$ is reachable from $\initconf$,
and $\negapprxvar$ (negative  approximation)
is an under-approximation
of the probability with which $\lbl$ is {\it not} reachable from $\initconf$.
$\posapprxvar$ serves as a lower bound on $\probval$, while, $1-\negapprxvar$ serves as an upper bound: $\posapprxvar \leq \probval \leq 1 - \negapprxvar$.

\begin{figure}
\vspace{0.2cm}
\centering
\resizebox{0.8\textwidth}{!}{
\begin{minipage}{\textwidth}
\begin{algorithm}[H]
  \DontPrintSemicolon
  \SetAlgoRefName{}
  \KwIn{
    $\prog$: program;
    $\initconf\in\confsetof\prog$: configuration;
    $\lbl\in\lblset_\prog$: label;
    $\precision\in\preals$: precision.
  }
 \KwVar{
   $\posapprxvar,\negapprxvar\in\reals$: approximations, $\waitingvar\in\wordsover{(\confsetof\prog\times\reals)}$: queue\;
 }
  $\posapprxvar:=0$; $\negapprxvar:=0$; 
  $\waitingvar\assigned\tuple{\initconf,1}$\;\label{quant:reach:fig:init:line}
  \While{$\posapprxvar+\negapprxvar<1-\precision$}{\label{quant:reach:fig:while:line}
    $\tuple{\conf,\fiprobval}:=\headof{\waitingvar}$; $\waitingvar\assigned\tailof{\waitingvar}$\;
    \lIf{$\lbl\in\conf$}{
      $\posapprxvar\assigned\posapprxvar+\fiprobval$
    }
    \lElseIf{$\neg(\conf\starmovesto\prog\lbl)$}{
      \label{quant:reach:fig:else:if:line}
      $\negapprxvar\assigned\negapprxvar+\fiprobval$
    }
    \Else{ \lFor{{\bf each} $\conf'$ with $\conf\movesto\prog\pconf$}{
        $\waitingvar\assigned\waitingvar\app\tuple{\pconf,\fiprobval\cdot\prmpmtrxof\prog{\conf,\pconf}}$
      }
    }
  }
  \KwRet{$\posapprxvar$}
  \caption{\quantreach}
\end{algorithm}
\end{minipage}}
\caption{The quantitative reachability algorithm.}
\label{quant:reach:alg:fig}
\end{figure}


The algorithm iteratively improves these approximations until we reach a point where
their sum is within $\precision$ from 1 (line \algolinenum{4}).
%
In such a case, the desired value of $\probval = \posapprxvar$ is an $\precision$-precise
approximation.

To calculate the approximations, the 
algorithm performs forward reachability analysis
starting from the initial configuration $\initconf$.
It generates the set of $\initconf$-paths in a breadth-first manner, using the 
$\waitingvar$ FIFO queue.
For each generated path $\pth$ it also calculates the probability
of $\pth$.
Instead of the whole path $\pth$, 
$\waitingvar$ only stores the last configuration, $\conf$, of $\pth$ and
the probability of $\pth$, $\fiprobval$, as a pair $\langle\conf, \fiprobval\rangle$.
%
%

The approximation variables are initialized (line \algolinenum{3}) to zero,
and $\waitingvar$ queue is initialized to contain a single pair,
$\langle\initconf, 1\rangle$, representing the initial configuration $\initconf$ (which occurs with
probability one).
The while-loop executes until we achieve the desired precision.
At each iteration, we check whether we already have
reached the desired precision.
If not, the algorithm pops the pair 
$\tuple{\conf,\fiprobval}$ from the $\waitingvar$-queue.
There are three possibilities depending on $\conf$:
\begin{enumerate}
  \item If $\lbl\in\conf$ (if-branch, line \algolinenum{6}),
  the current path reaches $\lbl$ and, consequently,
  we increment $\posapprxvar$ by $\fiprobval$, the weight 
  of the current path.
  \item If $\lbl$ is not reachable from $\conf$ 
  (else-if branch, line \algolinenum{7}),
  the measure of runs that reach $\lbl$ starting from $\conf$ 
  is zero, and hence we increment $\negapprxvar$ by $\fiprobval$.
  \item If neither of the above hold (line \algolinenum{10}),
  the current path needs to be explored further, 
  we enqueue all successors $\pconf$ of $\conf$ into the queue. 
  The probability of the new path to $\pconf$ is 
  $\fiprobval\cdot\prmpmtrxof\prog{\conf,\pconf}$.
\end{enumerate}
%
%
%
%
%
%

To show correctness of the algorithm, let
$\posapprxvarof\ii$ and $\negapprxvarof\ii$ represent the value
of $\posapprxvar$ and $\negapprxvar$ prior to performing the 
$i^{\it th}$ iteration.
We show that in the limit as $i \rightarrow \infty$, the value of
$\posapprxvarof\ii + \negapprxvarof\ii$ tends to $1$.
Technically this follows by Lemma~\ref{finite:attractor:lemma}.
By this lemma, any $\initconf$-run almost surely either
(i) reaches a plain configuration from which $\lbl$ is not reachable,
or (ii) repeatedly reaches a plain configuration from which $\lbl$ is reachable.
In case (ii) it will almost surely reach $\lbl$.
This implies that 
$\measureof{\prog}
{\initconf\models(\eventually(\lbl\vee\neg\exists\eventually\lbl))}=1$,
i.e., an $\initconf$-run will almost surely either
reach $\lbl$ or reach a configuration from which $\lbl$ is not reachable,
implying that $\posapprxvarof\ii + \negapprxvarof\ii$ tends to $1$.
Finally, by Lemma~\ref{tso:lbl:reachability:lemma} we can effectively check
the condition of the if-statement, and hence the algorithm terminates.

The correctness of the approximation on termination follows by the property that
$\posapprxvarof\ii$ and $\negapprxvarof\ii$ are under-approximations of the
reach and non-reach probabilities.
This follows from the following invariants:
\begin{align*}
\posapprxvarof\ii\leq\measureof{\prog}
{\initconf\models\eventually\lbl} &\qquad
\negapprxvarof\ii\leq\measureof{\prog}
{\initconf\models\eventually\forall\always\neg\lbl} \\
%
\measureof{\prog}
{\initconf\models\eventually\lbl}&\leq
1-\measureof{\prog}
{\initconf\models\eventually\forall\always\neg\lbl} \\
%
\posapprxvarof\ii+\negapprxvarof\ii&>1-\precision \text{ holds on termination}
\end{align*}

%
%
These imply that, on termination,
$\posapprxvar$ is within $\precision$-precision	
of $\probval$.
%
%
%
\begin{restatable}{thm}{thmQuantReachDec}
\label{quantreach:theorem}
\quantreach is solvable.
\end{restatable}

\subsection{Approximate Quantitative Repeated Reachability}
\label{quan:repeated:reachability:section}


%
In the case of the approximate quantitative repeated reachability problem, 
we are interested in approximating the probability 
of visiting a given label $\lbl$ infinitely often.
We develop an algorithm that uses an iterative approximation scheme
similar to the reachability case.
%
%
%
%
We defer full details of this algorithm to the supplementary material and 
instead give an intuitive explanation on how
it differs from Sec.\ref{quan:plain:reachability:section}.

This algorithm too maintains approximations $\posapprxvar$ and $\negapprxvar$
and iteratively narrows the error margin until it is smaller than $\precision$.
The main difference is in the condition at line \algolinenum{6} of 
Figure \ref{quant:reach:alg:fig}. 
In the case of reachability the lower estimate $\posapprxvar$, is increased when
$\lbl \in \conf$.
In the repeated reachability case, this is not sufficient;
we need to ensure that there is no state $\conf'$ that is reachable 
from the current state $\conf$ and such that $\lbl$ is not reachable from $\conf'$.
The existence of such a $\pconf$ implies existence of a non-zero measure continuation of
the current run in which $\lbl$ is not reached infinitely often.
Hence, the conditional of the if-statement is modified to:
 $ \forall \conf' \in \bplainvar.~~(\conf\starmovesto\prog\pconf)\Rightarrow(\pconf\starmovesto\prog\lbl)$. 

We note that naively we would have to check the above condition for all
configurations $\pconf \in \confsetof\prog$, which is infeasible since
$\confsetof\prog$ is an infinite set.
We address this by using Lem. \ref{bplain:finite:attractor:lemma}, 
which shows that runs from all configurations eventually
reach a B-plain configuration. Hence it is sufficent to only 
check the condition for the (finitely many) B-plain configurations, 
which are precomputed in $\bplainvar$.

%
%

%
%
%
\begin{restatable}{thm}{thmQuantRepReachDec}
\label{quantrepreach:theorem}
\quantrepreach is solvable.
\end{restatable}

\section{Expected Average Costs}
\label{cost:section}
%
In this section, we develop a cost model for concurrent
programs where we assign a cost to 
the execution of each instruction,
the goal begin to approximate the expected
cost of runs that reach a given label.
%
%

\subsection{Computing costs over runs}

A {\it cost function}
$\funtype\cost{\lblsetof\prog}\pnat$
for program $\prog$
defines for each label $\lbl\in\lblsetof\prog$ the cost of executing the instruction at $\lbl$.
%
A particular way to define the function is to assign
a cost to each instruction in the programming language,
so that $\costof\lbl$ depends only on $\stmtof\lbl$ and not on $\lbl$ itself. 
But we consider the general case.
%
%
We extend $\cost$ to runs as follows.
Consider configurations $\conf=\conftuple$ and $\pconf$
such that $\conf\progmovesto\prog{}\pconf$.
If $\conf\progmovesto\prog{\proc}\pconf$, for
process $\proc$, then we
define 
$\ccostof\conf{\pconf}:=\costof{\labelingof\proc}$.
In other words, it is the cost of the instruction executed
by $\proc$.
%
Recall from Sec.~\ref{semantics:section} that $\proc$ is unique
and therefore the function is well-defined.
If $\disabledof{\conf}$ or if $\neg(\conf\progmovesto\prog{}\pconf)$ then we define
$\ccostof\conf{\pconf}:=0$.
Consider a run $\run\in \{\runsetof\conf ~|~ \run\modelswrt\prog\neventually{=\ii}\lbl\}$,
i.e. a $\conf$-run that reaches $\lbl$ for the first time at step $\ii$.
We define
$\runcostof\run\lbl=\sum_{1\leq\jj\leq\lengthof\ii-1}\ccostof{\run[\jj]}{\run[\jj+1]}$,
i.e, the sum of costs of all executed instructions along
$\run$ up to the first visit to $\lbl$.

%
%
\smallskip
\smallskip

%
%

For a configuration $\conf$,
a label $\lbl$, and
a cost function $\cost$,
we define a random variable
$\funtype{\rcost\conf\lbl\cost}\Om\reals$ over support $\Om=\conf\app\prminfwordsover\confset\mchain$ as follows:
\begin{equation*}
\rcostof\conf\lbl\cost\run=
\begin{cases}
  0 & \run\not\in\{\runsetof\conf ~|~ \run\modelswrt\prog\neventually{=\ii}\lbl\} \\
  \rcostof\conf\lbl\cost\run=\runcostof\run\lbl & \text{otherwise }	
\end{cases}
\end{equation*}
%
%

\begin{figure}[h]
	\centering
\begin{mdframed}[backgroundcolor=black!5,linecolor=black]
	\resizebox{\textwidth}{!}{
	\begin{tabular}{l}	
	\textbf{Given:} program $\prog$, configuration $\initconf \in \plainconfsetof\prog$, cost function $\funtype\cost{\lblsetof\prog}\pnat$,  \\
	 label $\lbl \in \lblsetof\prog$ s.t. $\initconf\models\eventually\lbl$, precision value $\epsilon \in \mathbb{R}^{+}$ \\[0.2cm]
	\expaveragecost:$\quad$ \textbf{Determine} $\probval$ s.t. $\cecostof\initconf\lbl\cost \in [\probval, \probval + \precision]$ 
	\end{tabular}
	}
\end{mdframed}
	\vspace{-0.2cm}
\end{figure}
\noindent\textbf{The expected average cost problem}
$\ecostof\conf\lbl\cost$ is defined as the expected 
cost of reaching $\lbl$ from $\conf$ and
$\cecostof\conf\lbl\cost$ as the conditional expectation
over runs that reach $\lbl$.
%
If $\neg(\conf\modelswrt\prog\exists\eventually\lbl)$ then
the expected cost is not defined.
If however $\conf\modelswrt\prog\exists\eventually\lbl$
then $\cecostof\conf\lbl\cost=
{\ecostof\conf\lbl\cost}/{\measureof\prog{\conf\modelswrt\prog\eventually\lbl}}$, 
which follows since for the non-reaching runs, the cost is zero.
We present the expected average cost problem, in the figure above, 
where we want to approximate $\cecostof\conf\lbl\cost$ to $\precision$-precision.
%
%

\subsection{Eagerness}
%
%
Our solution to \expaveragecost relies on the fact that $\mchainof\prog$  satisfies 
an {\it eagerness} property in the sense of 
\cite{DBLP:conf/atva/AbdullaHMS06}.
In  our setting, eagerness means that the probability
of avoiding the target label $\lbl$ decreases exponentially with the number
of steps.
Concretely,
we show that there are two constants:
the {\it eagerness degree} $\eagernessof\prog\in\preals$, and
the {\it eagerness threshold} $\eagernessthresholdof\prog\in\preals$
satisfying the following:
\begin{equation*}
  \forall\conf\in\smallconfsetof\prog ~\forall\lbl\in\lblsetof\prog
  ~\forall\nn\geq\eagernessthresholdof\prog \quad
  \conf\modelswrt\prog\exists\eventually\lbl \Rightarrow
\measureof\prog{\conf\modelswrt\prog\neventually
{\geq\nn}\lbl}\leq\left(\eagernessof\prog\right)^\nn
\end{equation*}
i.e. for $\nn \geq\eagernessthresholdof\prog$, 
the probability of avoiding $\lbl$ during
the first $\nn$ steps decreases exponentially with $\nn$. 
The following lemma forms the crux of this section.
\begin{restatable}[{\bf Eagerness Lemma}]{lem}{lemEagerness}
\label{eagerness:lemma}
$\eagernessof\prog$  and $\eagernessthresholdof\prog$ exist and are computable. 
\end{restatable}

We devote this sub-section to give an overview
of the the proof of Lemma~\ref{eagerness:lemma}
(the formal proof is provided in the supplementary material).
We consider the behavior of runs with respect to the small 
and large configurations,
exploiting the fact that the runs of the system 
tend to gravitate towards the small configurations.
However here we use a property, called
{\it left-biasedness} (defined in Sec.~\ref{gravity:subsubsection}), 
that is stronger than the left-orientedness property
of Sec.~\ref{plain:attractor:subsection}.
%


To prove Lemma~\ref{eagerness:lemma}, we show that, 
for a small configuration $\conf\in\smallconfsetof\prog$,
the runs from $\conf$ 
satisfy the following three properties
with a high probability:
(i) 
they make their first return to $\smallconfsetof\prog$
 within a small number of steps,
(ii)
they return to 
$\smallconfsetof\prog$ multiple times,
within a small number of steps, and
(iii) 
if they eventually reach $\lbl$ then they will do that
within a few steps.
We collect these results to obtain the 
proof of Lemma~\ref{eagerness:lemma}.

\subsubsection{Gravity: First Return}
\label{gravity:subsubsection}
We recall that buffer sizes can increase by at most one
during process transitions, and that any number of messages can be flushed
to the memory during an update transition
(Sec.~\ref{semantics:section} and Sec.~\ref{plain:attractor:subsection}).
Based on this, we show {\it left-biasedness}, defined
as follows:

\begin{quote}
  \textbf{Left-biasedness} $\forall \conf\in\largeconfsetof\prog$ the probability of moving from $\conf$ to a smaller configuration is bounded below by 2/3 and that of moving to a larger configuration is bounded above by 1/3, regardless of $\prog$.
\end{quote}

Using left-biasedness, we
show that the set  $\smallconfsetof\prog$  
has a {\it gravity} property, namely,
a run starting from a small configuration will,
with a high probability, return to the set $\smallconfsetof\prog$
(for the first time) within a  few number of steps.
Formally, we define the {\it gravity parameter} $\gravityof\prog$
as follows: $\qqstar:=2/3$, $\ppstar:=1/3$, and
 $\gravityof\prog:=2\sqrt{\qqstar\cdot\ppstar}=\frac{2\cdot\sqrt2}3$.
We prove the following lemma.
%

\begin{restatable}[{\bf Gravity Lemma}]{lem}{lemGravity}
\label{gravity:lemma}
$\measureof\prog{\conf\modelswrt\prog\nxt\neventually{\geq\nn}\smallconfsetof\prog}
\leq\left(\gravityof\prog\right)^\nn$,
for all $\conf\in\smallconfsetof\prog$ and all $\nn\in\nat$ .
\end{restatable}

The lemma states that, starting from a small configuration, the probability
that a run avoids $\smallconfsetof\prog$ in the next
$\nn$ steps decreases exponentially with $\nn$.
%
%
%
%
%
%

\subsubsection{Multiple Revisits}
\label{multiple:revisits:subsubsection}
Notice that the gravity lemma is concerned with 
the {\it first} return to
the set of small configurations.
We will now apply this argument repeatedly to
conclude that, with high probability, {\it multiple} re-visits
to small configurations take place ``quickly''. 
That is, the set of runs starting from $\smallconfsetof\prog$
and {\it frequently} re-visiting $\smallconfsetof\prog$ has a high measure.
%
%
To formalize these arguments, we make the following definition.
For $\mm,\nn:1\leq\mm\leq\nn$,
we define
$\visitof\prog\nn\mm$ to be the set of runs that visit the set
$\smallconfsetof\prog$
exactly $\mm$ times in their first 
$\nn-1$ steps\footnote{For technical convenience, 
we use $\nn-1$ instead of $\nn$ in the definition of $\visit$.
This allows us to avoid some corner cases in the proofs.}.
We use the $\visit$ predicate
to partition the set of $\conf$-runs,
depending on how often they return to $\smallconfsetof\prog$ 
during their first $\nn$ steps.
We distinguish these as
{\it Sporadic-Runs (S-Runs)}: runs
that visit the $\smallconfsetof\prog$ {\it sporadically}
during their first $\nn$ steps, and
{\it Frequent-Runs (F-Runs)}:
runs that visit $\smallconfsetof\prog$
{\it frequently} during their first $\nn$ steps.
We will derive a constant $\border\in\nat$ (see below)
that delineates the {\it border} between these sets.
We formally define:
\begin{eqnarray*}
\srunsetof\conf\nn&:=&
\cup_{1\leq\mm\leq\left\lfloor\frac\nn\border\right\rfloor}
\setcomp{\run\in\runsetof\conf}{\run\models\visitof\prog\nn\mm}
\\
\frunsetof\conf\nn&:=&
\cup_{\left\lfloor\frac\nn\border\right\rfloor+1\leq\mm\leq\nn}
\setcomp{\run\in\runsetof\conf}{\run\models\visitof\prog\nn\mm}
\end{eqnarray*}
\begin{figure}[h]
\centering
\includegraphics[scale=0.23]{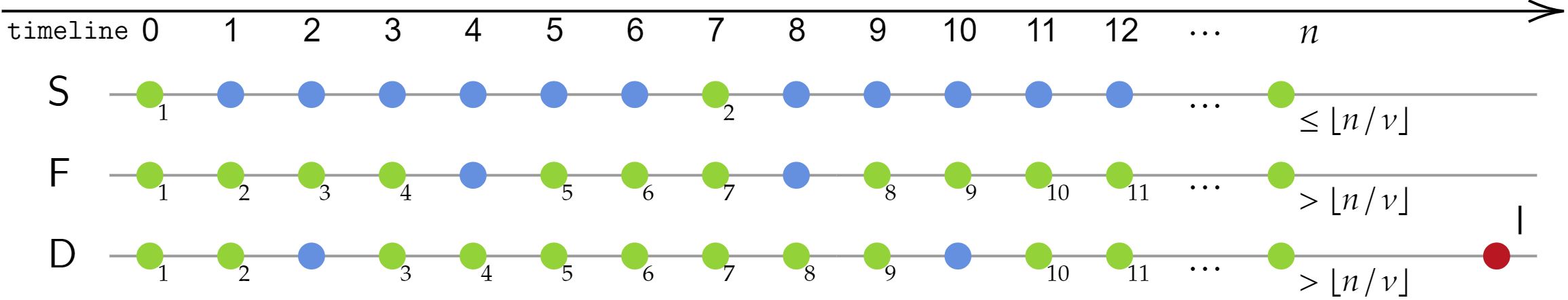}
\vspace{-0.3cm}
\caption{
Figure depicting configuration sequences of S, F and D runs.
Green dots represent small configurations, blue dots represent large configurations.
All runs start in a small (plain) configuration. 
Within the first $n$ configurations: 
the S-run visits $\smallconfsetof\prog$ at most $\lfloor\frac{n}{\border}\rfloor$ times,
the F, D runs visit $\smallconfsetof\prog$ at least $\lfloor\frac{n}{\border}\rfloor+1$ times.
A D-run is a special case of an F-run which does not visit label $\lbl$ (red dot) in the first $n$
steps.
}
\label{eagerness:fig}
\end{figure}

The value of $\nn/\border$ distinguishes the S-Runs from the F-Runs.
%
Our goal is to give an upper bound on the measure of the S-Runs.
%
For a prefix path $\pth$
of length $\nn$, 
there are
$\genfrac(){0pt}{2}{\nn-1}{\mm-1}$
ways to choose the $\mm-1$ indices along $\pth$ 
at which $\smallconfsetof\prog$ is reached (since the run starts from $\smallconfsetof\prog$).
Each of the $\mm-1$ path fragments between these indices represents one consecutive revisit of $\smallconfsetof\prog$.
%
By Lemma~\ref{gravity:lemma},
the measure of the set of such runs
is bounded by 
$\left(\gravityof\prog\right)^{\nn-\mm}=\left(\frac{2\sqrt2}3\right)^{\nn-\mm}$, giving 
\begin{equation*}
\resizebox{0.95\textwidth}{!}{$
\measureof\prog{\srunsetof\conf\nn}
\,\leq\,\sum_{\mm=1}^{\lfloor\frac{\nn}\border\rfloor}
\genfrac(){0pt}{2}{\nn-1}{\mm-1}\cdot
\gravityof\prog^{\nn-\mm}
\;\leq\;
\left(
\sqrt{\frac83}\cdot
\left(\frac\border{\border-1}\right)
\cdot
\left(
2+\sqrt3\cdot\border
\right)^{\lfloor\frac{1}\border\rfloor}
\right)^\nn$}
\end{equation*}
%
under the condition that $4\leq2\cdot\border\leq\nn$.
The second inequality is obtained through algebraic manipulations
using 
$\gravityof\prog=\frac{2\cdot\sqrt2}3$.
Define 
$\fun(\xx):=
\sqrt{\frac83}\cdot
\left(\frac\xx{\xx-1}\right)
\cdot
\left(
2+\sqrt3\cdot\xx
\right)^{\lfloor\frac{1}\xx\rfloor}
$.
We have $\fun(150)=0.986<1$.
Hence, for parameter $\border:=150$, 
defining $\seagernessof\prog:=\fun(\border)$,
we have the following lemma, where the bound decays exponentially with $\nn$ since $\seagernessof\prog < 1$.
\begin{restatable}[{\bf S-Run Bound}]{lem}{lemSRun}
\label{sruns:lemma}
$\measureof\prog{\conf\modelswrt\prog\srunsetof\conf\nn}\leq\left(\seagernessof\prog\right)^\nn$,
for all $\conf\in\smallconfsetof\prog$ and all 
$\nn$ such that $300=2\cdot\border\leq\nn$.  
\end{restatable}


\subsubsection{Reaching the label $\lbl$}
\label{reaching:target:subsubsection}
We now turn our attention to the set of F-Runs.
Our goal is to show that if an F-Run
reaches $\lbl$ then, with a high probability,  it
 will reach  $\lbl$ ``quickly''.
To that end, we consider the opposite  scenario and introduce a subset of 
the F-Runs which we call {\it Delayed Runs (D-Runs)}:
\[
\drunsetof\conf\lbl\nn:=
\cup_{\mm=\lfloor\frac{\nn}\border\rfloor+1}^\nn
\setcomp{\run\in\runsetof\conf}
{\run\modelswrt\prog\neventually{=\nn}\lbl\land \visitof\prog\nn\mm}
\]
A D-Run is an F-Run that {\it delays}
its first visit to 
the label $\lbl$ until the $\nn^{\it th}$ step
for some $n$.
We show that the measure of D-Runs decreases $\nn$ increases.    
Note that $\lbl$ is reachable from all configurations 
from a path that ends at $\lbl$. 
%
Therefore, we consider the set
$\aaset:=\setcomp{\conf\in\smallconfsetof\prog}{\conf\modelswrt\prog\exists\eventually\lbl}$,
of small configurations from which $\lbl$ is reachable.
We analyze how often a run starting from a small configuration,
visits $\aaset$ before finally visiting the label
$\lbl$.
For sets of configurations $\confs_1,\confs_2\subseteq\confsetof\prog$,
 a run $\run$, and $\mm\in\nat$, we write
$\run\models\confs_1\kbefore\mm\confs_2$ to denote that
$\run$ visits the set $\confs_1$ 
at least $\mm$ times before visiting $\confs_2$
for the first time.
Notice 
\begin{eqnarray}
\label{drunset:mbefore:upper:bound:eqn}
\drunsetof\conf\lbl\nn&\subseteq&
\bigcup_{\mm=\lfloor\frac{\nn}\border\rfloor+1}^\nn
\setcomp{\run\in\runsetof\conf}
{\run\modelswrt\prog\aaset\kbefore\mm\lbl}
\end{eqnarray}
To upper bound the measure of D-Runs, 
we start by upper bounding the measure of the set
$\setcomp{\run\in\runsetof\conf}{\run\modelswrt\prog\aaset\kbefore\mm\lbl}$, 
i.e. $\conf$-runs making $\mm$ visits to $\aaset$
before visiting $\lbl$.
%
We consider the probability that a run from a small configuration $\conf$
{\it does} visit $\lbl$ before returning to $\conf$.
We can compute a $\mu$ such that
\begin{eqnarray}
\label{drunset:mu:eqn}
 &&0<\mu\leq\min_{\conf\in\aaset}\measureof\prog{\conf\models\nxt(\lbl\kbefore1\conf)}
\end{eqnarray}
Hence $\mu$ is a lower bound on the measure of runs that
start from some configuration in $\conf\in\aaset$ and visit $\lbl$
before returning to $\conf$.
%
%
%
%
To obtain an upper bound on the measure of D-Runs, we show the following inequality:
%
\begin{equation*}
\resizebox{\textwidth}{!}{
$\measureof\prog{\drunsetof\conf\lbl\nn}
\leq
\sum\limits_{\mm=\lfloor\frac{\nn}\border\rfloor+1}^\nn
\sum\limits_{\pconf\in\aaset}
(1-\mu)^{\left\lceil\frac\mm{\sizeof\aaset}\right\rceil-1}
\leq
\frac{\sizeof{\aaset}
  }{(1-\mu)\cdot\left(1-(1-\mu)^{\frac{1}{\sizeof\aaset}}\right)}\cdot
  \left((1-\mu)^{
  \frac{\nn}{\border\cdot\sizeof\aaset}}
  \right)
$}
\end{equation*}
The first inequality follows from 
formulas~\ref{drunset:mbefore:upper:bound:eqn}~and~\ref{drunset:mu:eqn},
while the second is obtained through algebraic techniques.
Define 
$\deagernessof\prog$ such that
$(1-\mu)^{\frac1{\border\cdot\sizeof\aaset}}<\deagernessof\prog<1$.
Such an $\deagernessof\prog$ is computable since 
$\border$, $\aaset$, $\mu$ are computable.
Since $(1-\mu)^{\frac1{\border\cdot\sizeof\aaset}}<\deagernessof\prog$
it follows that there is a natural number, denoted by
$\deagernessthresholdof\prog$, such that 
$
\frac{\sizeof{\aaset}}{(1-\mu)\cdot(1-(1-\mu)^{\frac1{\sizeof\aaset}})}\cdot\left((1-\mu)^{\frac{1}{\border\cdot\sizeof\aaset}}\right)^\nn \leq (\deagernessof\prog)^\nn$
for all $\nn\geq\deagernessthresholdof\prog$.
This gives the following lemma.

\begin{restatable}[{\bf D-Run Bound}]{lem}{lemDRun}
\label{druns:lemma}
$\measureof\prog{\drunsetof\conf\lbl\nn}
\leq\left(\deagernessof\prog\right)^\nn$,
for all $\conf\in\smallconfsetof\prog$ and all 
$\nn\geq\deagernessthresholdof\prog$.  
\end{restatable}

%
%

\newcommand{\probvar}{\phi}

\subsubsection{Proof of Lemma~\ref{eagerness:lemma}}
\label{eagerness:subsubsection}
We now give a sketch of the proof of the eagerness property.


Choose a value $\sdeagernessof\prog$ such that,
$\max(\seagernessof\prog,\deagernessof\prog) < \sdeagernessof\prog < 1$.
From Lemma~\ref{sruns:lemma} and
Lemma~\ref{druns:lemma} it follows that
for some constant
$\sdeagernessthresholdof\prog > \max(\deagernessthresholdof\prog, 300)$,
$\measureof\prog{\conf\modelswrt\prog\neventually{=\nn}\lbl}\leq
(\sdeagernessof\prog)^n$,
for all $\nn > \sdeagernessthresholdof\prog$ (sufficiently large).
%
%
%
The final step 
is to extend the argument to the set of $\conf$-runs
that reach $\lbl$ in $\nn$ {\it or more} steps
(as required by Lemma~\ref{eagerness:lemma}).
\[
\begin{array}{l}
\measureof\prog{\conf\modelswrt\prog\neventually{\geq\nn}\lbl}
=
\sum_{\kk=\nn}^\infty\measureof\prog{\conf\modelswrt\prog\neventually{=\nn}\lbl}
\leq
\sum_{\kk=\nn}^\infty\left(\sdeagernessof\prog\right)^\kk
=
\frac{\left(\sdeagernessof\prog\right)^\nn}{1-\sdeagernessof\prog}
\end{array}
\]
Choose $\eagernessof\prog$, (exists since $\sdeagernessof\prog < 1$) such that $\sdeagernessof\prog < \eagernessof\prog < 1$.
There exists an $\eagernessthresholdof\prog$ such that
$\frac{\left(\sdeagernessof\prog\right)^\nn}{1-\sdeagernessof\prog} \leq \left(\eagernessof\prog\right)^\nn$
for all $\nn\geq\eagernessthresholdof\prog$, and hence
$\measureof\prog{\conf\modelswrt\prog\neventually{\geq\nn}\lbl}
\geq\left(\eagernessof\prog\right)^\nn$
for all $\nn\geq\eagernessthresholdof\prog$ (sufficiently large). This gives us the result.
%



\subsection{The Algorithm}
%
%
Now we proceed to describe the algorithm.
The goal is to approximate
$\cecostof\initconf\lbl\cost$.
The scheme followed by the algorithm is similar to the quantitative section:
it iteratively improves an approximations until it is $\precision$-precise.
However, the implementation is much more challenging since we need to maintain 
error margins on both the cost and the probabilities.
It performs forward reachability analysis, starting
from $\initconf$, and generating,
successively longer $\initconf$-paths,
in a breadth-first manner.
%
%

The variable $\waitingvar$ contains triples of form $\langle\conf, \psi, \probvar\rangle$ 
corresponding to $\initconf$-paths waiting to be analysed.
For such a path $\pth$, $\conf$ is the last configuration of $\pth$, 
$\psi$ is the cost of $\pth$, and 
$\probvar$ is the probability of taking $\pth$.
We initialize $\waitingvar$ to contain a triple
corresponding to the empty path from $\initconf$: $\langle \initconf, 0, 1\rangle$.
Prior to the $\ii^{\it th}$ iteration loop (line \algolinenum{10}), $\waitingvar$ contains
triples corresponding to paths of length $\ii$.
At each loop iteration the triples in $\waitingvar$ are analysed
and the triples for paths one step deeper are generated for the next iteration.
%
%
%
%
\begin{figure}
\centering
\vspace{0.3cm}
\resizebox{0.8\textwidth}{!}{
\begin{minipage}{\textwidth}
\begin{algorithm}[H]
  \SetAlgoRefName{}
  \KwIn{
    $\prog$: program;
    $\initconf\in\confsetof\prog$: configuration
    $\lbl\in\lblset_\prog$: label with 
    $\initconf\models\exists\eventually\lbl$; \;
    $\funtype\cost{\instrsetof\prog}\reals$: cost function;
    $\precision\in\preals$: precision;
   }
  \KwVar{
    $\waitingvar,\pwaitingvar\in\wordsover{(\confsetof\prog\times\reals\times\reals)}$: queues\;
    $\costapprxvar\in\reals$: approximation of $\ecostof\conf\lbl\cost$\;
    $\probapprxvar\in\reals$: under-approximation of 
    $\measureof{\mchainof\prog}{\conf\modelswrt\prog\eventually\lbl}$\;
    $\costerrorvar\in\reals$, $\proberrorvar\in\reals$: over-approximations of errors\;
       $\kvar,\nvar\in\nat$;
  }
  $\kvar\assigned\maxcostof\cost$; $\nvar\assigned0$\;
  $\costapprxvar\assigned0$; $\probapprxvar\assigned0$; $\waitingvar\assigned\tuple{\initconf,0,1}$\;
  $\costerrorvar\assigned\frac{\kvar}{\left(1-\eagernessof\prog\right)^2}$; 
  $\proberrorvar\assigned\frac{1}{1-\eagernessof\prog}$\; 
  \Repeat{$\left(
    \frac{\costapprxvar+\costerrorvar}{\probapprxvar}
    -
    \frac{\costapprxvar}{\probapprxvar+\proberrorvar}<\precision
    \right) \land (\proberrorvar>0)\land (\nn\geq\eagernessthresholdof\prog)$}{
    $\nvar\assigned\nvar+1$; $\pwaitingvar\assigned\emptyset$\;
    \For{$\ii=1$ {\bf to} $\lengthof\waitingvar$}{
      $\tuple{\conf,\costval,\probvar}\assigned\waitingvar[\ii]$\;
      \If{$\lbl\in\conf$}{
        $\costapprxvar\assigned\costapprxvar+\costval\cdot\probvar$; $\probapprxvar\assigned\probapprxvar+\probvar$\;
      }
      \Else{
        \For{{\bf all} $\pconf:\conf\progmovesto\prog{}\pconf$}{
          $\pwaitingvar\assigned\pwaitingvar\app\tuple{\conf',\costval+\ccostof\conf\pconf,\probvar\cdot\prmpmtrxof\prog{\conf,\pconf}}$\;
        }
      }
    }
    $\costerrorvar\assigned\costerrorvar\cdot\eagernessof\prog$; $\proberrorvar\assigned\proberrorvar\cdot\eagernessof\prog$\;
    $\waitingvar\assigned\pwaitingvar$\;
  }
  \KwRet{$\frac{\costapprxvar}{\probapprxvar+\proberrorvar}$}
\caption{Solving \expaveragecost}
\end{algorithm}
\end{minipage}}
\caption{The expected average cost algorithm.}
\label{cost:alg:fig}
\end{figure}

The iterations calculate increasingly
precise approximations of 
$\ecostof{\initconf}\lbl\cost$, and of
$\measureof\prog{\initconf\modelswrt\prog\eventually\lbl}$,
maintained in variables $\costapprxvar$ and
$\probapprxvar$, respectively.
%
%
%
%
We maintain two additional variables ($\costerrorvar$ and $\proberrorvar$)
that help us to provide an upper bound on the estimation errors.
%
%
Defining $\maxcostof\cost:=\max\setcomp{\costof\lbl}{\lbl\in\lblsetof\prog}$,
we explain the correctness of the algorithm with a number of invariants.
%
%
%
\begin{restatable}{lem}{lemCostInv}
\label{cost:invariant:lemma}
The algorithm maintains the following invariants where invariants (\ref{costapprx:invariant},\ref{probapprx:invariant},\ref{cerror:invariant},\ref{perror:invariant}) hold for all $\ii>0$ and invariants (\ref{ecost:costapprx:invariant},\ref{prob:probapprx:invariant}) hold for all $\ii\geq\eagernessthresholdof\prog$.
\end{restatable}

\begin{enumerate}
\item
\label{costapprx:invariant}
$\costapprxvarof\ii=
\sum\limits_{\{\run\in\runsetof{\initconf}~|~\run\models\neventually{\leq\ii}\lbl\}} \costof\run\cdot\probof\prog\run$:
%

%
\item
\label{probapprx:invariant}
$\probapprxvarof\ii=\measureof\prog{\initconf\models\neventually{\leq\ii}\lbl}$:
%

\item
\label{ecost:costapprx:invariant}
$\costapprxvarof\ii\leq\ecostof\conf\lbl\cost\leq\costapprxvarof\ii+\costerrorvarof\ii$.
%
\item 
\label{prob:probapprx:invariant}
$\probapprxvarof\ii\leq\measureof\prog{\conf\modelswrt\prog\eventually\lbl}\leq\probapprxvarof\ii+\proberrorvarof\ii$.
\item
\label{cerror:invariant}
$\costerrorvarof\ii=
\maxcostof\cost\cdot\frac{\eagernessof\prog^\ii}
{(1-\eagernessof\prog)^2}$.
\item
\label{perror:invariant}
$\proberrorvarof\ii=
\frac{\eagernessof\prog^\ii}
{1-\eagernessof\prog}$.
\end{enumerate}

%
Invariants~\ref{cerror:invariant} and \ref{perror:invariant} imply
that as $i \rightarrow \infty$ 
$\costerrorvarof\ii$ and
$\proberrorvarof\ii$
tend to 0.
%
%
Hence,
$\lim_{i\rightarrow\infty}
\left(
\frac{\costapprxvarof\ii+\costerrorvarof\ii}{\probapprxvarof\ii}
-
\frac{\costapprxvarof\ii-\costerrorvarof\ii}{\probapprxvarof\ii+\proberrorvarof\ii}
\right)=0$
%
implying termination.
Since $n\geq \eagernessthresholdof\prog$ when the algorithm terminates,
%
by invariants \ref{ecost:costapprx:invariant} and \ref{prob:probapprx:invariant} it follows that 
$
\costapprxvarof\nn\leq
\ecostof\conf\lbl\cost\leq
\costapprxvarof\nn+\costerrorvarof\nn$ 
%
and $\probapprxvarof\nn\leq
\measureof\prog{\conf\modelswrt\prog\eventually\lbl}\leq
\probapprxvarof\nn+\proberrorvarof\nn
$.
%
%
Combining these two inequalities and the termination condition of the algorithm,
we get the following:
\begin{align*}
\resizebox{0.9\textwidth}{!}{$
\frac{\costapprxvarof\nn}
{\probapprxvarof\nn+\proberrorvarof\nn}\leq
\frac{\ecostof\conf\lbl\cost}
{\measureof\prog{\conf\modelswrt\prog\eventually\lbl}}
<
\frac{\costapprxvarof\nn}
{\probapprxvarof\nn+\proberrorvarof\nn}
+\precision
$}
\end{align*}
Hence on termination, 
$\probval:=
\frac{\costapprxvarof\nn}
{\probapprxvarof\nn+\proberrorvarof\nn}
$ is within $\precision$-precision 
of the true value, implying correctness of the algorithm.
%
We get the following theorem.


\begin{restatable}{thm}{thmCost}
\label{cost:solvable:theorem}
The above algorithm solves \expaveragecost.
\end{restatable}

\textbf{Related Work}
Only recently there has been an increased interest in the 
formulation and verification of liveness properties for 
weak memory models. In \cite{Lahav2020MakingWM}, they
factor the system into a process and memory subsystems
and define notions of fairness for either. This is reminiscent
of our approach, where
we consider probabilistic policies for process scheduling and memory update.
Their model on the other hand is non-probabilistic and 
they have weaker fairness guarantees, which we describe in more detail in
Sec. \ref{plain:attractor:subsection}.
The liveness verification problem for TSO has been considered in 
\cite{DBLP:journals/corr/abs-2107-09930}, where they show undecidability
for various liveness properties. However, once again work with 
non-probabilistic notions of fairness. We show in this paper, that 
with stronger (probabilistic) fairness, reachability and 
repeated reachability problems become decidable.


%
In \cite{DBLP:conf/popl/AtigBBM10}, they show the undecidability 
of the repeated reachability problem,
{\it without} fairness conditions,
for finite-state programs running under the TSO semantics.
In contrast, we show that checking
repeated reachability qualitatively is decidable 
(Sec.~\ref{qual:plain:repeated:reachability:section}),
and that we can even compute the measure of runs satisfying
the property with arbitrary precision
(Sec.~\ref{quan:repeated:reachability:section}).

There has been a huge amount of work on the verification of 
{\it finite-state} Markov chains 
(see, e.g., \cite{PMC:book,DBLP:conf/cav/KwiatkowskaNP11}).
Since the  buffers in TSO are unbounded, we however, get
an infinite-state Markov chain. 
%
There is also a substantial literature on the verification of 
infinite-state Markov chains, where
specialized techniques are developed
for particular classes of systems.
Several works have considered
probabilistic push-down automata and probabilistic
recursive machines
\cite{DBLP:journals/jacm/EtessamiY15,DBLP:journals/jcss/BrazdilKKV15,DBLP:conf/lics/EsparzaKM04}.
However, these techniques don't apply in our case since
push-down automata cannot encode
the FIFO store-buffer data-structure.
%
%
%

%
Works such as \cite{DBLP:conf/atva/BrazdilC00V19,AbdullaHM07,DBLP:conf/csl/BrazdilKKNK14,DBLP:journals/jacm/BrazdilKK14}
develop algorithmic and complexity results for checking termination and reachability
for systems such as probabilistic VASS, probabilistic Petri nets, probabilistic multi-counter systems.
Again, these models
are different from ours and cannot encode FIFO queues.

The works closest to ours are those on probabilistic lossy channel systems
\cite{AbdullaHM07,DBLP:conf/atva/AbdullaHMS06}.
%
These works also rely on the frameworks 
of decisive and eager Markov chains.
However, lossy channel systems and TSO are fundamentally different, 
and the manner in which we instantiate the frameworks 
of decisive/eager Markov chains differs.
The decidability of verification for probabilistic extensions
of lossy channels is sensitive to the definition of the message losses.
In the case of lossy channel systems,
if messages are only allowed to be lost at one end of the channel
(a model that is close to our notion of message updates), then all non-trivial 
verification problems become undecidable for probabilistic
lossy channel systems ~\cite{DBLP:journals/iandc/AbdullaBIJ05}.
Therefore, although there is a reduction from TSO to lossy channel systems
in the case of non-probabilistic models \cite{DBLP:conf/popl/AtigBBM10},
we know of no such reduction between the corresponding probabilistic models.

Finally, the concept of decisiveness has been
extended to more general models 
such as generalized semi-Markov processes, stochastic timed automata
\cite{DBLP:conf/icalp/0001BBC16}, and
lossy channel-based stochastic games
\cite{DBLP:conf/fossacs/AbdullaHAMS08}.

\section{Conclusions, Discussions, and Perspectives}
\label{conclusions:section}
We presented {\it PTSO}, a probabilistic extension of the classical
TSO semantics.
We have shown decidability/computability results
 for a wide a range of properties
such as quantitative and qualitative reachability/repeated
reachability and expected average costs.
As far as we know, this is the first 
study of probabilistic verification for weak memory models,
%
%
and opens many avenues for future work.
\smallskip

\noindent\textit{Refined Probability Distributions.}
For ease of presentation, we developed our results in the
context of specific scheduling and update policies. 
However, we emphasize that our results
carry-over to policies satisfying faithfulness and
left-orientedness, which are fairly weak conditions. 
Hence we believe that developing more refined models
that better capture behaviours of TSO implementations, 
using techniques such as parameter estimation, is interesting future work.

\noindent\textit{General Cost Models}
Similar can be said for cost models: our algorithm works for all 
cost functions such that the cost of a path is exponentially bounded by its length.
In particular, developing cost models that closely mimic usage of processor resources, 
e.g. cost based on read from local store-buffer vs. read from memory, can be
useful to gain a better understanding of the implementation.

\noindent\textit{Other Memory Models}
Finally, we are interested in extending our approach to other weak memory models such
as RA/SRA, POWER, ARM.

\bibliographystyle{unsrt}
\bibliography{bibdatabase}

\vfill
{\small\medskip\noindent{\bf Open Access} This chapter is licensed under the terms of the Creative Commons\break Attribution 4.0 International License (\url{http://creativecommons.org/licenses/by/4.0/}), which permits use, sharing, adaptation, distribution and reproduction in any medium or format, as long as you give appropriate credit to the original author(s) and the source, provide a link to the Creative Commons license and indicate if changes were made.}

{\small \spaceskip .28em plus .1em minus .1em The images or other
third party material in this chapter are included in the\break
chapter's Creative Commons license, unless indicated otherwise in a
credit line to the\break material.~If material is not included in
the chapter's Creative Commons license and\break your intended use
is not permitted by statutory regulation or exceeds the
permitted\break use, you will need to obtain permission directly
from the copyright holder.}

\medskip\noindent\includegraphics{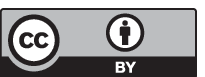}

\newpage
\appendix

\section{Glossary of Notation}

In this section we provide a glossary of notation.
\smallskip

\begin{table}
    \centering
    \setlength\tabcolsep{0.5cm}
\resizebox{\textwidth}{!}{
    \begin{tabular}{l | l | l}
        Notation & Meaning & Reference \\ \hline\hline
        \multicolumn{3}{l}{Transition Systems} \\ \hline
        $\conf$ and $\confset$ & One and a set of configuration(s) & \S\ref{prels:section} \\
        $\plainmovesto{}, \plainmovesto{*}$, $\plainmovesto{\kk}$ & Single, multi and $k$-step reachability & \S\ref{prels:section} \\
        \multicolumn{3}{l}{Temporal Logic} \\ \hline
        $\rho\modelswrt\ts\neventually{\kk}\confs$ & $\rho$ reaches $\confs$ first at the $k^{th}$ step & \S\ref{prels:section} \\
        $\rho\modelswrt\ts\nnxt{\kk}\confs$ & $\rho$ reaches $\confs$ at the $k^{th}$ step (possibly before) & \S\ref{prels:section} \\
        $\plainmovesto{}$, $\plainmovesto{\kk}$ & Simple and $k$-step reachability & \S\ref{prels:section} \\
        \multicolumn{3}{l}{Concurrent Programs} \\ \hline
        $\prog$ & A program & \S\ref{section:concprograms} \\
        $\proc$, $\procset$ & A process, set of processes & \S\ref{section:concprograms} \\
        $\regsetof\proc$, $\regsetof\prog$ & Registers of a process, registers of a program & \S\ref{section:concprograms} \\
        $\lblsetof\proc$, $\lblsetof\prog$ & Labels of a process, registers of a program & \S\ref{section:concprograms} \\
        \multicolumn{3}{l}{Operational Semantics} \\ \hline
        \multirow{2}{*}{$\conf = \conftuple$} & Labelling, Register, Buffer, Memory & \multirow{2}{*}{\S\ref{confs:semantics:section}} \\
        & components of $\conf$ & \\
        $\sizeof\conf$ & Size of (buffers of) a configuration  & \S\ref{confs:semantics:section} \\
        $\confsetof\prog$ & All configurations of $\prog$ & \S\ref{confs:semantics:section} \\
        $\plainconfsetof\prog$ & Plain (empty buffer) configurations of $\prog$ & \S\ref{confs:semantics:section} \\
        $\tsdenotationof\prog$ & Transition system of $\prog$ & \S\ref{TSO:semantics:section} \\
        $\mcdenotationof\prog$ & Markov chain for $\prog$ & \S\ref{mc:semantics:section} \\
        $\procmovesto{\proc}, \updatemovesto{}, \progmovesto{\prog}{}$ & Process, Update, Program transitions & \S\ref{TSO:semantics:section} \\
        $\weightof{\proc}, \rweightof\conf\proc$ & Weight and relative weight for scheduling & \S\ref{mc:semantics:section} \\
        \multicolumn{3}{l}{PTSO} \\ \hline
        $\smallconfsetof\prog$ & Small ($\sizeof\conf \leq 4$) configurations of $\prog$ & \S\ref{plain:attractor:subsection} \\
        $\largeconfsetof\prog$ & Large ($\sizeof\conf > 4$) configuration of $\prog$ & \S\ref{plain:attractor:subsection} \\
        \multicolumn{3}{l}{Costs} \\ \hline
        $\costof\lbl$ & Cost of instruction at $\lbl$ & \S\ref{cost:section} \\
        $\ccostof\conf{\pconf}$ & Single step cost & \S\ref{cost:section} \\
        $\runcostof\run\lbl$ & Cost of run $\run$ & \S\ref{cost:section} \\
        $\rcostof\conf\lbl\cost\run$ & Random variable for cost over runs & \S\ref{cost:section} \\ 
        $\eagernessof\prog$ & Eagerness parameter for $\prog$ & \S\ref{eagerness:subsubsection} \\
        $\eagernessthresholdof\prog$ & Eagerness bound for $\prog$ & \S\ref{eagerness:subsubsection} \\
        $\gravityof\prog$ & Gravity parameter for $\prog$ & \S\ref{gravity:subsubsection} \\
    \end{tabular}
}
\end{table}

\newpage
We now prove Lemma \ref{tso:prob:dist:lemma}.
\lemTSOProbDist*
\begin{proof}
 From any configuration $\conf$,  each 
transition consists of 2 parts : a process transition $\conf \procmovesto\proc \conf_p$ followed by an update transition $\conf_p\updatemovesto{}\pconf$. 
Assuming that the configuration $\conf$ is enabled, the process transition can be done by any enabled process. This is followed by considering all possible update transitions. Consider an enabled process $\proc$. Then $\conf \procmovesto\proc \conf_p$ happens with probability $\rweightof\conf\proc$. From $\conf_p$, 
we consider all sequences of processes which can result in an update.  
Let $S_{\conf_p,\pconf}=\{w \mid \conf_p\updatemovesto{w}\pconf\}$ be the set of sequences resulting in a fixed configuration $\pconf$ and let 
$S_{\conf_p}=\{w \mid \exists \pconf ~ \conf_p\updatemovesto{w}\pconf\}$ be all possible sequences
labelling an update from $\conf_p$.  Then the probability of reaching configuration $\conf'$  from $\conf_p$ after an update is 
$\frac{|S_{\conf_p,\pconf}|}{|S_{\conf_p}|}$. Thus, the probability to reach some configuration from $\conf_p$ after an update is $\sum_{\pconf} \frac{|S{\conf_p,\pconf}|}{|S_{\conf_p}|}=1$, since $\uplus_{\pconf} S_{\conf_p,\pconf} = S_{\conf_p}$.

\begin{enumerate}
\item 	
For an enabled configuration $\conf$, 
$\sum_{\proc \in \enabledof\conf}\sum_{\pconf\in\confsetof\prog}\procpmtrxof{\conf,\conf_p}\cdot\updatepmtrxof{\conf_p,\pconf}$,  where $\conf_p$ is the configuration such that $\conf \procmovesto\proc \conf_p\updatemovesto{}\pconf$  can be written as 

 $$\sum\limits_{\proc\in\enabledof\conf} \rweightof\conf\proc \sum\limits_{\pconf} \frac{|S_{\conf_p,\pconf}|}{|S_{\conf_p}|}=
 \sum_{\proc \in \enabledof\conf}\frac{\weightof\proc}
{\sum_{\pproc\in\enabledof\conf}\weightof{\pproc}}=1$$ 
%
%
 
%

\item For the case when $\conf$ is disabled, by definition, we have 
$\procpmtrxof{\conf,\conf}:=1$ and $\procpmtrxof{\conf,\conf_p}:=0$ for $\conf_p \neq \conf$. Further, from $\conf$, we can consider all possible update transitions 
resulting in a configuration $\pconf$. This gives us 
   $\sum_{\pconf\in\confsetof\prog}\procpmtrxof{\conf,\conf}\cdot\updatepmtrxof{\conf,\pconf}$ which can be seen to be 
    $$1. \sum\limits_{\pconf} \frac{|S_{\conf,\conf'}|}{|S_{\conf}|}=1$$

\end{enumerate}

Thus, in all cases, we have shown that 
$\progpmtrx\prog$
 is a probability distribution, and the induced transition system is a Markov Chain.

\end{proof}

Recall that in section \ref{TSO:semantics:section}, we introduced 
the notion of a transition system $\tsdenotationof\prog=\tuple{\confsetof\prog,\progmovesto\prog{}}$ given a program  $\prog$. 
The Markov chain associated to the program $\prog$ 
has been introduced in Section \ref{mc:semantics:section} as $\mcdenotationof\prog=\mchaintupleof{\prog}$. 
Now, we formally show that $\tsdenotationof\prog$ is the same as the transition system $\mctots{\left(\mcdenotationof\prog\right)}$ induced by 
$\mcdenotationof\prog$.

We begin with the proof for Lemma \ref{program:mchain:ts}, which says that the  transition system induced by the Markov chain is the same as the transition system induced by the program. 
\lemMChainTS*
\begin{proof}
First we show that if $\tuple{\conf,\pconf}\in\plainmovesto{}_{\mctots{\left(\mcdenotationof\prog\right)}}$ then $\tuple{\conf,\pconf} \in \plainmovesto{}_{\tsdenotationof\prog}$. 
Whenever $\tuple{\conf,\pconf}\in\plainmovesto{}_{\mctots{\left(\mcdenotationof\prog\right)}}$, we have  $\progpmtrx\prog{(\conf, \pconf)} > 0$. 

\begin{enumerate}
	\item Consider the case when $\conf$ is enabled. Then there exists some $\ppconf$ such that $\procpmtrxof{\conf,\ppconf}> 0$ and $\updatepmtrxof{\ppconf,\pconf}>0$. 
	Since $\procpmtrxof{\conf,\ppconf}> 0$, we have $\rweightof\conf\proc>0$ for some $\proc$ which resulted in obtaining $\ppconf$ from $\conf$.
 Hence, for process $\proc$ we have  $\conf\procmovesto\proc\ppconf$. 
 Similarly, there is a sequence $\word$ such that $\ppconf\updatemovesto\word\pconf$, i.e. $\ppconf\updatemovesto{}\pconf$. Composing the two, we obtain $\conf\plainmovesto{}_{\tsdenotationof\prog}\pconf$.

\item The second case is when $\conf$ is not enabled. Then $\procpmtrxof{\conf,\conf}=1$, by definition. An update transition can still be done from $\conf$ (empty update if all buffers are empty). In any case, the resultant configuration $\pconf$ after an update is such that $\updatepmtrxof{\conf,\pconf}>0$. Thus, as above, composing the two, we obtain $\conf\plainmovesto{}_{\tsdenotationof\prog}\pconf$ in ${\tsdenotationof\prog}$. 
\end{enumerate}

Next, we show that if $\tuple{\conf,\pconf} \in \plainmovesto{}_{\tsdenotationof\prog}$, then 
$\progpmtrx\prog{(\conf, \pconf)} > 0$. 
As above, there are two cases depending on whether $\conf$ is enabled or not.
 
 \begin{enumerate}
 	\item Assume $\conf$ is enabled. 
Then  $\plainmovesto{}_{\tsdenotationof\prog}$ is a composition of $\procmovesto{}$ and $\updatemovesto{}$. 
 There exists some process  $\proc$ and a sequence $\word$ such that $\conf\procmovesto\proc\ppconf\updatemovesto\word\pconf$. 
Hence $\procpmtrxof{\conf,\ppconf} = \rweightof\conf\proc>0$ and $\updatepmtrxof{\ppconf,\pconf} > 0$ since $\sizeof{\setcomp{\word}{\ppconf\updatemovesto\word\pconf}} > 0$. 
Hence $\progpmtrx\prog{(\conf, \pconf)} > 0$. 
\item If $\conf$ is not enabled, then all processes are disabled in $\conf$. In this case, the only 
transition in ${\tsdenotationof\prog}$ is $\conf\procmovesto{}\conf$, followed by an  update transition leading to some $\pconf$. Hence, $\progpmtrx\prog{(\conf, \pconf)} > 0$.

 \end{enumerate}
Thus we have shown that there is a  transition between a pair of configurations in 
 ${\tsdenotationof\prog}$ iff there exists a transition 
 between them in $\plainmovesto{}_{\mctots{\left(\mcdenotationof\prog\right)}}$.  
\end{proof}
%
%
%
%
%

Thanks to Lemma \ref{program:mchain:ts}, whenever 
 $\mctots{\left(\mcdenotationof\prog\right)}$ has a transition with non zero probability,  
 $\tsdenotationof\prog$ has the same transition. Thus,  it suffices to check 
reachability in $\tsdenotationof\prog$.

The reachability between plain configurations (those which have all buffers empty) follows from \cite{DBLP:conf/popl/AtigBBM10}. We prove
Lemma \ref{tso:conf:reachability:lemma} by reducing 
reachability from a given configuration to a plain configuration to the reachability problem between two plain configurations.
 Likewise, label reachability 
is
known to be decidable when starting from a plain configuration 
 in classical TSO semantics \cite{DBLP:journals/lmcs/AbdullaABN18}. We can prove 
Lemma \ref{tso:lbl:reachability:lemma}  in a similar manner (as in Lemma \ref{tso:conf:reachability:lemma}) by reducing 
the label reachability from a given configuration 
to label reachability from a plain configuration, and then invoking \cite{DBLP:journals/lmcs/AbdullaABN18}.

\lemTSOConfReachability*
\begin{proof}
	
Given a program $\prog$, and a configuration 
$\conf=\conftuple$, and a plain configuration $\pconf=\pconftuple$, can we reach $\pconf$ from $\conf$? 
Assume that there are $n$ processes 
$\proc_1, \dots, \proc_n$,  with shared variables 
$x_1, \dots, x_m$. 
 The bufferstate $\bufferstate$ consists of words $w_1, w_2, \dots, w_n$ where $w_i$ is the buffer content 
of process $i$ in $\bufferstate$. Note that 
each $w_i$ is a finite length word. Assume that $|w_i|=k_i$, and $w_i$ has the form $(x_{i1}, v_{i1})\dots (x_{ik_i}, v_{ik_i})$. 

We modify $\prog$ to a program $\prog'$ 
by (i) modifying the code of $\proc_1, \dots, \proc_n$ 
and by adding a new process $\proc_{n+1}$, (ii) we introduce two new local registers $r, r'$ 
in $\proc_1$, a new shared variable $flag$, initializing all of them to 0. We reduce the reachability of $\conf$ to $\pconf$ in $\prog$ 
to the reachability between two plain configurations in $\prog'$.

\begin{enumerate}
	\item Assume that $\memstate$ is given by 
	$x_i \mapsto u_i$, for all $1 \leq i \leq m$. 
		We alter $\proc_1$ by adding some instructions before 
		all the existing  instructions in $\proc_1$.

The new instructions added to $\proc_1$ 
are the following. 
 We begin with a while loop which checks $flag$ is 0. Inside the loop, we have the following.
    For $1 \leq i \leq m$, we add the instructions 
$r'=u_i; CAS(x_i, r, r')$. When we finish executing all $m$ of these instructions, 
this results in the memory state as given by $\memstate$. 
This is followed by writing $v_{11}$ to $x_{11}$,  $\dots,v_{1k_1}$ to $x_{1k_1}$, obtaining the buffer content 
of $\proc_1$ in $\bufferstate$ and also writing the appropriate values to all registers 
of $\proc_1$ as in $\regstate$.   This is followed by setting $flag$ to 1. The while loop is broken at this point. 
The next instruction checks if $flag$ is $n+1$, and if so, 
goes to the label $\ell_1$ (the control location of $\proc_1$ in $\labeling$). 

\item Processes $\proc_i$ for $2 \leq i \leq n$ are modified as follows. Add a new register $r_i$ 
\item for each $p_i$, and initialize to 0.  
 We begin with a while loop which checks $flag$ is $i-1$. Inside the loop, we have the following.
Using $r_i$, we write $v_{i1}$ to $x_{i1}$, and so on until we write $v_{ik_i}$ to $x_{ik_i}$, obtaining the buffer content 
of $\proc_i$ in $\bufferstate$. Then we write the appropriate values to all registers 
of $\proc_i$ as in $\regstate$.   This is followed by setting $flag$ to $i$. The while loop is broken at this point. 
The next instruction checks if $flag$ is $n+1$, and if so, 
goes to the label $\ell_i$ (the control location of $\proc_i$ in $\labeling$). 

\item For $\proc_{n+1}$, we begin with a while loop which checks if $flag$ is $n$. Inside the loop, it reads the memory and checks that it agrees with $\memstate$. This is possible as 
the buffer of $\proc_{n+1}$ is empty. After the check,  it changes the $flag$ to $n+1$, breaking the while loop and reaching the terminal instruction of $p_{n+1}$. 
\end{enumerate}

The modified program $\prog'$ has 
polynomially many extra instructions at the beginning of each process. Starting from the initial configuration $\initconf$ consisting 
of all initial labels of instructions 
in all processes $p_1, \dots, p_{n+1}$, with 
all variables and registers having value 0, and all empty buffers,  
 $\prog'$ 
first executes these extra instructions in all processes. The new shared variable $flag$ moves the processes 1 to $n+1$ in order until we obtain (i) the memory state as in $\memstate$,  (ii) buffer contents of all processes $\proc_1, \dots, \proc_n$ as in $\bufferstate$, (iii) local registers of all processes  $\proc_1, \dots, \proc_n$ as in $\regstate$. When $flag$ becomes $n+1$, all processes $\proc_1, \dots, \proc_n$  move to the 
control locations given by $\labeling$ in $\conf$. 
Ignoring $flag$ as well as the two new local registers added to $\proc_1$,  and $\proc_{n+1}$, the configuration of $\prog'$ at this point is $\conf=\conftuple$, the configuration 
given to us. 

Define a configuration $\ppconf=\ppconftuple$ as follows.
\begin{enumerate}
\item The labeling $\pplabeling$ agrees with $\plabeling$ for processes $p_1, \dots, p_n$,
\item 	$\ppbufferstate$ agrees with  $\pbufferstate$ in terms of buffer states of $p_1, \dots, p_n$, 
\item $\ppmemstate$ agrees with $\pmemstate$, the  memory state wrt $x_1, \dots, x_m$
\item $\ppregstate$ agrees with $\pregstate$, the  register state wrt the registers in $\prog$. In addition,  
\item the buffer of $p_{n+1}$ is empty in $\ppbufferstate$, 
\item The new shared variable $flag \mapsto n+1$ in the memorystate $\ppmemstate$, 
\item The new registers $r, r'$ of $p_1$ are such that 
$r$ has value 0 and $r'$ has value $u_m$ in $\ppregstate$,
\item The new register $r_i$ of $p_i$ for $2 \leq i \leq m$  has value $v_{ik_i}$ in $\ppregstate$.
\item The label of $p_{n+1}$ in $\pplabeling$ 
is the term instruction in $p_{n+1}$.  
\end{enumerate}

 We now ask the reachability query from $\initconf$ to $\ppconf$ in $\prog'$ 
which is known to be decidable \cite{DBLP:conf/popl/AtigBBM10}, since both $\initconf$ and $\ppconf$ are plain. 
Note that $\prog'$ starts simulating $\prog$ only 
when it reaches a configuration whose projection 
to the processes $\proc_1, \dots, \proc_n$, 
modulo the new registers and $flag$  is $\conf$. Indeed, if $\ppconf$ is reachable in $\prog'$ from $\initconf$,  then it must be that (i) Ignoring values of new registers and $flag$, 
$\conf$ is reachable from $\initconf$ in $\prog'$ (this follows by construction), and 
(ii) $\ppconf$ is reachable in $\prog'$ from $\conf$. Indeed since $\ppconf$ when projected to the old registers and shared variables and $p_1, \dots, p_n$ is $\pconf$, we obtain the reachability of $\pconf$ from $\conf$ in $\prog$.

\end{proof}

Finally we prove Lemma \ref{finite:attractor:lemma}.

\lemFiniteAttractor*
\label{app:basics:finiteattractor}

\begin{proof}
    \newcommand{\expectationof}[1]{\mathbb{E}(#1)}
    \newcommand{\astep}{t}
    \newcommand{\confsize}{\textrm{X}}
    \newcommand{\setofruns}{\mathsf{L}}

    We will show that $\plainconfsetof\prog$ is a finite attractor in the sense of \cite{AbdullaHM07}. 
    An attractor \cite{AbdullaHM07}, is a set of configurations which is eventually reached with probability 1 from every configuration in the Markov Chain $\mcdenotationof\prog$.  
    
    $\plainconfsetof\prog$ contains configurations with empty buffers. We intuitively want to show that the system behaviours tend to concentrate towards these. We make this notion precise through the concept of configuration size, $|\conf|$. It is sufficient to show that the expected value of $|\conf|$ at each step decreases (when transitioning from configurations with sufficiently large buffers). The expectation is over the possible transitions from $\conf$.

    We show this in two steps. We first show that
    \begin{equation*}
        \measureof\prog{\conf\models\always\eventually\lblconfsetof{\prog}{\leq k}}=1
    \end{equation*}
    for constant $k=4$ for all programs $\prog$ and configurations $\conf\in\confsetof\prog$. Recall that $\lblconfsetof{\prog}{\leq k}$ is the set of configurations with size (sum of buffer lengths) at most $k = 4$. Then we use this to prove the statement of the lemma. First to show that $\lblconfsetof{\prog}{\leq 4}$ is an attractor, we use the following result 
    from \cite{10.5555/1126294.1126298}. 

    \paragraph{Left-oriented Markov chains (\cite{10.5555/1126294.1126298})}
     Baier et al. consider (infinite) Markov chains where the state space $S$ is partitioned into non-negative integer labelled levels $\cup_{i \in \mathbb{N}}S_i$. For $s \in S$, the level of $s$, 
    $\levelfun(s)$ is $i$ if $s \in S_i$.  Then 
    $\expectationof{s}=\sum_{j=0}^\infty \measureof{}{s,S_j} \cdot j$ defines the 
    expected next level for state $s$. 
    The Markov chain is called left oriented 
iff there is a positive constant $\Delta$ such that 
$\expectationof{s} \leq \levelfun(s)-\Delta$  for all states $s \notin S_0$, that is, for all states $s$ at level 1 or more.  
 Theorem 2.1 from \cite{10.5555/1126294.1126298}
 shows  that for any left oriented Markov chain, the leftmost level 
$S_0$ is an attractor.  
%
%
%
%

We leverage Theorem 2.1 from \cite{10.5555/1126294.1126298} in our proof. To do this, we show that our Markov Chain 
$\mcdenotationof\prog$ is left oriented.  The ``levels'' in $\mcdenotationof\prog$  are the configuration sizes, except for the set $\lblconfsetof{\prog}{\leq 4}$. We formalize this as an abstraction function $\levelfun$ from the configuration set $\confsetof\prog$ to (non-negative) integers.
    
    \begin{align*}
        \levelfun(\conf) = 0 &\qquad \text{ if } \conf \in \lblconfsetof{\prog}{\leq 4} \\
        \levelfun(\conf) = |\conf| &\qquad \text{ otherwise }
    \end{align*}
    For a configuration $\conf$ in $\{\conf ~|~ \levelfun(\conf) > 0\}$, let $X_\conf$ be the random variable representing the next configuration obtained after a single transition from $\conf$. We then show that for all configurations in $\{\conf ~|~ \levelfun(\conf) > 0\}$, the single step expected change in the $\levelfun$ is negative. That is forall $\conf \in \lblconfsetof{\prog}{\geq 5}$, 
    $\expectationof{level(X_{\conf})} \leq |\conf| -\Delta$  
    where $\Delta$ is a positive constant
    where,
    \begin{equation*}
        \expectationof{level(X_{\conf})} = \sum_{\pconf} level(\pconf)\cdot\prmpmtrxof\prog{\conf, \conf'}
    \end{equation*}

    
    We show that $\sum_{\conf'} |\conf'| \cdot \prmpmtrxof\prog{\conf,\pconf} \leq |\conf| - \Delta$, (we replace $level(\conf')$ by $|\pconf|$ since $level(\conf') \leq |\pconf|$).
    Additionally, $|\conf|$ can be written as $|\conf| \cdot \sum_{\conf'}\prmpmtrxof\prog{\conf,\pconf}$, and hence we want to show:
    \begin{align*}
        \sum_{\conf'} (|\conf'| - |\conf|) \cdot \prmpmtrxof\prog{\conf,\pconf} < -\Delta
    \end{align*}
    Showing this helps us to conclude that $\mcdenotationof\prog$ is left oriented. Now, we focus on showing this. 

    \paragraph{$\mcdenotationof\prog$ is left-oriented}

    Consider a transition from the configuration $\conf^t$ (with $|\conf^t| \geq 5$), 
    to the configuration $\conf^{t+1}$ as a result of the \textit{process} ($\procmovesto{}$) and \textit{update} ($\updatemovesto{}$) sub-transitions. For a particular $\conf^{t+1}$, let $\pconf$ be the intermediate configuration satisfying $\conf^t\procmovesto{}\pconf\updatemovesto{}\conf^{t+1}$.  
    We can write the single step change in expectation when going from $\conf^t$ to $\conf^{t+1}$ as follows.
    \begin{align*}
        &\sum_{\conf^{t+1}} (|\conf^{t+1}| - |\conf^t|) \cdot \prmpmtrxof\prog{\conf^t,\conf^{t+1}} \\
        &= \sum_{\pconf} \procpmtrxof{\conf^t, \pconf} \cdot\left(\underbrace{|\pconf| - |\conf^t|}_{\procmovesto{}} + \sum_{\conf^{t+1}} \updatepmtrxof{\pconf, \conf^{t+1}}\cdot(\underbrace{|\conf^{t+1}|-|\pconf|}_{\updatemovesto{}}) \right)
    \end{align*}
    
    The first term $|\pconf| - |\conf^{t}|$ is the number of elements added to the buffer in $\conf^t\procmovesto{}\pconf$ and the expectation is over the $\procmovesto{}$ probabilities while the second term $|\conf^{t+1}| - |\pconf|$ is the number of elements removed from the buffer in $\pconf\updatemovesto{}\conf^{t+1}$ and the expectation is over the $\updatemovesto{}$ probabilities. 
    
    Now we see that (by Figure \ref{TSO:semantics:fig}) in any $\procmovesto{}$ step either one element is added to the buffer (in case of a \rulename{write} transition) or the buffer remains the same (for all other transitions). In particular, even when the \rulename{disabled} rule is taken, the configuration remains the same. We have the following.
    \begin{align*}
        \forall \conf' \qquad\conf^t\procmovesto{}\pconf &\implies 0 \leq |\pconf| - |\conf^{t}| \leqslant 1
    \end{align*}
    Since this holds for all $\conf, \pconf$ pairs, it certainly holds for expected values, and we can substitute in the above equation:
    \begin{align*}
        &\sum_{\conf^{t+1}} (|\conf^{t+1}| - |\conf^t|) \cdot \prmpmtrxof\prog{\conf^t,\conf^{t+1}} \\
        &\leq\qquad \sum_{\pconf} \procpmtrxof{\conf^t, \pconf} \cdot\left(1 + \sum_{\conf^{t+1}} \updatepmtrxof{\pconf, \conf^{t+1}} \cdot(|\conf^{t+1}|-|\pconf|) \right)
    \end{align*}

    As for the update part of the transition, we consider a uniform distribution over all possible update sequences from the state $\pconf$. Due to the combinatorial term involved, the probability of an update that propagates more writes to the memory is strictly greater than that of an update which propagates shorter lengths. Even this conservative reasoning implies that the update rule leads to a configuration with at most half the size with at least probability of $\frac{1}{2}$, giving us the following bound. Below, $\measureof{\updatemovesto{}}{x}$ denotes the probability of an update where $x$ specifies how many elements 
    from the buffer are pushed.

    \begin{align*}
        &\sum_{\conf^{t+1}}\updatepmtrxof{\pconf, \conf^{t+1}} \cdot(|\conf^{t+1}|-|\pconf|)
                 \\
        &= -\sum_{i=0}^{|\pconf|} i\cdot\measureof{\updatemovesto{}}{i \text{ elements of the buffer are updated}}  \\
        &\leq -\frac{|\pconf|}{2} \cdot\measureof{\updatemovesto{}}{\text{at least} \frac{|\pconf|}{2} \text{ elements are updated}} \leq -\frac{|\pconf|}{4}
    \end{align*}
    Here the first inequality follows from the fact that we are discounting for cases where less than $|\pconf|/2$ elements are updated. The second inequality follows from the fact that the probability of updating at least $|\pconf|/2$ elements is greater than $\frac{1}{2}$.
    We note that this holds for each configuration $\pconf$, and hence substituting this bound in the earlier expression, we get (noting that $|\conf'| \geq |\conf^t|\geq  5$),
    \begin{align*}
        \sum_{\conf^{t+1}} (|\conf^{t+1}| - |\conf^t|) \cdot \prmpmtrxof\prog{\conf^t,\conf^{t+1}}  &\leq \sum_{\pconf} \procpmtrxof{\conf^t, \pconf} \cdot\left(1 - \frac{|\pconf|}{4} \right) \\
        &\leq \sum_{\pconf} \procpmtrxof{\conf^t, \pconf} \cdot\left(1 - \frac{5}{4} \right) \\
        & \leq \left(1 - \frac{5}{4} \right) \cdot \sum_{\pconf} \procpmtrxof{\conf^t, \pconf} \leq -\frac{1}{4}  \\
    \end{align*}

    This proves that $\mcdenotationof\prog$ is left-oriented (for $\Delta = \frac{1}{4}$). Hence by invoking Theorem 2.1 from \cite{10.5555/1126294.1126298} we get that, the level 0 set is an attractor. That is $\measureof\prog{\conf\modelswrt\ts \always\eventually\lblconfsetof\prog{\leq 4}} = 1$.

    Now we want to show that $\measureof\prog{\conf\models\always\eventually\plainconfsetof\prog}=1$. But this follows directly from the notion of probabilistic fairness (for details see Theorem 10.25 of \cite{PMC:book}). We have (even for infinite Markov chains $\mathcal{M}$) and (finite) sets of configurations $T, T'$,
    \begin{equation*}
        \measureof{\mathcal{M}}{s\models\always\eventually T} = \measureof{\mathcal{M}}{s\models\bigwedge_{T' \subseteq Post^*(T)}\always\eventually T'}
    \end{equation*}
    We note that the sets $\plainconfsetof\prog$ and $\lblconfsetof\prog{\leq 4}$ are indeed finite (since the buffer sizes are bounded) and $\plainconfsetof\prog$ is reachable from $\lblconfsetof\prog{\leq 4}$, since the update rule can simply empty all buffers with non-zero probability. Hence instantiating $s = \conf$, $T = \lblconfsetof\prog{\leq 4}$, $T' = \plainconfsetof\prog$ we have that $\measureof\prog{\conf\models\always\eventually\plainconfsetof\prog}=1$ as desired.

\end{proof}

\section{Qualitative Reachability: Supplementary material for Sec. \ref{qual:reachability:section}}

In this section, we discuss the results for qualitative reachability and repeated reachability, filling in the 
details ommitted in the main paper. 
In particular, we provide algorithms for the almost-sure repeated reachability 
and almost-never repeated reachability.

Then, we go on to prove Theorems \ref{qual:decidability:theorem} and \ref{qual:complexity:theorem}.

\subsection{Almost-Sure Repeated Reachability}
The qualitative repeated reachability problem, differs 
from reachability in that now we are
interested in the property $\always\eventually\lbl$
i.e., whether $\lbl$ will be reached {\it infinitely often}.
%
In a similar manner
to the case of reachability,
the algorithm of Fig.~\ref{qual:rep:reach:algorithm:fig}
	analyzes 
the transition system $\tsdenotationof\prog$.

The difference between the two algorithms
is that we do not need to transform the program by removing the label $\lbl$ here, 
since reaching $\lbl$ a finite number of times 
does not affect repeated reachability.
Therefore, we perform analysis directly on the
input program $\prog$.
As before, the loop on line \algolinenum{1} 
generates all the plain configurations one by one,
and performs the same tests as in the qualitative reachability
algorithm of Fig.~\ref{qual:reach:algorithm:fig}.
More precisely,
the algorithm terminates and returns a negative answer
if it finds a plain configuration
that is reachable from $\initconf$ but that cannot reach $\lbl$.
Otherwise, it returns a positive answer.
\begin{wrapfigure}[11]{r}{0.45\textwidth}
	
	\resizebox{0.4\textwidth}{!}{
	\begin{minipage}{0.5\textwidth}
	\begin{algorithm}[H]
		\SetAlgoRefName{}
		\KwIn{
		$\prog$: program; 
		$\initconf\in\confsetof\prog$: configuration;
		$\lbl\in\lblset_\prog$: label.
		}
		\For{{\bf each} $\conf\in\plainconfsetof{\prog}$}{\label{qual:rep:reach:fig:for:line}
			\lIf{$\initconf\progmovesto\prog*\conf$ {\bf and} $\neg\left(\conf\progmovesto\prog*\lbl\right)$}{
			\KwRet{$\false$}
			}
		} 
		\KwRet{$\true$}
	\caption{\qualrepreach}
	\end{algorithm}
	\end{minipage}}
	\vspace{-0.2cm}
	\caption{Almost-sure repeated reachability.}
	\label{qual:rep:reach:algorithm:fig}
\end{wrapfigure}
To see the correctness of the algorithm, we observe that
it answers negatively only if it  finds a path with a positive probability 
from $\initconf$ to a plain configuration from which $\lbl$ is not reachable.
Using a 
similar reasoning to the case of reachability,
this implies that the measure of runs from $\initconf$ that reach 
$\lbl$ is smaller than one.
Therefore, the measure of runs from $\initconf$ that {\it repeatedly} reach 
$\lbl$ is also smaller than one.
In the other direction, if the algorithm answers positively then 
there is no
plain configuration
that is reachable from $\initconf$ but that cannot reach $\lbl$.
Using Lemma~\ref{finite:attractor:lemma},
as in Sec.~\ref{qual:plain:reachability:section}, we
conclude that any run 
$\run$ from $\initconf$ will almost surely repeatedly visit some
plain configuration $\conf$ from which $\lbl$ is reachable.
Consequently,
$\run$ almost surely visits $\lbl$ infinitely often.

\subsection{Almost-Never (Repeated) Reachability}
The almost-never variants 
of the (repeated) reachability problems, \neverqualreach resp.\ \neverqualrepreach, ask
%
whether the probabilities equal to $0$ rather than $1$.
%
%
The solution to \neverqualreach is straightforward, since 
$\measureof\prog{\initconf\models\eventually \lbl}=0$
iff $\neg(\initconf\progmovesto\prog*\lbl)$
the latter is decidable by Lemma~\ref{tso:lbl:reachability:lemma}.

\begin{wrapfigure}[14]{r}{0.5\textwidth}
\vspace{-0.1cm}
\resizebox{0.5\textwidth}{!}{
\begin{minipage}{0.6\textwidth}
\begin{algorithm}[H]
  \SetAlgoRefName{}
  \DontPrintSemicolon
  \KwIn{
    $\prog$: program; 
    $\initconf\in\confsetof\prog$: configuration;
    $\lbl\in\lblset_\prog$: label.
  }
    \For{{\bf each} $\conf\in\plainconfsetof{\prog}$}{
      \If{$\initconf\progmovesto\prog*\conf$}{
        \flagvar\assigned\true\;
            \For{{\bf each} $\pconf\in\plainconfsetof{\prog}$}{
              \lIf{$\conf\progmovesto\prog*\pconf$ {\bf and} $\neg\left(\pconf\progmovesto\prog*\conf\right)$}{
                \flagvar\assigned\false
              }
            }
            \lIf{$\flagvar=\true$ {\bf and} $\conf\progmovesto\prog*\lbl$}{
              \KwRet{$\false$}
            }
      }
    }
    \KwRet{$\true$}
\caption{\neverqualrepreach}
\end{algorithm}
\end{minipage}}
\vspace{-0.2cm}
\caption{Almost-never repeated reachability algorithm.}
\label{qual:never:rep:reach:algorithm:fig}
\end{wrapfigure}
  
We give the algorithm for solving the almost-never repeated
reachability problem in Fig.~\ref{qual:never:rep:reach:algorithm:fig}.
The algorithm searches for B-plain configurations
that are reachable from $\initconf$ and from which $\lbl$ is reachable.
If it detects such a configuration, it returns a negative answer.
Otherwise, it returns a positive answer.
%
%
The outer loop cycles through all plain configurations.
For each such configuration $\conf$ that is reachable
from $\initconf$,
the inner loop  
checks whether $\conf$ is a B-plain configuration 
(by trying to search for a plain configuration which is reachable but without a path back to $\conf$).
%

We intuitively explain the algorithm as follows.
Suppose there is a $\conf\in\bplainconfsetof\prog$ 
such
that $\initconf \progmovesto\prog* \conf \progmovesto\prog* \lbl$.
%
By Lemma~\ref{finite:attractor:lemma} and the fact that $\conf$
is a B-plain configuration, any run from $\conf$
will almost surely visit $\conf$ infinitely often.
Hence by the fairness property for Markov chains 
it follows that 
the run almost surely visits $\lbl$ infinitely often
and we return false.
%
%
Conversely, assume  there is no B-plain configuration 
satisfying $\initconf \progmovesto\prog* \conf \progmovesto\prog* \lbl$.
By Lemma~\ref{bplain:finite:attractor:lemma}, we know that
any run $\run$ from $\initconf$ will visits some 
B-plain configuration $\conf$ infinitely often.
Since $\lbl$ is not reachable from $\conf$ it follows that
$\run$ will almost never visit $\lbl$.

\subsection{Proofs for Sec. \ref{qual:reachability:section}}

\thmQualReachDec*

\paragraph{Reachability}

\begin{proof}
	Decidability follows by proving that the algorithm given in Figure \ref{qual:reach:algorithm:fig} gives the correct answer and terminates, which we now do.

	\paragraph{Correctness}

	(\textbf{Algorithm returns} \texttt{false}) When the algorithm returns \texttt{false}, we know that there exists a plain configuration $\conf$ and a finite length path $\initconf \progmovesto{\prog'}*\conf$, such that $\neg(\conf \progmovesto{\prog'}*\lbl)$. Since it is a finite path, it is taken with a non-zero probability, say \texttt{p}. Then we know that $\measureof{\prog}{\initconf\models\eventually\lbl} \leq 1-\texttt{p} < 1$ and we are done.
	
	\newcommand{\reachplain}{\confset^{\texttt{r-plain}}_\prog}
	\newcommand{\minprob}{\texttt{p}}

	(\textbf{Algorithm returns} \texttt{true}) Let the set of plain configurations (in the original program $\prog$) reachable from $\initconf$ be $\reachplain$.  Given that there are finitely many plain configurations, $|\reachplain|$ is also finite. Since the algorithm returned \texttt{true} we know that $\lbl$ is reachable from each configuration in $\reachplain$, with a finite length path and hence some non-zero probability. Let $\texttt{p}$ denote the minimum of these reachability probabilities over the (finite) set $\reachplain$. We must have $\texttt{p} > 0$, since $\reachplain\neq \phi$. Hence starting from any configuration in $\reachplain$, the probability of reaching $\lbl$ is at least $\texttt{p} > 0$. Now, Lemma \ref{finite:attractor:lemma} implies that the set $\reachplain$ is reached infinitely often, $\measureof{\prog}{\initconf\models \always\eventually\reachplain} = 1$. 

	Then the result follows from the fairness theorem for Markov chains which says,
	\begin{equation*}
		\measureof{\prog}{\conf \models \always\eventually \confset} = \measureof{\prog}{\conf \models \bigwedge_{
		\conf' \in Post^*(\confset)
		} \always\eventually \conf'}.
	\end{equation*}
	As a corollary, we get,
	\begin{equation*}
		\measureof{\prog}{\conf \models \always\eventually \confset} \leq \measureof{\prog}{\conf \models \always\eventually \confset'} \quad\quad \text{ for } \confset' \subseteq ~Post^*(\confset).
	\end{equation*}
	In particular, instantiating $\conf = \initconf, \confset = \reachplain, \confset' = \{\conf' ~|~ \lbl \in \conf'\}$, we get
	\begin{align*}
		\measureof{\prog}{\initconf \models \always\eventually \reachplain} = 1 &\leq \measureof{\prog}{\initconf \models \always\eventually \lbl} \\
		&\leq \measureof{\prog}{\initconf \models \eventually \lbl}.
	\end{align*}
	Both the inequalities must be equalities which completes the proof for correctness.

	\paragraph{Termination}
	The set of plain configurations is finite, and hence the loop performs finite iterations. At each iteration, both of the queries are decidable as discussed in Lemma \ref{tso:lbl:reachability:lemma} and Lemma \ref{tso:conf:reachability:lemma}. This shows termination and proves the theorem.

\end{proof}

\paragraph{Repeated Reachability}

\begin{proof}
\newcommand{\reachplain}{\confset^{\texttt{r-plain}}_\prog}
	\newcommand{\minprob}{\texttt{p}}

	This proof is similar to its reachability counterpart. The algorithm returns true
	only when all plain configurations 
	$\conf \in \reachplain$ reachable from $\initconf$, 
	are such that $\lbl$ is also reachable from $\conf$. By Lemma \ref{finite:attractor:lemma} 
	we know that any run 
	from $\initconf$ visits almost surely, some plain configuration 
	infinitely often. In our case, any run from $\initconf$ visits some configuration from $\conf \in \reachplain$ infinitely often; since $\lbl$ is reachable from $\conf$, 
	the run from $\initconf$ which reaches $\lbl$ will visit almost surely, $\conf$ again and again, and hence $\lbl$ also, infinitely often, almost surely.
	In particular, we show correctness and termination for the algorithm in Figure \ref{qual:rep:reach:algorithm:fig}.

	\paragraph{Correctness}
		Proof of correctness is identical to the earlier. In particular, at the last step we also showed
		\begin{equation*}
			\measureof{\prog}{\initconf \models \always\eventually \reachplain} = 1 \leq \measureof{\prog}{\initconf \models \always\eventually \lbl}
		\end{equation*}
		where the inequality must be an equality, owing to the probability values.
	
	\paragraph{Termination}
	The set of plain configurations is finite, and hence the loop performs finite iterations. At each iteration, both of the queries are decidable as discussed in Lemma \ref{tso:lbl:reachability:lemma} and Lemma \ref{tso:conf:reachability:lemma}. This shows termination and proves the theorem.
\end{proof}

\paragraph{Almost-Never Repeated Reachability}
\begin{proof}
\newcommand{\reachplain}{\confset^{\texttt{r-plain}}_\prog}
	\newcommand{\minprob}{\texttt{p}}

\emph{Correctness}. 	(\textbf{Algorithm returns} \texttt{false}) When the algorithm returns \texttt{false}, we know that there exists a B-plain configuration $\conf$ reachable from $\initconf$  
and a finite length path $\conf \progmovesto{\prog}* \lbl$. Since it is a finite path, it is taken with a non-zero probability, say $\texttt{p}$. By Lemma \ref{finite:attractor:lemma} and the fact that $\conf$ can reach back to itself, we know that any run from  
 $\conf$ visits $\conf$ infinitely often. 
 Since $\lbl$ is reachable from $\conf$ with probability $\texttt{p}>0$, we can reach $\lbl$ infinitely often with probability $>0$. Thus, 
  $\measureof{\prog}{\initconf\models \Box \eventually\lbl} > 0$ and we are done.

(\textbf{Algorithm returns} \texttt{true}) 
Assume that all B-plain configurations $\conf$ reachable from $\initconf$ are such that $\neg(\conf \progmovesto{\prog}* \lbl)$. Then we know 
by Lemma \ref{finite:attractor:lemma}, $\conf$ is visited infinitely often on any run from $\initconf$;
 since $\lbl$ is not reachable from $\conf$, it is not repeatedly reachable along any run from $\initconf$ (
 if $\lbl$ were repeatedly reachable on some run from $\initconf$, then since some B-plain configuration $\conf$ is also visited infinitely often along that run, it would result in reaching $\lbl$ from $\conf$, contradicting 
 the non reachability of $\lbl$  
from $\conf$). Thus, the probability of repeatedly reaching $\lbl$ from $\initconf$ is 0 :  
$\measureof{\prog}{\initconf\models \always\eventually\lbl} = 0$. 

\paragraph{Termination}
	
	The set of plain configurations is finite, and hence the loop performs finite iterations. At each iteration, both of the queries 
(checking if a plain configuration $\conf$ is B-plain : that is, it is reachable from itself, as well as whether $\lbl$ is reachable from $\conf$)	
	are decidable as discussed in Lemma \ref{tso:lbl:reachability:lemma} and Lemma \ref{tso:conf:reachability:lemma}. This shows termination and proves the theorem.
\end{proof}

\thmQualReachNPR*

\paragraph{Reachability}

\begin{proof}[Proof]
	The proof  follows from a reduction from the (non-probabilistic) control-state reachability problem for TSO, which is known to be non-primitive recursive \cite{DBLP:conf/popl/AtigBBM10}.
	
	 Concretely, consider an instance program $\prog$ of the (non-probabilistic) reachability problem for TSO. We ask whether $\initconf \progmovesto\prog*\lbl^*$ for some label $\lbl^* \in \lblsetof{\proc_r}$
	  (we ask for reachability of process $\proc_{r}$, and other processes can be in arbitrary labels).

	\paragraph{Reduction construction}
	For the reduction, we construct a program  $\prog'$ obtained from $\prog$ by modifying process $\proc_r$, and adding a new process $\proc_{new}$. All other processes of $\prog$ remain unchanged. $\prog'$ has all the shared variables and local registers as $\prog$,
	and a new shared variable $x$, as well as 
	a new local register $a$. As always, all shared variables and local registers are initialized to 0. Assume the finite data domain of $\prog$ (and $\prog'$) consists 
	of values $\{v_0,v_1, \dots, v_n\}$, where $v_0=0$. 
		 
\begin{enumerate}
	\item 
		 $\proc_{new}$ consists of a sequence of instructions which starts by checking 
		 if $a$ is $v_0$, and if so,  
		 goes to the next two instructions which are  
		 $a:=v_1;x:=a$. This is followed by an instruction which checks if $a$ is $v_1$, and if so, goes to the next two instructions which are  
		 $a:=v_2;x:=a$. This is continued till we 
		 reach the instruction which checks if $a$ is $v_n$ and if so, goes to the first instruction. Thus, $\proc_{new}$ is a loop which repeatedly writes values $v_0, \dots, v_n$ to $x$. 
		    
		   \item  Now we discuss the modifications 
		   in $\proc_r$. 
		   
		   \begin{itemize}
		   \item We add two fresh instruction labels 
		   $\lbl_{win}$ and $\lbl_{lose}$ to $\proc_r$ such that 
		  $$\lbl_{win} : {\tt{if}}~(a\geq 0)~{\tt{then}}~~\lbl_{win}, ~~
		  		   \lbl_{lose} : {\tt{if}}~(a\geq 0)~{\tt{then}}~~\lbl_{lose}$$

		   	\item Immediately before 
		     each instruction label $\lbl \neq \lbl^*$ in 
		    $\proc_r$, we add two fresh instructions 
		   labeled  $\lbl_{pre1}$ and $\lbl_{pre2}$ as follows. 
		   $$\lbl_{pre1}: a:=x,~~~~\lbl_{pre2} : {\tt{if}}(a=v_i)~{\tt{then}}~~\lbl_{win}$$ 
		  
		  Thus, if the value of $a$ is not $v_i$, control moves to $\lbl$; otherwise to $\lbl_{win}$. 
		   		   
		   \item Immediately before $\lbl^*$, we 
		    add  two fresh instructions 
		   labeled  $\lbl^*_{pre1}$ and $\lbl^*_{pre2}$ as follows. 
		   $$\lbl^*_{pre1}: a:=x,~~~~\lbl^*_{pre2} : {\tt{if}}(a=v_i)~{\tt{then}}~~ \lbl_{lose}$$ 
		   
		  Thus, if the value of $a$ is not $v_i$, control moves to $\lbl^*$; otherwise to $\lbl_{lose}$. 

		   \end{itemize}

\end{enumerate}

%
%
%
%
%
%
%
	We see that $\neg(\lbl_{win}\progmovesto{\prog'}{*}\lbl_{lose})$ and $\neg(\lbl_{lose}\progmovesto{\prog'}*\lbl_{win})$. 

	\paragraph{Equivalence}
	For this program $\prog'$,  we ask the probabilistic qualitative reachability problem: $\measureof{\prog'}{\initconf\models\eventually\lbl_{win}}=1?$ We claim that an answer to this question, allows us to decide reachability to $\lbl^*$ in $\prog$. 

	($\Rightarrow$) If the answer to this is yes, we know that $\neg(\initconf \progmovesto{\prog}*\lbl^*)$, since a (finite-length) path to $\lbl^*$, implies a finite length path to  $\lbl^*_{pre1}, \lbl^*_{pre2}$. This implies a finite 
		and hence non-zero probability path to reach $\lbl_{lose}$, from which there is no path to $\lbl_{win}$.

	\newcommand{\winconf}{\confset^{win}_{\prog'}}

	($\Leftarrow$) On the other hand, if 
	$\lbl^*$ is not reachable in $\prog$, we have
	 $\initconf\models \always 
	(\lblsetof{\proc_{r}} \setminus\{\lbl^*\})$ and hence 
				$\initconf\models \always (\lblsetof{\proc'_{r}} \setminus\{\lbl_{lose}, \lbl^*, \lbl^*_{pre1}, \lbl^*_{pre2}\})$ in $\prog'$. In the 
				extended program $\prog'$, define $\winconf$ as the set of configurations with $\labeling(\proc_r') = \lbl_{win}$. Additionally, we have that all reachable configurations $\lblsetof{\proc'_{r}} \setminus
				\{ \lbl_{lose}, \lbl^*, \lbl^*_{pre1}, \lbl^*_{pre2}\}$ can reach $\winconf$ in a single step (with a non-zero probability). Thus, $\lbl_{win}$ is reachable in a single step with non zero probability from all configurations which are reachable from $\initconf$. 
		This implies that the probability to reach $\winconf$ from $\initconf$ is 1, which proves the lemma.
\end{proof}


\paragraph{Repeated Reachability}
\begin{proof}[Proof]
	This proof is identical to its reachability counterpart. In particular note that $\lbl_{win}, \lbl_{lose}$ are associated with self loop transitions, so, reachability and repeated reachability for $\lbl_{win}, \lbl_{lose}$ are equivalent problems for program $\prog'$.
\end{proof}

\paragraph{Never Reachability}
\begin{proof}[Proof]
We can use the construction above, once again.  
Note that by construction, reaching $\lbl_{win}$ 
is same as never reaching label $\lbl_{lose}$. We have already established the reduction 
from reachability to some $\lbl$ to the problem 
 $\measureof{\prog'}{\initconf\models\eventually\lbl_{win}}=1?$.
 Since $\measureof{\prog'}{\initconf\models\eventually\lbl_{win}}=1?$
  iff 
$\measureof{\prog'}{\initconf\models\eventually\lbl_{lose}}=0?$, we now have the reduction from reachability in classical TSO to the never reachability  
in probabilistic TSO. 
	
\end{proof}

\paragraph{Never Repeated Reachability}
\begin{proof}[Proof]
Once again, 
by construction, reaching $\lbl_{win}$ 
is same as never repeatedly reaching label $\lbl_{lose}$. 
 Thus,  $\measureof{\prog'}{\initconf\models\eventually\lbl_{win}}=1?$
  iff 
$\measureof{\prog'}{\initconf\models\Box \eventually\lbl_{lose}}=0?$, we now have the reduction from reachability in classical TSO to the never repeated reachability  
in probabilistic TSO. 
	
\end{proof}


\section{Quantitative Reachability: Supplementary material for Sec. \ref{quan:reachability:section}}

In this section, we provide the algorithm for the approximate repeated reachability problem
and then provide proofs of correctness and termination.

\subsection{Algorithmic details for Approximate Quantitative Repeated Reachability}
\label{app:quan:repeated:reachability:section}

%
In the case of the approximate quantitative repeated reachability problem, 
\quantrepreach, our task is to approximate the probability 
of visiting a given label infinitely often.
We provide an algorithm for approximate quantitative reachability in 
Figure~\ref{quant:repreach:alg:fig}.

\begin{wrapfigure}[30]{r}{0.58\textwidth}
\centering
\resizebox{0.55\textwidth}{!}{
\begin{minipage}{0.73\textwidth}
\begin{algorithm}[H]
  \SetAlgoRefName{}
  \KwIn{
    $\prog$: program;
    $\initconf\in\confsetof\prog$: configuration;
    $\lbl\in\lblset_\prog$: label;
    $\precision\in\preals$: precision.
  }
  \KwVar{
    $\flagvar,\posflagvar\in\bool$: Boolean flags\;
    $\bplainvar\subseteq\plainconfsetof\prog$: subset of the plain configurations\;
    $\posapprxvar,\negapprxvar\in\reals$: under-approximations\;
    $\waitingvar\in\wordsover{(\confsetof\prog\times\reals)}$: queue\;
  }
  $\posapprxvar:=0$; $\negapprxvar:=0$\; 
  $\bplainvar\assigned\emptyset$;
  $\waitingvar\assigned\tuple{\initconf,1}$\;
  \For{{\bf each} $\conf\in\plainconfsetof{\prog}$}{
    \flagvar\assigned\true\;
    \For{{\bf each} $\pconf\in\plainconfsetof{\prog}$}{
      \If{$(\conf\progmovesto\prog*\pconf)$ {\bf and} $\neg(\pconf\progmovesto\prog*\conf)$}{
        \flagvar\assigned\false\;}}
    \If{$\flagvar=\true$}{
      $\bplainvar\assigned\bplainvar\cup\set\conf$}
  }           
  \While{$\posapprxvar+\negapprxvar<1-\precision$}{
    $\tuple{\conf,\fiprobval}:=\headof\waitingvar$; $\waitingvar\assigned\tailof\waitingvar$\;
    $\posflagvar\assigned\true$\;
    \For{{\bf each} $\pconf\in\bplainvar$}{
      \lIf{$(\conf\starmovesto\prog\pconf)\land\neg(\pconf\starmovesto\prog\lbl)$}{$\posflagvar\assigned\false$}
      }
    \lIf{$\posflagvar=\true$}{$\posapprxvar\assigned\posapprxvar+\fiprobval$}
    \lElseIf{$\neg(\conf\starmovesto\prog\lbl)$}{$\negapprxvar\assigned\negapprxvar+\fiprobval$}
    \Else{ 
      \lFor{{\bf each} $\conf'$ with $\conf\movesto\prog\pconf$}
           {
             $\waitingvar\assigned\waitingvar\app\tuple{\pconf,\fiprobval\cdot\prmpmtrxof\prog{\conf,\pconf}}$
           }
    }
  }
  \KwRet{$\posapprxvar$}
  \caption{\quantrepreach}
\end{algorithm}
\end{minipage}}
\vspace{-0.2cm}
\caption{Quantitative repeated reachability algorithm.}
\label{quant:repreach:alg:fig}
\end{wrapfigure}


%
%
%
%

The algorithm for repeated reachability is very similar to the 
one for reachability.
The main difference compared to algorithm of
Fig.~\ref{quant:reach:alg:fig} is the condition
of the if-statement (line \algolinenum{19}).
Instead of checking whether we have reached label $\lbl$,
we now increase the value
of $\posapprxvar$  if there is no $\conf$-path to a configuration from which $\lbl$ is not reachable.
To check this condition, we first compute 
set of B-plain configurations and store them in $\bplainvar$.
We go through the B-plain configurations that are reachable from
the current configuration $\conf$ one by one.
We increase the value of $\posapprxvar$ if
$\lbl$ is reachable from all such configurations.

Again, we let $\posapprxvarof\ii$ and $\negapprxvarof\ii$ represent the value
of $\posapprxvar$ resp.\ $\negapprxvar$ prior to performing the 
$\ii^{\it th}$ iteration.
The partial correctness of the algorithm, follows from
the following properties of the algorithm:
(i)
The value of $\posapprxvar$ increases only 
by weights of $\initconf$-paths that visit configuration
$\conf$ from which all reachable B-plain
configurations can in turn reach $\lbl$.
We argue that any $\conf$-run will almost surely
repeatedly reach $\lbl$.
To see that, we know by Lemma~\ref{bplain:finite:attractor:lemma}
that $\run$ will almost surely visit the set of B-plain
configurations.
By finiteness of the set, $\run$ will almost surely visit
a particular B-plain configuration  $\pconf\in\bplainconfsetof\prog$ 
infinitely often.
%
%
Since $\pconf\models\eventually\lbl$ it follows
that $\run$ will almost surely visit $\lbl$ infinitely often.
It follows that 
$\posapprxvarof\ii\leq\measureof{\prog}{\initconf\models\always\eventually\lbl}$.
(ii)
We increase the value of $\negapprxvar$ only 
by weights of $\initconf$-paths that end up at a configuration
$\conf$ from which $\lbl$ is not reachable.
Since $(\conf\models\eventually\lbl)=\emptyset$, and
hence also $(\conf\models\always\eventually\lbl)=\emptyset$.
Therefore, 
$\negapprxvarof\ii\leq\measureof{\prog}{\initconf\models\neg\always\eventually\lbl}$.
(iii)
If the algorithm terminates after the $\ii^{\it th}$ iteration, the condition
of the while-loop implies that
$\posapprxvarof\ii+\negapprxvarof\ii>1-\precision$.
From (i), (ii),  and (iii), it follow that
if the termination point is $\ii$ then:
\[
\posapprxvarof\ii
\leq
\measureof{\prog}
{\initconf\models\always\eventually\lbl}
\leq
1- \negapprxvarof\ii
<
\posapprxvarof\ii+\precision
\]
Therefore, on termination, $\posapprxvar$
is within $\precision$-precision of 
$\probval$. 
%
%
%

\subsection{Proofs of correctness for Sec. \ref{quan:reachability:section}}

\thmQuantReachDec*
\begin{proof}
	Decidability follows by proving that the algorithm given in Figure \ref{quant:reach:alg:fig} gives the correct answer and terminates, which we now do.

	\paragraph{Correctness}
	We have that $\posapprxvarof{i}$ is monotone in $i$ and that 
	\begin{equation*}
		\posapprxvarof{i} \leq \measureof\prog{\initconf\models\eventually\lbl}
	\end{equation*}
	since we only accumulate probabilities of distinct paths reaching $\lbl$ in $\posapprxvarof{i}$. On the other hand, for $\negapprxvarof{i}$ we have the following inequality,
	\begin{equation*}
		\negapprxvarof{i} \leq \measureof\prog{\initconf\models \neg\eventually\lbl} = 1 - \measureof\prog{\initconf\models \eventually\lbl}.
	\end{equation*}
	This in turn follows from the fact that $\negapprxvarof{i}$ accumulates probabilities of distinct (infinite) paths which never will reach $\lbl$. Hence, $\negapprxvarof{i} + \posapprxvarof{i} \geq 1 - \precision$ implies that
	\begin{equation*}
		\posapprxvarof{i} \leq \measureof\prog{\initconf\models\eventually\lbl} \leq 1 - \negapprxvarof{i} \leq \posapprxvarof{i} + \precision
	\end{equation*}
	showing that $\posapprxvarof{i}$ approximates $\measureof\prog{\initconf\models \eventually\lbl}$ to $\precision$ precision and proving correctness of the algorithm.

	\newcommand{\reachplain}{\confset^{\texttt{r-plain}}_\prog}
	\newcommand{\minprob}{\texttt{p}}
	\newcommand{\sumapprx}{\texttt{UndetApprx}}
	\newcommand{\sumapprxvarof}[1]{\sumapprx^{(#1)}}
	\newcommand{\depthofiteration}[1]{\mathsf{depth}(#1)}
	\newcommand{\iterationatdepth}[1]{\mathsf{maxiter}(#1)}
	\newcommand{\undetpaths}[1]{\mathsf{Undet}(#1)}
	\newcommand{\plainpos}{\Gamma_\prog^\texttt{pos-plain}}
	\newcommand{\plainneg}{\Gamma_\prog^\texttt{neg-plain}}

	\paragraph{Termination}
	This proof crucially uses the existence of a finite set $\plainconfsetof\prog$ which is reached repeatedly with probability one and the fairness theorem on Markov chains. The fairness theorem says that for a Markov chain with (set of) states $S_1$ and $S_2$, 
	if there is a non-zero probability path from $S_1$ to $S_2$,  then the probability of taking a path which reaches $S_1$ infinitely often but which never reaches $S_2$ is zero.

	As introduced earlier in the main text, we denote by $\posapprxvarof{i}, \negapprxvarof{i}$ the iterates at the $i$ iteration of the while-loop and additionally we define $$\sumapprxvarof{i} = 1 - \posapprxvarof{i} - \negapprxvarof{i}$$ 
	 If we can show that $\lim_{i\rightarrow \infty} \sumapprxvarof{i} = 0$ we are done since we have that
	\begin{equation*}
		\posapprxvarof{i} \leq \measureof\prog{\initconf\models\eventually\lbl} \leq \posapprxvarof{i} + \sumapprxvarof{i}.
	\end{equation*}

	We now work towards this goal. First we observe that the algorithm performs a breadth-first traversal of the space of configurations. Hence at each loop iteration, the configuration which is dequeued from the $\waitingvar$-queue is associated with a certain depth of search. We denote this depth by $\depthofiteration{i}$ for iteration $i$. Conversely, for each depth $d$, there exists a maximal loop-iteration $i$ that considers a configuration at that depth, denoted by $\iterationatdepth{d} = \max_{i} \{\depthofiteration{i} = d\}$. At each depth, there are only finitely many configurations which are considered, and hence, $\max$ is over a finite set, and is well defined.  The finiteness of the number of configurations at each depth follows from the observation that the $Post$ set of each configuration $\conf$ is finite (concretely it can be represented as a polynomial in the size of the configuration, $|\conf|$). 
	This implies that $\lim_{i\rightarrow\infty} \depthofiteration{i} = \infty$. Hence, the limits, whether taken over the loop iteration count $i$ or over the depth of search $\depthofiteration{i}$ directly are equal. Hence we switch to the limits over the depth of search that is more convenient to reason about.
	\begin{equation*}
		\lim_{i\rightarrow \infty} \sumapprxvarof{i} = \lim_{d \rightarrow \infty} \sumapprxvarof{\iterationatdepth{d}}
	\end{equation*}
	
	Now for a depth $d$, consider the set of all paths of length $d$: $\initconf\app\confset^{d}$. A path $\run \in\initconf\app\confset^{d}$ can be one of three types: (1) which have reached $\lbl$, i.e $\exists i, \pth[i] = \lbl$ (2) which cannot reach $\lbl$: $\neg(\pth[d] \progmovesto\prog{} \lbl)$ and (3) undetermined (where none of (1,2) hold). The paths from (1) and (2) have probabilities corresponding to $\posapprxvarof{\iterationatdepth{d}}$ and $\negapprxvarof{\iterationatdepth{d}}$. The probability that a path belongs to (3) on the other hand is given by $\sumapprxvarof{\iterationatdepth{d}}$. 
	
	For a given $d$, denote the set of these undetermined paths from (3) as $\undetpaths{d}$. Let $U = \{\pth ~|~ \forall d, \pth[0]\cdots \pth[d] \in \undetpaths{d}\} \subseteq \initconf\app\infwordsover\confset$. These are the set of infinite paths from $\initconf$, for which all finite prefixes are undetermined w.r.t reachability to $\lbl$. Now we know that $\measureof\prog{\initconf \models \always\eventually \plainconfsetof\prog} = 1$. Partition $\plainconfsetof\prog = \plainpos \uplus \plainneg$ into two: (1) $\plainpos = \{\conf ~|~ \conf\progmovesto\prog{} \lbl\}$ and (2) $\plainneg = \{\conf ~|~ \neg(\conf\progmovesto \prog{} \lbl)\}$. We have the following since $\plainconfsetof\prog$ is reached repeatedly with probability one.
	\begin{equation*}
		\measureof\prog{\initconf \models U} = \measureof\prog{\initconf \models (U \land \always\eventually\plainconfsetof\prog)}
	\end{equation*}
	However, any path satisfying $\pth \models \eventually\plainneg$ cannot belong to $U$ (if it belongs to $U$, each finite prefix is undetermined wrt $\lbl$, contradicting $\eventually\plainneg$ which is determined to not reach $\lbl$), we must have the following.
	\begin{equation*}
		\measureof\prog{\initconf \models U} = \measureof\prog{\initconf \models (U \land \always\eventually\plainpos)}
	\end{equation*}
	Since $\plainconfsetof\prog$ is finite so it $\plainpos$ and we have
	\begin{equation*}
		\measureof\prog{\initconf \models U} \leq \sum_{\conf \in \plainpos}\measureof\prog{\initconf \models (U \land \always\eventually\conf)}
	\end{equation*}
	where the sum is over a finite set. Now $U$ consists of paths which never reach $\lbl$. However these paths reach $\plainpos$ infinitely often. Consequently, since each $\conf \in \plainpos$ has a finite path (with non-zero probability) to $\lbl$, and by the fairness theorem on Markov chains we must have, for all $\conf \in \plainpos$,
	\begin{equation*}
		\measureof\prog{\initconf \models (U \land \always\eventually\conf)} = 0,
	\end{equation*}
	and hence,
	\begin{equation*}
		\measureof\prog{\initconf \models U} \leq 0
	\end{equation*}

	To finish the proof note that the approximation margin term $\sumapprxvarof{d}$ approaches $\measureof\prog{\initconf \models U}$ as $d$ approaches infinity (since at a given value of $d$ it expresses the probability of taking paths which are undetermined for $d$ steps). Hence we have 
	\begin{equation*}
		\lim_{d \rightarrow \infty} \sumapprxvarof{\iterationatdepth{d}} = \measureof\prog{\initconf \models U} = 0.
	\end{equation*}
	Proving the requisite claim and hence the thorem.






\end{proof}

\thmQuantRepReachDec*
\begin{proof}
	Decidability follows by proving that the algorithm given in Figure \ref{quant:repreach:alg:fig} gives the correct answer and terminates, which we now do.

	\paragraph{Correctness}
	The proof of correctness is similar to that for Theorem \ref{quantreach:theorem}. In particular, the only difference is that $\posapprxvarof{i}$ and $\negapprxvarof{i}$ now estimate (from below) the probabilities $\measureof\prog{\initconf \models \always\eventually\lbl}$ and $\measureof\prog{\initconf \models \neg\always\eventually\lbl}$. The remaining analysis follows replacing reachability by repeated reachability.

	\newcommand{\reachplain}{\confset^{\texttt{r-plain}}_\prog}
	\newcommand{\minprob}{\texttt{p}}
	\newcommand{\sumapprx}{\texttt{UndetApprx}}
	\newcommand{\sumapprxvarof}[1]{\sumapprx^{(#1)}}
	\newcommand{\depthofiteration}[1]{\mathsf{depth}(#1)}
	\newcommand{\iterationatdepth}[1]{\mathsf{maxiter}(#1)}
	\newcommand{\undetpaths}[1]{\mathsf{RepUndet}(#1)}
	\newcommand{\plainpos}{\Gamma_\prog^\texttt{pos-plain}}
	\newcommand{\plainneg}{\Gamma_\prog^\texttt{neg-plain}}

	\paragraph{Termination}
	The proof is similar to the termination argument for Theorem \ref{quantreach:theorem}. There are two main differences. First the definition of the undetermined set of infinite paths ($U$ in Theorem \ref{quantreach:theorem}) changes: we call this set $V$ here. Secondly we must use a stronger variant of the fairness theorem on Markov chains, which says that for two (sets of) states $S_1, S_2$ if we have a non-zero probability path from $S_1$ to $S_2$ then the probability of taking infinite paths which reach $S_1$ infinitely often, but reach $S_2$ only finitely often is zero. 
	
	To begin, we once again define $\sumapprxvarof{i} = 1 - \posapprxvarof{i} - \negapprxvarof{i}$ and invoke the finite branching of the transition system to go from limit over the iteration count to limit over the depth of search. The following two relations hold.
	\begin{align*}
		\posapprxvarof{i} &\leq \measureof\prog{\initconf\models\eventually\lbl} \leq \posapprxvarof{i} + \sumapprxvarof{i} \\
		\lim_{i\rightarrow \infty} \sumapprxvarof{i} &= \lim_{d \rightarrow \infty} \sumapprxvarof{\iterationatdepth{d}}
	\end{align*}
	Then once again it remains to prove: $\lim_{d \rightarrow \infty} \sumapprxvarof{\iterationatdepth{d}} = 0$. 
	
	Each path $\run \in\initconf\app\confset^{d}$ (from the ones considered upto depth $d$) fall into three (disjoint) sets: (1) which henceforth will reach $\lbl$ infinitely often, i.e $\exists i\in [0\cdots d], \pth[i] \models \forall\always\exists\lozenge \lbl$ (2) which cannot reach $\lbl$: $\neg(\pth[d] \progmovesto\prog{} \lbl)$ and (3) undetermined (in none of the sets (1,2)). 
	The probability that a path belongs to (3) is given by $\sumapprxvarof{\iterationatdepth{d}}$. 

	For a given $d$, denote the set of these undetermined paths from (3) as $\undetpaths{d}$. Let $V = \{\pth ~|~ \forall d, \pth[0]\cdots\pth[d] \in \undetpaths{d}\} \subseteq \initconf\app\infwordsover\confset$. These are infinite paths from $\initconf$, for which all finite prefixes are undetermined w.r.t repeated reachability to $\lbl$. Invoking $\measureof\prog{\initconf \models \always\eventually \plainconfsetof\prog} = 1$ we partition $\plainconfsetof\prog = \plainpos \uplus \plainneg$: (1) $\plainpos = \{\conf ~|~ \conf\progmovesto\prog \lbl\}$ and (2) $\plainneg = \{\conf ~|~ \neg(\conf\progmovesto\prog{} \lbl)\}$. We have the following since $\plainconfsetof\prog$ is reached repeatedly with probability one.
	\begin{equation*}
		\measureof\prog{\initconf \models V} = \measureof\prog{\initconf \models (V \land \always\eventually\plainconfsetof\prog)}
	\end{equation*}
	However, any path satisfying $\pth \models \eventually\plainneg$ cannot belong to $V$ (as it is determined to not reach $\lbl$, it cannot be in $V$), we must have the following.
	\begin{equation*}
		\measureof\prog{\initconf \models V} = \measureof\prog{\initconf \models (V \land \always\eventually\plainpos)}
	\end{equation*}
	Since $\plainconfsetof\prog$ is finite so it $\plainpos$ and we have
	\begin{equation*}
		\measureof\prog{\initconf \models V} \leq \sum_{\conf \in \plainpos}\measureof\prog{\initconf \models (V \land \always\eventually\conf)}
	\end{equation*}
	where the sum is over a finite set. Now $V$ consists of paths which never reach $\lbl$. However these paths reach $\plainpos$ infinitely often. Consequently, since each $\conf \in \plainpos$ has a finite path (with non-zero probability) to $\lbl$, and by the (extended) fairness theorem on Markov chains we must have the probability of repeatedly reaching $\conf \in \plainpos$ yet reaching $\lbl$ only finitely many times is zero:
	\begin{equation*}
		\measureof\prog{\initconf \models (V \land \always\eventually\conf)} = 0,
	\end{equation*}
	and hence,
	\begin{equation*}
		\measureof\prog{\initconf \models V} \leq 0
	\end{equation*}
	Again note that the approximation margin term $\sumapprxvarof{d}$ approaches $\measureof\prog{\initconf \models V}$ as $d$ approaches infinity (since at a given value of $d$ it expresses the probability of taking paths which are undetermined for $d$ steps). Hence we have 
	\begin{equation*}
		\lim_{d \rightarrow \infty} \sumapprxvarof{\iterationatdepth{d}} = \measureof\prog{\initconf \models V} = 0.
	\end{equation*}

\end{proof}
\section{Gravitation to small configurations: Proofs for Sec.\ref{cost:section}}
We give the full details of the proof
of Lemma~\ref{eagerness:lemma}.
First, we recall the classical {\it Gambler's ruin} problem.
We show Lemma~\ref{gravity:lemma}
by comparing the probability measures of runs in the 
gambler's ruin problem with 
probability measures of runs from small configurations.

\subsection{The Gambler's Ruin Problem}

We consider the family of {\it Gambler's Ruin}
Markov chains $\gmchainof{\pp,\qq}=\gmchaintupleof{\pp,\qq}$.
The family is 
parameterized by two positive real numbers
$\pp,\qq\in\preals$ such $\pp+\qq=1$.
For each instantiation of the parameters, we get a
concrete Markov chain.
The set of configurations is the set of natural numbers,
and the probability matrix is parameterized by $\pp$ and
$\qq$.
More precisely, we have 
$\gpmtrxof{\pp,\qq}(\ii,\ii+1)=\pp$ for $\ii>0$,
$\gpmtrxof{\pp,\qq}(\ii,\ii-1)=\qq$ for $\ii>0$, and
$\gpmtrx(0,0)=1$.
In other words, the left-most configuration $0$ is a sink (a configuration
which we cannot leave).
In configurations different from $0$, 
we move right with probability $\pp$ and move left
with probability $\qq$.
If $\qq\geq\pp$ we say that the Markov chain is
``left-oriented''; otherwise
we say it ``right-oriented''.
The following lemma is classical.
\begin{lemm}
\label{gambler:ruin:prob:lemma}
If $\pp\leq\qq$ then $\measureof{\gmchainof{\pp,\qq}}{\kk\models\eventually 0}=1$
for all $\kk\in\nat$.
\end{lemm}
Lemma~\ref{gambler:ruin:prob:lemma} tells us that 
if the Markov is left-oriented then, from any configuration,
we will almost surely reach the sink state.
%

%
The following lemma states that if the Markov chain is left-oriented then
the probability of reaching a left segment of the chain,
within a given number of steps,
is higher if we are closer to the left.
\begin{lemm}
\label{gambler:ruin:close:lemma}
If $\pp\leq\qq$ then
$\measureof{\gmchainof{\pp,\qq}}{\kk\models\neventually{\geq\nn}0}\leq
\measureof{\gmchainof{\pp,\qq}}{\kk+1\models\neventually{\geq\nn}0}$
for all $\kk,\nn\in\nat$.
\end{lemm}
\begin{proof}
We use induction on $\nn$.
For the base case, with $\nn=0$, we know that
$\measureof{\gmchainof{\pp,\qq}}{\kk+1\models\neventually{\geq0}0}=\measureof{\gmchainof{\pp,\qq}}{\kk+1\models\eventually0}$.
Hence, Lemma~\ref{gambler:ruin:prob:lemma}  
immediately implies the result.
For the induction step, we consider two sub-cases.
If $\kk=0$ then 
$\measureof{\gmchainof{\pp,\qq}}{\kk\models\neventually{\geq\nn}0}=0$,
and the result holds trivially.
In  the second case, we assume that
$\kk>0$.
\[
\begin{array}{lll}
& \measureof{\gmchainof{\pp,\qq}}{\kk\models\neventually{\geq\nn+1}0} & \\
& = \qq\cdot\measureof{\gmchainof{\pp,\qq}}{\kk-1\models\neventually{\geq\nn}0}+
\pp\cdot\measureof{\gmchainof{\pp,\qq}}{\kk+1\models\neventually{\geq\nn}0} &
\textreason{}{Definition of $\gmchain$} \\
& \leq 
\qq\cdot\measureof{\gmchainof{\pp,\qq}}{\kk\models\neventually{\geq\nn}0}+
\pp\cdot\measureof{\gmchainof{\pp,\qq}}{\kk+2\models\neventually{\geq\nn}0}
& \textreason{}{Induction Hypothesis} \\
&= \measureof{\gmchainof{\pp,\qq}}{\kk+1\models\neventually{\geq\nn+1}0} & \textreason{}{Definition of $\gmchain$}
\end{array}
\]
\end{proof}

\begin{corollary}
\label{gambler:ruin:close:corollary}
$\pp\leq\qq$ and $\kk_1\leq\kk_2$
imply
$\measureof{\gmchainof{\pp,\qq}}{\kk_1\models\neventually{\geq\nn}0}\leq
\measureof{\gmchainof{\pp,\qq}}{\kk_2\models\neventually{\geq\nn}0}$.
\end{corollary}

The following lemma is an instantiation of
equation (4.14), page 352,  in \cite{Feller:book}.
It gives an upper bound on the probability of avoiding the sink
in the $\nn^{\it th}$ step,
starting from position $1$.
\begin{lemm}
\label{Fuller:lemma}
$\measureof{\gmchainof{\pp,\qq}}{1\models\neventually{=\nn}0}
=
\frac1{\nn}
\cdot\left(\begin{array}{c}\nn\\\frac{\nn+1}2\end{array}\right)\cdot\pp^{\frac{\nn-1}2}\cdot
\qq^{\frac{\nn+1}2}
$,
if $\nn$ is odd, and 
$\measureof\gmchain{1\models\neventually{=\nn}}=0$,
if  $\nn$ is even.
\end{lemm}
We use Lemma~\ref{Fuller:lemma} to give 
an upper bound on the probability of avoiding the sink
in the next $\nn$ steps,
starting from position $1$.

\begin{lemm}
\label{gambler:ruin:lemma}
$\measureof{\gmchainof{\pp,\qq}}{1\models\neventually{\geq\nn}0}\leq\frac{3\cdot\qq}{\sqrt\pi}\cdot(4\cdot\pp\cdot\qq)^{\left\lfloor\frac\nn2\right\rfloor}$,
for all $\nn\geq2$, $\pp$, and $\qq$.
\end{lemm}

\begin{proof}
\[
\begin{array}{l}
\measureof{\gmchainof{\pp,\qq}}{1\models\neventually{\geq\nn}0}
\\
=\frac1{\nn}
\cdot\left(\begin{array}{c}\nn\\\frac{\nn+1}2\end{array}\right)\cdot\pp^{\frac{\nn-1}2}\cdot
\qq^{\frac{\nn+1}2}
\textreason{}{Lemma~\ref{Fuller:lemma}}
\\
=\sum_{\mm=\left\lfloor\frac\nn2\right\rfloor}^\infty
\frac1{2\cdot\mm+1}\cdot
\left(\begin{array}{c}2\cdot\mm+1\\\mm+1\end{array}\right)
\cdot\pp^\mm\cdot\qq^{\mm+1}
\textreason{}{$\nn\geq2$}
\\
=\sum_{\mm=\left\lfloor\frac\nn2\right\rfloor}^\infty
\frac1{\mm+1}\cdot
\left(\begin{array}{c}2\cdot\mm\\\mm\end{array}\right)
\cdot\pp^\mm\cdot\qq^{\mm+1}
\textreason{}{Algebra}
\\
\leq\sum_{\mm=\left\lfloor\frac\nn2\right\rfloor}^\infty
\frac1{\mm+1}\cdot
\frac1{\sqrt{\pi\cdot\mm}}\cdot2^{2\cdot\mm}
\cdot\pp^\mm\cdot\qq^{\mm+1}
\textreason{}{\cite{Binomial:2001}}
\\
\leq \frac\qq{\sqrt\pi}\cdot
\sum_{\mm=\left\lfloor\frac\nn2\right\rfloor}^\infty
\frac1{\mm\cdot\sqrt{\mm}}\cdot
(4\cdot\pp\cdot\qq)^\mm
\textreason{}{Algebra}
\\
\leq \frac\qq{\sqrt\pi}\cdot
\left(
\frac1{\left\lfloor\frac\nn2\right\rfloor\cdot\sqrt{\left\lfloor\frac\nn2\right\rfloor}}\cdot
(4\pp\qq)^{\left\lfloor\frac\nn2\right\rfloor}
+
\int_{\left\lfloor\frac\nn2\right\rfloor}^\infty
\frac1{\mm\cdot\sqrt{\mm}}\cdot
(4\pp\qq)^\mm\, d\mm
\right)
\textreason{}{Approximating $\sum$ by $\int$}
\\
\leq \frac\qq{\sqrt\pi}\cdot
\left(
\frac1{\left\lfloor\frac\nn2\right\rfloor\cdot\sqrt{\left\lfloor\frac\nn2\right\rfloor}}\cdot
(4\pp\qq)^{\left\lfloor\frac\nn2\right\rfloor}
+
2 \cdot \left\lfloor\frac\nn2\right\rfloor^{-\frac12}
\cdot(4\pp\qq)^{\left\lfloor\frac\nn2\right\rfloor}
\right)
\\
\textreason{}{Overapprox, since $\int \frac{e^{-x}}{x\sqrt{x}} dx 
= -2\frac{e^{-x}}{x\sqrt{x}} -2\int \frac{e^{-x}}{\sqrt{x}} dx$}
\\
\leq\frac{3\cdot\qq}{\sqrt\pi}\cdot(4\cdot\pp\cdot\qq)^{\left\lfloor\frac\nn2\right\rfloor}
\textreason{}{$\nn\geq2$}
\end{array}
\]
\end{proof}

\subsection{Gravity}
In this  sub-section, we give the details of 
Lemma~\ref{gravity:lemma}.

We define 
\[
\qq_\prog\,:=\,\min_{\conf\in\numconfsetof\prog{=5}}
\sum_{\pconf\in\numconfsetof\prog{\leq4}}\progpmtrxof\prog{\conf,\pconf}
\]

In other words, it is the smallest probability by which a configuration
of size $5$ will decrease its buffer size in the next transition step,
and thus moves to a small configurations.

In Lemma \ref{threshold:prob:ptso:lemma}, we first show that $\qq_\prog$ is always (independent of the program $\prog$) bounded from below 
$\qqstar=2/3$.
We will then use this bound $\qqstar$ in our further development. 

\begin{lemm}
\label{threshold:prob:ptso:lemma}
For all programs $\prog$, $\qq_\prog \geq \qqstar$.
\end{lemm}
\begin{proof}

The transition from $\conf \in \numconfsetof\prog{=5}$ is composed of two parts, the $\procmovesto{}$ and $\updatemovesto{}$ transitions. The $\procmovesto{}$ transition can lead to a configuration $\conf'$ of size either 5 (when the process does not take a \rulename{write} transition) or 6 (when the process does take a \rulename{write} transition). Then $\updatemovesto{}$ transition essentially pushes writes from the buffer such that all possible update words are given equal weight. We treat the cases with configuration size 5, 6 separately. 

\paragraph{$|\conf'| = 5$}
We consider all possible distributions of buffer sizes across the processes. Depending upon the number of processes, ($|\procset|$), we have different cases. We only need to consider atmost 5 processes since beyind this, the remaining proceses must have empty buffers. The possible distributions are as follows (since order is immaterial, we represent distribution as a set): $\{1, 1, 1, 1, 1\}, \{2, 1, 1, 1, 0\}, \{2, 2, 1, 0, 0\}, \{3, 1, 1, 0, 0\}, \{3, 2, 0, 0, 0\}, \{4, 1, 0, 0, 0\}, \{5, 0, 0, 0, 0\}$.

For each of these distributions, the number of non-empty update words is clearly greater than the number of empty update words (a singleton set, $\epsilon$), i.e. 1. Since we choose uniformly across all update words, we choose the non-empty word with probability greater than 0.5 and hence reach a configuration in $\numconfsetof\prog{\leq 4}$ w.p. greater than 0.5.

\paragraph{$|\conf'| = 6$}
Now it suffices to consider 6 processes. The possible distributions of the buffer sizes across these are as follows (since order is immaterial, we represent distribution as a set): $\{1, 1, 1, 1, 1, 1\}, \{2, 1, 1, 1, 1, 0\}, \{2, 2, 1, 1, 0, 0\}, \{2, 2, 2, 0, 0, 0\}, \{3, 1, 1, 1, 0, 0\}, \{3, 2, 1, 0, 0, 0\}$, $\{3, 3, 0, 0, 0, 0\}, \{4, 1, 1, 0, 0, 0\}, \{4, 2, 0, 0, 0, 0\}, \{5, 1, 0, 0, 0, 0\}, \{6, 0, 0, 0, 0, 0\}$.

Once again the number of update words of length less or equal to 1 for each of these distributions are 7, 6, 5, 4, 5, 4, 3, 4, 3, 3, 2 respectively. This is clearly less than half the total number of update words of each configuration. On choosing an update word of length greater than one, we reach a configuration in $\numconfsetof\prog{\leq 4}$. Since we choose uniformly amongst these words, and since words longer than 1 outnumber those less or equal to 1, we reach a configuration in $\numconfsetof\prog{\leq 4}$ w.p. greater than 0.5.

Since we show this without making any assumption on the intermediate configuration $\conf'$ (except for the size), we can conclude that $\qq_\prog > 0.5$. On enumerating the exact update word counts for each of the above cases, we verify
that $\qqstar = \frac{2}{3}$ satisfies the needed constraints.

\end{proof}
Henceforth, we will continue to use the symbol $\qqstar$, instead of the concrete value, to make the terms in the presentation clearer to understand. However, we highlight that we the concrete value of $\frac{2}{3}$ that can be substituted in place of $\qqstar$.
The next lemma states that probability of decreasing the size of the buffer
is at least $\qqstar$ for all configuration of size at least $4$.
\begin{lemm}
\label{larger:than:threshold:prob:ptso:lemma}
For any 
$\conf\in\numconfsetof\prog{\geq5}$, we have
$\sum_{\pconf\in\confset^{<|\conf|}}\progpmtrxof\prog{\conf,\pconf}\geq\qqstar$.
\end{lemm}
\begin{proof}

By a similar reasoning as the earlier, we see that starting from the configuration $\conf$, following the $\procmovesto{}$ transition, the intermediate configuration $\conf'$ has size $\conf' \in \{|\conf|, |\conf|+1\}$. We can consider the update words for both possibilities.

\paragraph{$|\conf'| = |\conf|$}
For this the only the empty update word leads to a configuration in $\numconfsetof\prog{=|\conf|}$. On the other hand, the total number of update words is atleast greater than the configuration size (atleast one possible update word for each number of single updates). Hence we have that the probability to reach $\numconfsetof\prog{<|\conf|}$ from $\conf'$ is $>\frac{|\conf|}{|\conf|+1} \geq \frac{5}{6}$ (since $|\conf| \geq 5$). 
This gives us the following for all $\conf\in\numconfsetof\prog{\geq 5}$.
\begin{equation*}
	\sum_{\pconf\in \numconfsetof\prog{< |\conf|}}\updatepmtrxof{\conf,\pconf} > \underbrace{2/3}_{\qqstar} 
\end{equation*}

\paragraph{$|\conf'| = |\conf|+1$}
Let the distribution of buffer contents across processes be $\{b_0, b_1, \cdots\}$. Then the number of update words of length $|\conf'| = \sum b_i$ is given by the multinomial coefficient ${|\conf'| \choose {b_0, b_1, \cdots}}$.
For this case update words of length 0 or 1 lead to a configuration not in $\numconfsetof\prog{<|\conf|}$. The number of these words is $1 + \sum_i \textbf{1}_{b_i \neq 0}$, where $\textbf{1}$ is the indicator function. It is clear that the multinomial coefficient is greater than the this expression by atleast a factor of 2 for $|\conf'| \geq 5$. This follows from the fact that under the constraints $\sum_i b_i = |\pconf|$ and $\sum_i \textbf{1}_{b_i \neq 0} = c$ for some fixed $c$, the largest value of $\prod_i b_i!$ (and hence the smallest value of the multinomial coefficient) is given by the distribution $\{b_i\} = \{|\pconf|-c+1, \underbrace{1, \cdots, 1}_{c-1 \text{ times}}, 0, 0, \cdots\}$

Hence the probability that the length of the update word is atmost 1 is less than $\frac{1}{3}$, and we have the following.
\begin{align*}
	\sum_{\pconf\in \numconfsetof\prog{< |\conf|-1}}\updatepmtrxof{\conf,\pconf} > \underbrace{2/3}_{\qqstar} 
\end{align*}


Since we showed the above two inequalities for all configurations, this holds for any possible $\procmovesto{}$ transition and hence we have the result as desired.

\end{proof}

We define specific Gambler's Ruin's Markov chain,
induced by the program $\prog$, namely
$\gmchainof\prog:=\gmchainof{\ppstar,\qqstar}$.
From Lemma~\ref{gambler:ruin:prob:lemma} and
Lemma~\ref{threshold:prob:ptso:lemma} we get the following lemma.
\begin{lemm}
\label{prog:gambler:ruin:markov:chain:one:lemma}
$\measureof{\gmchainof\prog}{\kk\models\eventually0}=1$,
for all $\kk\in\nat$ and $\ell\geq5$.
\end{lemm}

We consider the probability of reaching
the set $\smallconfsetof\prog$ of small configurations.
To that end, we define the  function 
\[
\funtype\yyfun{\nat\times\nat}\reals 
\;\; \mbox{where}\;\;
\yfunof\kk\nn\,:=\,
\max_{\conf\in\numconfsetof\prog{=\kk}}
\measureof\prog{\conf\models\neventually{\geq\nn}{\smallconfsetof\prog}}
\]
In other words, it is the maximum of the probability measures by which
runs from  configurations of size $\kk$ 
can avoid small configurations in the next $\nn$ steps.
The following lemma relates this probability with the corresponding 
probability in the Gambler Ruin's problem.
Essentially, the lemma abstracts the set of configurations of
$\mcdenotationof\prog$ to the configurations
in $\gmchainof\prog$ as given by the following $\levelfun$ function, 
which was first introduced in \ref{app:basics:finiteattractor}:
$\levelfun(\conf) = 0 \quad\text{ if }\quad \conf \in \lblconfsetof{\prog}{\leq 4}$, and $\levelfun(\conf) = |\conf|$ otherwise.


The following lemma follows directly from the fact
that the size of the configuration will never increase by more than
one in PTSO (which is the case when a \rulename{write} $\procmovesto{}$ transition is taken, and no element of any buffer is pushed to the memory).
\begin{lemm}
\label{fattractor:prob:one:ptso:lemma}
$\sum_{\pconf\in\numconfsetof\prog{\leq\nn+1}}\prmpmtrxof\prog{\conf,\pconf}=1$, 
for all $\conf\in\numconfsetof\prog\nn$ for all $\nn$.
\end{lemm}

   
\newpage 
\begin{lemm}
\label{ptso:gambler:ruin:lemma}
$\yfunof{\kk+4}{\nn}\leq\measureof{\gmchainof\prog}{\kk\models\neventually{\geq\nn}0}$, for all $\kk,\nn\in\nat$.
\end{lemm}
\begin{proof}
We use induction on $\nn$.
In the base case, we have $\nn=0$.
By Lemma~\ref{gambler:ruin:prob:lemma} and Lemma~\ref{threshold:prob:ptso:lemma}
it follows that
$\measureof{\gmchainof\prog}{\kk\models\neventually{\geq\nn}0}=1$, and
the result follows immediately.

For the induction step
we consider two cases, namely when $\kk=1$ and when $\kk\geq1$.
If $\kk=1$ then $\yfunof{\kk+4}{\nn+1}=0$ and the results follows
immediately.
If $\kk\geq2$, we fix 
$\conf\in\numconfsetof\prog{=\kk+4}$, where $\kk\geq1$, such that
$\yfunof{\kk+3}{\nn+1}=\measureof\prog{\conf\modelswrt\prog\neventually{\geq\nn+1}\smallconfsetof\prog}$.
Such a configuration exists by the definition of $\yyfun$.
\[
\begin{array}{ll}
\yfunof{\kk+4}{\nn+1}
&\textreason={Definition of $\conf$}
\\
&\measureof\prog{\conf\modelswrt\prog\neventually{\geq\nn+1}\smallconfsetof\prog} \\
&\textreason={Lemma~\ref{fattractor:prob:one:ptso:lemma}} \\
& \sum_{\jj=0}^{\kk+3}
\sum_{\pconf\in\numconfsetof\prog{=\jj}}
\progpmtrxof\prog{\conf,\pconf}
\cdot
\measureof\prog{\pconf\modelswrt\prog\neventually{\geq\nn}\smallconfsetof\prog} \\
& \quad +
\sum_{\pconf\in\numconfsetof\prog{=\kk+4}}
\progpmtrxof\prog{\conf,\pconf}
\cdot
\measureof\prog{\pconf\modelswrt\prog\neventually{\geq\nn}\smallconfsetof\prog} \\
&\quad +
\sum_{\pconf\in\numconfsetof\prog{=\kk+5}}
\progpmtrxof\prog{\conf,\pconf}
\cdot
\measureof\prog{\pconf\modelswrt\prog\neventually{\geq\nn}\smallconfsetof\prog} \\
& \textreason\leq{Definition of $\yyfun$} \\
& \sum_{\jj=0}^{\kk+3}
\sum_{\pconf\in\numconfsetof\prog{=\jj}}
\progpmtrxof\prog{\conf,\pconf}
\cdot
\yfunof\jj{\nn} \\
& \quad +
\sum_{\pconf\in\numconfsetof\prog{=\kk+4}}
\progpmtrxof\prog{\conf,\pconf}
\cdot
\yfunof{\kk+4}{\nn} \\
& \quad +
\sum_{\pconf\in\numconfsetof\prog{=\kk+5}}
\progpmtrxof\prog{\conf,\pconf}
\cdot
\yfunof{\kk+5}{\nn} \\
& \textreason={Algebra} \\
& \sum_{\jj=0}^{\kk+3}
\yfunof\jj{\nn}
\cdot\left(
\sum_{\pconf\in\numconfsetof\prog{=\jj}}
\progpmtrxof\prog{\conf,\pconf}
\right) \\
& \quad +
\yfunof{\kk+4}{\nn}\cdot\left(
\sum_{\pconf\in\numconfsetof\prog{=\kk+4}}
\progpmtrxof\prog{\conf,\pconf}
\right) \\
& \quad +
\yfunof{\kk+5}{\nn}\cdot\left(
\sum_{\pconf\in\numconfsetof\prog{=\kk+5}}
\progpmtrxof\prog{\conf,\pconf}
\right) \\
& \textreason\leq{Induction Hypothesis} \\
& \sum_{\jj=0}^{\kk+3}
\measureof{\gmchainof\prog}{\jj-4\models\neventually{\geq\nn}0}
\cdot\left(
\sum_{\pconf\in\numconfsetof\prog{=\jj}}
\progpmtrxof\prog{\conf,\pconf}
\right) \\
& \quad +
\measureof{\gmchainof\prog}{\kk\models\neventually{\geq\nn}0}
\cdot\left(
\sum_{\pconf\in\numconfsetof\prog{=\kk+4}}
\progpmtrxof\prog{\conf,\pconf}
\right) \\
& \quad +
\measureof{\gmchainof\prog}{{\kk+1}\models\neventually{\geq\nn}0}
\cdot\left(
\sum_{\pconf\in\numconfsetof\prog{=\kk+5}}
\progpmtrxof\prog{\conf,\pconf}
\right) \\
& \textreason\leq{Corollary~\ref{gambler:ruin:close:corollary}} \\
& \measureof{\gmchainof\prog}{\kk-1\models\neventually{\geq\nn}0}
\cdot\left(
\sum_{\jj=0}^{\kk+3}
\sum_{\pconf\in\numconfsetof\prog{=\jj}}
\progpmtrxof\prog{\conf,\pconf}
\right) \\
& \quad +
\measureof{\gmchainof\prog}{\kk+1\models\neventually{\geq\nn}0}
\cdot\left(
\sum_{\pconf\in\numconfsetof\prog{=\kk+4}}
\progpmtrxof\prog{\conf,\pconf}
\right) \\
& \quad +
\measureof{\gmchainof\prog}{{\kk+1}\models\neventually{\geq\nn}0}
\cdot\left(
\sum_{\pconf\in\numconfsetof\prog{=\kk+5}}
\progpmtrxof\prog{\conf,\pconf}
\right) \\
& \textreason={Lemma~\ref{fattractor:prob:one:ptso:lemma}} \\
& \measureof{\gmchainof\prog}{\kk-1\models\neventually{\geq\nn}0}
\cdot\left(
\sum_{\pconf\in\numconfsetof\prog{\leq\kk+3}}
\progpmtrxof\prog{\conf,\pconf}
\right) \\
& \quad + 
\measureof{\gmchainof\prog}{\kk+1\models\neventually{\geq\nn}0}
\cdot\left(
1-
\sum_{\pconf\in\numconfsetof\prog{\leq\kk+3}}
\progpmtrxof\prog{\conf,\pconf}
\right) \\
& \textreason\leq{
Lemma~\ref{threshold:prob:ptso:lemma} and 
Lemma~\ref{larger:than:threshold:prob:ptso:lemma}}
\\
& \qqstar\cdot\measureof{\gmchainof\prog}{\kk-1\models\neventually{\geq\nn}0}
+
\ppstar\cdot\measureof{\gmchainof\prog}{\kk+1\models\neventually{\geq\nn}0} \\
& \textreason={Definition of $\gmchainof\prog$} \\
& \measureof{\gmchainof\prog}{\kk\models\neventually{\geq\nn+1}0}
\end{array}
\]
\end{proof}
\newpage

From Lemma~\ref{ptso:gambler:ruin:lemma} and
Lemma~\ref{gambler:ruin:lemma} we  get the following lemma.

\begin{lemm}
\label{gravity:yyfun:upper:bound:lemma}
$\yfunof{\kk+4}\nn\leq
\frac{3\cdot\qqstar}{\sqrt\pi}\cdot
(4\cdot\ppstar\cdot\qqstar)^{\left\lfloor\frac\nn2\right\rfloor}$, 
for all $\nn\geq2$.
\end{lemm}

Now we have the ingredients to formally prove Lemma~\ref{gravity:lemma}.
\lemGravity*

\begin{proof}
If $\nn=0$ then 
$\measureof\prog{\conf\modelswrt\prog\nxt\neventually{\geq0}\smallconfsetof\prog}\leq1\leq\left(\gravityof\prog\right)^0$

If $\nn=1$ then, by Lemma~\ref{gravity:yyfun:upper:bound:lemma}, we have
\[
\begin{array}{ll}
\measureof\prog{\conf\modelswrt\prog\nxt\neventually{\geq0}\smallconfsetof\prog} & \\
\leq \ppstar 
& \textreason{}{Definition of $\ppstar$} \\
\leq \sqrt{4\cdot\ppstar\cdot\qqstar} 
& \textreason{}{Since $\qqstar>\ppstar$} \\
\leq \gravityof\prog &
\end{array}
\]

If $\nn\geq2$.
\[
\begin{array}{ll}
\measureof\prog{\conf\modelswrt\prog\nxt\neventually{\geq\nn}\smallconfsetof\prog}
& 
\\
\leq\ppstar\cdot\yfunof4{\nn+1} & 
\textreason{}{Definition of $\yyfun$ and $\ppstar$}
\\
 \leq \ppstar\cdot\measureof{\gmchainof\prog}{1\models\neventually{\geq\nn+1}0} &
\textreason{}{Algebra and $4\cdot\ppstar\cdot\qqstar<1$} 
\\
 \leq \frac{3\cdot\qqstar\cdot\ppstar}{\sqrt\pi}\cdot
(4\cdot\ppstar\cdot\qqstar)^{\frac\nn2} & 
\textreason{}{Lemma~\ref{ptso:gambler:ruin:lemma}} 
\\
 \leq (4\cdot\ppstar\cdot\qqstar)^{\frac\nn2} & 
\textreason{}{$3<\pi$ and $\ppstar\cdot\qqstar<1$} 
\\
\leq (\gravityof\prog)^{\nn} &
\textreason{}{Definition of $\gravityof\prog$}
\end{array}
\]
\end{proof}

\newpage
\subsection{S-Runs and F-Runs}
\label{app:subsec:sandf}

We first define the predicate $\visit$ formally.
For a natural number $1\leq\mm\leq\nn$, let
$\nn\partition\mm\subseteq\kwordsover\mm{\left(\pnat\right)}$ 
be the set of words 
$\word=\ii_1,\ldots,\ii_\mm$ of length $\mm$, over the set of positive 
natural numbers, such that
$\ii_1+\cdots+\ii_\mm=\nn$.
Notice that 
\[
\sizeof{\nn\partition\mm}=
\left(\begin{array}{c}\nn-1\\\mm-1\end{array}\right)
\]
We define $\nn^\partition:=\cup_{1\leq\mm\leq\nn}\nn\partition\mm$.
For $\word\in\nn^\partition$,
we define
$\visitof\prog\nn\word$ to be the set of runs 
of the form
$\conf_0\app\pth_0\app\conf_1\app\cdots\app\conf_\mm\app\pth_\mm\app\prun$,
such that the following conditions are satisfied
\renewcommand\labelitemi{$\triangleright$}
\begin{itemize}
\item
$\mm=\lengthof\word$.
\item
$\forall\ii:0\leq\ii\leq\mm:\wordof\ii=\lengthof{\pth_\ii}+1$.
\item
$\forall\ii:0\leq\ii\leq\mm:\conf_\ii\in\smallconfsetof\prog$.
\item
  $\forall\ii:0\leq\ii\leq\mm:\forall\jj:1\leq\jj\leq\lengthof{\pth_\ii}:
\pth_\ii[\jj]\not\in\smallconfsetof\prog$.
\end{itemize}
We define 
$\visitof\prog\nn\mm:=\cup_{\word\in(\nn\partition\mm)}\visitof\prog\nn\word$.
Intuitively, $\visitof\prog\nn\mm$ is the set of runs
whose prefixes of length $\nn$ visit the set of small configurations exactly
$\mm$ times.
%


%
\begin{lemm}
\label{visit:measure:lemma}
For every $\mm,\nn:1\leq\mm\leq\nn$,
$\word\in\left(\nn\partition\mm\right)$, and
$\conf\in\smallconfsetof\prog$, we have
$
\measureof\prog{\conf\modelswrt\prog\visitof\prog\nn\word}
\leq
\left(\gravityof\prog\right)^{\nn-\mm}
$
\end{lemm}
\begin{proof}
We use induction on $\mm$.

The base case corresponds to $\mm=1$, i.e.
$\word\in(\nn\partition1)$.
\[
\begin{array}{rl}
\measureof\prog{\conf\modelswrt\prog\visitof\prog\nn\word}
&=\measureof\prog{\conf\modelswrt\prog\nxt\neventually{\geq\nn-1}\smallconfsetof\prog}
\textreason{}{Definition of $\visit$}
\\
&\leq\left(\gravityof\prog\right)^{\nn-1}
\textreason{}{Lemma~\ref{gravity:lemma}}
\end{array}
\]

Let 
$\word=
\conf_0\app\pth_0\app\conf_1\app\cdots\app\conf_\mm\app\pth_\mm\app\prun$.
Define $\vord:=\conf_1\app\cdots\app\conf_\mm\app\pth_\mm\app\prun$.
We know that
\[
\begin{array}{l}
\measureof\prog{\conf\modelswrt\prog\visitof\prog\nn\word}
\\
=\sum_{\pconf\in\smallconfsetof\prog}
\measureof\prog{\conf\modelswrt\prog\nxt\left(\neventually{=\lengthof{\pth_0}}\pconf\right)}
\cdot
\measureof\prog{\pconf\modelswrt\prog\visitof\prog{\nn-\lengthof{\pth_0}-1}\vord} \\
\qquad\qquad\textreason{}{Definition of $\visit$}
\\
\leq\sum_{\pconf\in\smallconfsetof\prog}
\measureof\prog{\conf\modelswrt\prog\nxt\left(\neventually{=\lengthof{\pth_0}}\pconf\right)}
\cdot
\left(\gravityof\prog\right)^{\nn-\lengthof{\pth_0}-1-(\mm-1)} \textreason{}{Induction Hypothesis}
\\
\leq\left(\gravityof\prog\right)^{\nn-\lengthof{\pth_0}-\mm}
\cdot
\sum_{\pconf\in\smallconfsetof\prog}
\measureof\prog{\conf\modelswrt\prog\nxt\left(\neventually{=\lengthof{\pth_0}}\pconf\right)}
\textreason{}{Algebra}
\\
\leq\left(\gravityof\prog\right)^{\nn-\lengthof{\pth_0}-\mm}
\cdot
\measureof\prog{\conf\modelswrt\prog\nxt\left(\neventually{=\lengthof{\pth_0}}\smallconfsetof\prog\right)}
\textreason{}{Definition of $\measure\prog$}
\\
\leq\left(\gravityof\prog\right)^{\nn-\lengthof{\pth_0}-\mm}
\cdot
\left(\gravityof\prog\right)^{\lengthof{\pth_0}} \textreason{}{Lemma~\ref{gravity:lemma}}
\\
\leq\left(\gravityof\prog\right)^{\nn-\mm}\textreason{}{Algebra}
\end{array}
\]

\end{proof}

\newpage

We now recall and give the proof of Lemma~\ref{sruns:lemma}.
\lemSRun*

\begin{proof}

\[
\begin{array}{l}
\measureof\prog{\srunsetof\conf\nn}

\\
=
\sum_{\mm=1}^{\lfloor\frac{\nn}\border\rfloor}
\measureof\prog{\conf\modelswrt\prog\neventually{=\nn}\lbl\land\visitof\prog\nn\mm}
\textreason{}{Definition of s-runs}
\\
\leq\sum_{\mm=1}^{\lfloor\frac{\nn}\border\rfloor}
\measureof\prog{\conf\modelswrt\prog\visitof\prog\nn\mm}
\textreason{}{Definition of $\measure\prog$}
\\
=
\sum_{\mm=1}^{\lfloor\frac{\nn}\border\rfloor}
\sum_{\word\in\nn\partition\mm}
\measureof\prog{\conf\modelswrt\prog\visitof\prog\nn\mm}
\textreason{}{Definition of $\partition$}
\\
=\sum_{\mm=1}^{\lfloor\frac{\nn}\border\rfloor}
\sum_{\word\in\nn\partition\mm}
\gravityof\prog^{\nn-\mm}
\textreason{}{Lemma~\ref{visit:measure:lemma}}
\\
\leq
\sum_{\mm=1}^{\lfloor\frac{\nn}\border\rfloor}
\left(\begin{array}{c}\nn-1\\\mm-1\end{array}\right)
\cdot
\gravityof\prog^{\nn-\mm}
\textreason{}{Algebra}
\\
=
\gravityof\prog^\nn\cdot
\left(\sum_{\mm=1}^{\lfloor\frac{\nn}\border\rfloor}
\left(\begin{array}{c}\nn-1\\\mm-1\end{array}\right)
\cdot \gravityof\prog^{-\mm}
\right)
\textreason{}{Algebra}
\\
\leq
\gravityof\prog^\nn\cdot
\left(
\sum_{\mm=1}^{\lfloor\frac{\nn}\border\rfloor}
\left(\begin{array}{c}\nn\\\mm\end{array}\right)
\cdot \gravityof\prog^{-\mm}
\right)
\textreason{}{Algebra}
\\
\leq
\gravityof\prog^\nn\cdot
\left(
\sum_{\mm=1}^{\lfloor\frac{\nn}\border\rfloor}
\left(\begin{array}{c}\nn\\\lfloor\frac\nn\border\rfloor\end{array}\right)
\cdot
\left(\begin{array}{c}\lfloor\frac\nn\border\rfloor\\\mm\end{array}\right)
\cdot
\frac{\left(\nn-\left\lfloor\frac\nn\border\right\rfloor\right)!\cdot
\left(\left\lfloor\frac\nn\border\right\rfloor-\mm\right)!}{\left(\nn-\mm\right)!}
\cdot \gravityof\prog^{-\mm}
\right)
\textreason{}{Algebra}
\\
\leq
\gravityof\prog^\nn\cdot
\left(\begin{array}{c}\nn\\\lfloor\frac\nn\border\rfloor\end{array}\right)
\cdot
\left(
\sum_{\mm=1}^{\lfloor\frac{\nn}\border\rfloor}
\left(\begin{array}{c}\lfloor\frac\nn\border\rfloor\\\mm\end{array}\right)
\cdot
\frac{\left(\nn-\left\lfloor\frac\nn\border\right\rfloor\right)!\cdot
\left(\left\lfloor\frac\nn\border\right\rfloor-\mm\right)!}{\left(\nn-\mm\right)!}
\cdot \gravityof\prog^{-\mm}
\right)
\textreason{}{Algebra}
\\
\leq
\gravityof\prog^\nn\cdot
\left(\begin{array}{c}\nn\\\lfloor\frac\nn\border\rfloor\end{array}\right)
\cdot
\left(
\sum_{\mm=1}^{\lfloor\frac{\nn}\border\rfloor}
\left(\begin{array}{c}\lfloor\frac\nn\border\rfloor\\\mm\end{array}\right)
\cdot
\left(\prod_{\ii=1}^{\left\lfloor\frac\nn\border\right\rfloor}\frac\ii{n-\frac\nn\border+\ii}\right)
\cdot \gravityof\prog^{-\mm}
\right)
\textreason{}{Algebra}
\\
\leq
\gravityof\prog^\nn
\cdot
\left(\begin{array}{c}\nn\\\lfloor\frac\nn\border\rfloor\end{array}\right)
\cdot
\left(
\sum_{\mm=1}^{\lfloor\frac{\nn}\border\rfloor}
\left(\begin{array}{c}\lfloor\frac\nn\border\rfloor\\\mm\end{array}\right)
\cdot \left(\frac1\border\right)^{\left\lfloor\frac\nn\border\right\rfloor-\mm}
\cdot \gravityof\prog^{-\mm}
\right)
\textreason{}{Algebra}
\\
\leq
\gravityof\prog^\nn
\cdot
\left(\begin{array}{c}\nn\\\lfloor\frac\nn\border\rfloor\end{array}\right)
\cdot
\left(\frac1\border\right)^{\left\lfloor\frac\nn\border\right\rfloor}
\cdot
\left(
\sum_{\mm=1}^{\lfloor\frac{\nn}\border\rfloor}
\left(\begin{array}{c}
\lfloor\frac\nn\border\rfloor\\\mm
\end{array}\right)
\cdot \left(\frac\border{\gravityof\prog}\right)^{\mm}
\right)
\textreason{}{Algebra}
\\
\leq\gravityof\prog^\nn
\cdot
\left(\begin{array}{c}\nn\\\lfloor\frac\nn\border\rfloor\end{array}\right)
\cdot
\left(\frac1\border\right)^{\left\lfloor\frac\nn\border\right\rfloor}
\cdot
\left(1+\left(\frac\border{\gravityof\prog}\right)^{\lfloor\frac{\nn}\border\rfloor}
\right)
\textreason{}{Algebra: $\sum_{\bb=0}^\infty\left(\begin{array}{c}\gravity\\\bb\end{array}\right)\cdot\xx^\bb
=(1+x)^\gravity$}

\\
\leq
\gravityof\prog^\nn
\cdot
\frac{1}{\sqrt{2\cdot\pi}}
\cdot
\frac{\nn^{\frac{\nn+1}{2}}}{\left(\nn-\lfloor\frac\nn\border\rfloor\right)^{\nn-\lfloor\frac\nn\border\rfloor+\frac{1}{2}}\cdot\left(\lfloor\frac\nn\border\rfloor\right)^{\lfloor\frac\nn\border\rfloor+\frac{1}{2}}}
\cdot
\left(1+\left(\frac\border{\gravityof\prog}\right)^{\lfloor\frac{\nn}\border\rfloor}
\right)
\textreason{}{\cite{Binomial:2001}}
\\
\leq
\gravityof\prog^\nn
\cdot
\left(
\frac{\nn}{\nn-\left\lfloor\frac\nn\border\right\rfloor}
\right)^\nn
\cdot
\left(
\frac{\nn-\left\lfloor\frac\nn\border\right\rfloor}{\left\lfloor\frac\nn\border\right\rfloor}
\right)^{\left\lfloor\frac\nn\border\right\rfloor}
\cdot
\sqrt{\frac\nn{2\cdot\pi\cdot\left(\nn-\left\lfloor\frac\nn\border\right\rfloor\right)\cdot\left\lfloor\frac\nn\border\right\rfloor}}
\cdot
\left(
\frac1\border+\frac1{\gravityof\prog}
\right)^{\lfloor\frac{\nn}\border\rfloor}
\textreason{}{Algebra}
\\
\leq
\gravityof\prog^\nn
\cdot
\left(
\frac{\nn}{\nn-\frac\nn\border}
\right)^\nn
\cdot
\left(
\frac{\nn-\left\lfloor\frac\nn\border\right\rfloor}{\left\lfloor\frac\nn\border\right\rfloor}
\right)^{\left\lfloor\frac\nn\border\right\rfloor}
\cdot
\sqrt{
\frac\nn
{2\cdot\pi\cdot\left(\nn-\frac\nn\border\right)\cdot\frac\nn\border}
}
\cdot
\left(
\frac1\border+\frac1{\gravityof\prog}
\right)^{\lfloor\frac{\nn}\border\rfloor}
\textreason{}{Algebra}
\\
\leq
\gravityof\prog^\nn
\cdot
\left(
\frac{\border}{\border-1}
\right)^\nn
\cdot
\left(
\frac{\nn-\left\lfloor\frac\nn\border\right\rfloor}{\left\lfloor\frac\nn\border\right\rfloor}
\right)^{\left\lfloor\frac\nn\border\right\rfloor}
\cdot
\sqrt{
\frac\border
{2\cdot\pi\cdot\left(\nn-\frac\nn\border\right)
}
}
\cdot
\left(
\frac1\border+\frac1{\gravityof\prog}
\right)^{\lfloor\frac{\nn}\border\rfloor}
\textreason{}{Algebra}
\\
=
\gravityof\prog^\nn
\cdot
\left(
\frac{\border}{\border-1}
\right)^\nn
\cdot
(2\cdot\border)^{\lfloor\frac{\nn}\border\rfloor}
\cdot
\left(
\frac1\border+\frac1{\gravityof\prog}
\right)^{\lfloor\frac{\nn}\border\rfloor}
\textreason{}{$4\leq2\cdot\border\leq\nn$}
\\
\leq
\gravityof\prog^\nn
\cdot
\left(\left(\frac\border{\border-1}\right)(2\cdot\border)^{\frac1\border}\right)^\nn
\cdot
\left(
\frac1\border+\frac1{\gravityof\prog}
\right)^{\lfloor\frac{\nn}\border\rfloor}
\textreason{}{Algebra}
\\
\leq
\left(\left(\frac\border{\border-1}\right)(2\cdot\border)^{\frac1\border}
\cdot
\left(
\frac1\border+\frac1{\gravityof\prog}
\right)^{\lfloor\frac{1}\border\rfloor}
\cdot 
 \gravityof\prog
\right)^\nn
\textreason{}{Algebra}
\\
\leq
\left(\seagernessof\prog\right)^\nn
\textreason{}{Definition}
\end{array}
\]

\end{proof}

\newpage

Next, we define a bound on $\drunsetof\conf\lbl\nn$.
We characterize the set of runs that visit the set
of small configurations ``many times'' before visiting $\lbl$.
For sets of configurations $\confs_1,\confs_2\subseteq\confsetof\prog$,
 a run $\run$, and $\mm\in\nat$, we write
$\run\models\confs_1\kbefore\mm\confs_2$
to denote that 
$\run=\pth\app\prun$ for some path $\pth$ and run
$\prun$, $\pth$ is of the form
$\pth_1\app\conf_1\app\cdots\app\pth_\mm\app\conf_\mm$,
and the following conditions are satisfied
\begin{itemize}
\item
$\conf_\ii\in\confs_1$ for all $\ii:1\leq\ii\leq\mm$.
\item
$\pth[\ii]\not\in\confs_2$ for all 
$\ii:0\leq\ii\leq\lengthof\pth$.
\end{itemize}
In other words $\run$ visits the set $\confs_1$ 
at least $\mm$ times before visiting $\confs_2$
for the first time.
We usually write $\before$ instead of $\kbefore1$, and
write $\conf\kbefore\kk\confs$ instead of $\set\conf\kbefore\kk\confs$.
Define the set $\aaset$ of small configurations from which
$\lbl$ is reachable.
\[ 
\aaset:=
\smallconfsetof\prog\cap (\conf\modelswrt\prog\exists\eventually\lbl))
\]
Consider a $\mu$ satisfying (well defined since $\aaset$ is finite),
\[
0<\mu\leq\min_{\conf\in\aaset}\measureof\prog{\conf\models\nxt(\lbl\onebefore\conf)}.
\]
This means that $\mu$ is a lower bound on the  measure of runs that
start from some configuration in $\conf\in\aaset$ and  visit $\lbl$
before visiting $\conf$.

\begin{lemm}
\label{aaset:measure:lemma}
$\measureof\prog{\conf\models\conf\kbefore\mm\lbl}\leq(1-\mu)^{\mm-1}$
for each $\conf\in\aaset$.
\end{lemm}
\begin{proof}
By induction on $\mm$.
The base case, when $\mm=1$ is trivial.

For the induction step, we observe that, by definition,
we have 
\[
\begin{array}{l}
\measureof\prog{\conf\models\conf\kbefore2\lbl}
=
\measureof\prog{\conf\models\nxt(\conf\onebefore\lbl)}
\leq
(1-\mu)
\end{array}
\]
By the induction hypothesis we know that
\[
\begin{array}{l}
\measureof\prog{\conf\models\conf\kbefore{\mm-1}\lbl}
\leq
(1-\mu)^{\mm-2}
\end{array}
\]
We obtain
\[
\begin{array}{rl}
\measureof\prog{\conf\models\conf\kbefore{\mm}\lbl} &=
\measureof\prog{\conf\models\conf\kbefore{2}\lbl}
\cdot
\measureof\prog{\conf\models\conf\kbefore{\mm-2}\lbl} \\
&\leq (1-\mu)\cdot(1-\mu)^{\mm-2}
= (1-\mu)^{\mm-1}
\end{array}
\]
\end{proof}

\newpage

We now recall and give the proof of Lemma~\ref{druns:lemma}.
\lemDRun*
\begin{proof}
There are two possible cases:
(i)
$\conf\in
\smallconfsetof\prog-\aaset$.
From the definitions, it follows that
$\sum_{\mm=\lfloor\frac{\nn}\border\rfloor+1}^\nn
\measureof\prog{\conf\modelswrt\prog\neventually{=\nn}\lbl\land \visitof\prog\nn\mm}=0$.
(ii)
$\conf\in\aaset$.
We analyze this case below.

For any $\mm$ we have that
\[
\begin{array}{l}
\measureof\prog{\drunsetof\conf\lbl\nn}
\\
= \sum_{\mm=\lfloor\frac{\nn}\border\rfloor+1}^\nn
\measureof\prog{\conf\modelswrt\prog\neventually{=\nn}\lbl\land \visitof\prog\nn\mm}
\textreason{}{Definition of F-Runs}
\\
\leq
\sum_{\mm=\lfloor\frac{\nn}\border\rfloor+1}^\nn
\measureof\prog{
\conf\models\aaset\kbefore\mm\lbl}
\textreason{}{Definitions of $\neventually\mm$, $\visit$, and $\before$}

\\
\leq \sum_{\mm=\lfloor\frac{\nn}\border\rfloor+1}^\nn
\sum_{\pconf\in\aaset}
\measureof\prog{
\conf\models\pconf\kbefore{\left\lceil\frac\mm{\sizeof\aaset}\right\rceil}\lbl} \\
\qquad\qquad \textreason{}{Finiteness of $\aaset$ and pigeonhole principle}
\\
\leq
\sum_{\mm=\lfloor\frac{\nn}\border\rfloor+1}^\nn
\sum_{\pconf\in\aaset}
\measureof\prog{
\conf\models\pconf\onebefore\lbl}
\cdot
\measureof\prog{
\pconf\models\pconf\kbefore{\left\lceil\frac\mm{\sizeof\aaset}\right\rceil}\lbl} \\
\qquad\qquad\textreason{}{Definition of $\before$}
\\
\leq
\sum_{\mm=\lfloor\frac{\nn}\border\rfloor+1}^\nn
\sum_{\pconf\in\aaset}
\measureof\prog{
\pconf\models\pconf\kbefore{\left\lceil\frac\mm{\sizeof\aaset}\right\rceil}\lbl}
\textreason{}{$\measureof\prog{\cdot}\leq1$}

\\
\leq
\sum_{\mm=\lfloor\frac{\nn}\border\rfloor+1}^\nn
\sum_{\pconf\in\aaset}
(1-\mu)^{\left\lceil\frac\mm{\sizeof\aaset}\right\rceil-1}
\textreason{}{Lemma~\ref{aaset:measure:lemma}}

\\
\leq
\sum_{\mm=\lfloor\frac{\nn}\border\rfloor+1}^\nn
\sum_{\pconf\in\aaset}
(1-\mu)^{\frac\mm{\sizeof\aaset}}
\textreason{}{Algebra}\textreason={Algebra}
\\
\leq
\frac{\sizeof{\aaset}}{1-\mu}\cdot\sum_{\mm=\lfloor\frac{\nn}\border\rfloor+1}^\nn
(1-\mu)^{\frac\mm{\sizeof\aaset}}
\\
=
\frac{\sizeof{\aaset}}{1-\mu}\cdot
\frac{(1-\mu)^{
\frac{\left\lfloor\frac\nn\border\right\rfloor+1}{\sizeof\aaset}}
-
(1-\mu)^{\frac{\nn+1}{\sizeof\aaset}}
}
{1-(1-\mu)^{\frac1{\sizeof\aaset}}}
\textreason{}{Algebra}
\\
\leq
\frac{\sizeof{\aaset}}{1-\mu}\cdot
\frac{(1-\mu)^{
\frac\nn{\border\cdot\sizeof\aaset}}
-
(1-\mu)^{\frac{\nn+1}{\sizeof\aaset}}
}
{1-(1-\mu)^{\frac1{\sizeof\aaset}}}
\textreason{}{Algebra}
\\
=
\frac{\sizeof{\aaset}}{1-\mu}\cdot
\frac{(1-\mu)^{
\frac\nn{\border\cdot\sizeof\aaset}
}
}
{1-(1-\mu)^{\frac1{\sizeof\aaset}}}
\textreason{}{Algebra}

\\
=
\frac{\sizeof{\aaset}}{(1-\mu)\cdot\left(1-(1-\mu)^{\frac1{\sizeof\aaset}}\right)}\cdot
\left((1-\mu)^{
\frac{1}{\border\cdot\sizeof\aaset}}
\right)^\nn
\textreason{}{Algebra}
\\
=
\left(\deagernessof\prog\right)^\nn
\textreason{}{Definition}
\end{array}
\]

\end{proof}

\newpage
\subsection{Eagerness: Existence and Computability}

The results from \ref{app:subsec:sandf} give us all the ingredients that were necessary to proof Lemma \ref{eagerness:lemma}. 
We briefly discussed the proof idea for this lemma in the main paper. We give here the proof
with full details.

\lemEagerness*

\paragraph{Existence} We start off by showing the existence of $\eagernessof\prog$ and $\eagernessthresholdof\prog$.
\begin{proof}

For values $\seagernessof\prog, \deagernessof\prog < 1$, we have,
\[
\begin{array}{rl}
&\text{For }~\nn \geq 300:~~\sum\limits_{\mm=1}^{\left\lfloor\frac\nn\border\right\rfloor}
\measureof\prog{\conf\modelswrt\prog\neventually{=\nn}\lbl
\land\visitof\prog\nn\mm} \leq (\seagernessof\prog)^n \qquad\textreason{}{Lemma \ref{sruns:lemma}} \\[0.2cm]
&\text{For }~\nn\geq\deagernessthresholdof\prog:~~\sum\limits_{\mm=\left\lfloor\frac\nn\border\right\rfloor}^\nn
\measureof\prog{\conf\modelswrt\prog\neventually{=\nn}\lbl
\land\visitof\prog\nn\mm} \leq (\deagernessof\prog)^n \qquad\textreason{}{Lemma \ref{druns:lemma}}	
\end{array}
\]

\newcommand{\threshold}{\eta}

Choose (exists since $\seagernessof\prog, \deagernessof\prog < 1$) a value $\sdeagernessof\prog$ such that,
$\max(\seagernessof\prog,\deagernessof\prog) < \sdeagernessof\prog < 1$.
%
It follows that for some value $\threshold^1$, $\measureof\prog{\conf\modelswrt\prog\neventually{=\nn}\lbl}\leq
(\sdeagernessof\prog)^n$ for all $n \geq \threshold^1$. 
Define $\sdeagernessthresholdof\prog = \max(\deagernessthresholdof\prog, 300, \threshold^1)$.
From the earlier two results, we get the following,

\begin{equation*}
\text{For }~\nn\geq\sdeagernessthresholdof\prog:~~\measureof\prog{\conf\modelswrt\prog\neventually{=\nn}\lbl} \leq\left(\sdeagernessof\prog\right)^\nn
\end{equation*}

The final step 
is to extend the argument to the set of $\conf$-runs
that reach $\lbl$ in $\nn$ {\it or more} steps
(as required by Lemma~\ref{eagerness:lemma}).
\[
\begin{array}{l}
\measureof\prog{\conf\modelswrt\prog\neventually{\geq\nn}\lbl}
=
\sum_{\kk=\nn}^\infty\measureof\prog{\conf\modelswrt\prog\neventually{=\nn}\lbl}
\leq
\sum_{\kk=\nn}^\infty\left(\sdeagernessof\prog\right)^\kk
=
\frac{\left(\sdeagernessof\prog\right)^\nn}{1-\sdeagernessof\prog}
\end{array}
\]
Choose (exists since $\sdeagernessof\prog < 1$) $\sdeagernessof\prog < \eagernessof\prog < 1$.
There exists an $\eagernessthresholdof\prog \geq \sdeagernessthresholdof\prog$ such that
$\frac{\left(\sdeagernessof\prog\right)^\nn}{1-\sdeagernessof\prog} \leq \left(\eagernessof\prog\right)^\nn$
for all $\nn\geq\eagernessthresholdof\prog$, and hence we have,

\begin{equation*}
\text{For }~\nn\geq\eagernessthresholdof\prog:~~\measureof\prog{\conf\modelswrt\prog\neventually{\geq\nn}\lbl}\geq\left(\eagernessof\prog\right)^\nn
\end{equation*}
This gives us the result.
%
\end{proof}

\paragraph{Computability}
Now we show that computatbility of these terms. 
We proceed systematically along the dependencies 
and illustrate how each term can be computed, not just for our model but
for arbitrary models.

\begin{proof}

\begin{itemize}
\item
$\qqstar:=\min_{\conf\in\numconfsetof\prog{=5}}\sum\limits_{\pconf\in\smallconfsetof\prog}\progpmtrxof\prog{\conf,\pconf}$.
Computable since the sets $\numconfsetof\prog{=4}$
and $\numconfsetof\prog{\leq3}$ are finite, and for any two configurations
$\conf,\pconf\in\confset$, we can compute 
$\progpmtrxof\prog{\conf,\pconf}$. In fact, for us, this has the constant value of $2/3$.
\item
$\ppstar:=1-\qqstar$.
Computable since $\qqstar$ is computable.
\item
$\eagernessof\prog:=2\sqrt{\ppstar\cdot\qqstar}$.
Computable since $\qqstar$ and $\ppstar$ are computable.
\item
$\aaset$ is computable
since 
$\aaset=\smallconfsetof\prog\cap \{\run\in\runsetof\conf ~|~ \run\modelswrt\prog\exists\eventually\lbl))$, 
the set $\smallconfsetof\prog$ is finite (and explicitly given),
and the property $\conf\modelswrt\prog\exists\eventually\lbl$
is decidable by Lemma~\ref{tso:lbl:reachability:lemma}.
\item
Compute $\border$ such that
\[
\left(\frac\border{\border-1}\right)(2\cdot\border)^{\frac1\border}
\cdot
\left(
\frac1\border+\frac1{\gravityof\prog}
\right)^{\lfloor\frac{1}\border\rfloor}
\cdot 
\gravityof\prog
\leq1
\]
This is possible since the function is monotone and its limit approaches $\gravityof\prog<1$, 
i.e.,

\[
\lim_{\border\rightarrow\infty}
\left(\frac\border{\border-1}\right)(2\cdot\border)^{\frac1\border}
\cdot
\left(
\frac1\border+\frac1{\gravityof\prog}
\right)^{\lfloor\frac{1}\border\rfloor}
\cdot 
\gravityof\prog
=\gravityof\prog<1
\]
\item
Define 
\[
\neagernessof1\prog:=
\left(\frac\border{\border-1}\right)(2\cdot\border)^{\frac1\border}
\cdot
\left(
\frac1\border+\frac1{\gravityof\prog}
\right)^{\lfloor\frac{1}\border\rfloor}
\cdot 
\gravityof\prog
\]
This is possible since both 
$\gravityof\prog$ and $\border$ are computable. 
In fact for our model, the constant value of $\border = 150$ suffices.
\item
We can find a $\mu$ such that
$0<\mu\leq\min_{\conf\in\aaset}\measureof\prog{\conf\models\nxt(\lbl\onebefore\conf)}$,
using the following procedure.
We explore the paths that start from $\conf$ in a breadth-first manner,
until we find a path $\pth$ that end with $\lbl$.
Such a $\pth$ exists since $\conf\models\exists\eventually\lbl$ holds
by assumption.
We define $\mu$ to be the probability of $\pth$.

\item
Define 
$\neagernessof2\prog$ such that
$(1-\mu)^{\frac1{\border\cdot\sizeof\aaset}}<\neagernessof2\prog<1$.
We can compute $\neagernessof2\prog$
since $\border$, $\aaset$, $\mu$ are computable.
Since $(1-\mu)^{\frac1{\border\cdot\sizeof\aaset}}<\neagernessof\prog2$
it follows that there is a natural number, which we denote
by $\keagernessthresholdof\prog2$ such that 
$
(\neagernessof2\prog)^\nn\leq\frac{\sizeof{\aaset}}{(1-\mu)\cdot1-(1-\mu)^{\frac1{\sizeof\aaset}}}\cdot\left((1-\mu)^{\frac{1}{\border\cdot\sizeof\aaset}}\right)^\nn$
for all $\nn\geq\keagernessthresholdof\prog2$.

\item
Define 
$\eagernessof\prog$ such that
$\max(\neagernessof1\prog,\neagernessof2\prog)<\eagernessof\prog<1$.
It follows that there is  a natural number, which we denote
by $\eagernessthresholdof\prog$ such that 
$(\eagernessof\prog)^\nn<
\left(\max(\neagernessof1\prog,\neagernessof2\prog)\right)^\nn$ ,
for all $\nn\geq\eagernessthresholdof\prog$.

\end{itemize}

This concludes the proof of computatbility and hence gives us Lemma \ref{eagerness:lemma}.
\end{proof}



\newpage

\subsection{Proving the invariants}
We now prove Lemma \ref{cost:invariant:lemma} that states the validity of the invariants. 

\lemCostInv*

The invariants 
(\ref{costapprx:invariant}),
(\ref{probapprx:invariant}),
(\ref{cerror:invariant}), and
(\ref{perror:invariant}) follow directly from the definitions.
Below, we show Invariant (\ref{ecost:costapprx:invariant}) and
(\ref{prob:probapprx:invariant}).

Invariant (\ref{ecost:costapprx:invariant}): Let $\kvar = \maxcostof\cost$.
\[
\def\arraystretch{1.5}
\begin{array}{l}
\ecostof\conf\lbl\cost - \costapprxvarof\nn
\\
=\sum_{i=0}^\infty
\sum\limits_{\run\in\{\run \in\runsetof{\initconf}|\run\models\neventually{=\ii}\lbl\}}
\costof\run\cdot\probof\prog\run
- 
\costapprxvarof\nn
\textreason{}{Definition}
\\
=\sum_{i=0}^\infty
\sum\limits_{\run\in\{\run \in\runsetof{\initconf}|\run\models\neventually{=\ii}\lbl\}}
\costof\run\cdot\probof\prog\run
- 
\sum_{i=0}^\nn
\sum\limits_{\run\in\{\run \in\runsetof{\initconf}|\run\models\neventually{=\ii}\lbl\}}
\costof\run\cdot\probof\prog\run \\
\qquad \qquad\textreason{}{Invariant~\ref{costapprx:invariant}}
\\
\leq\kvar\cdot\sum\limits_{i=0}^\infty
\ii\cdot
\sum\limits_{\run\in\{\run \in\runsetof{\initconf}|\run\models\neventually{=\ii}\lbl\}}
\probof\prog\run
- 
\kvar\cdot\sum\limits_{i=0}^\nn
\ii\cdot
\sum\limits_{\run\in\{\run \in\runsetof{\initconf}|\run\models\neventually{=\ii}\lbl\}}
\probof\prog{\initconf\models\neventually{=\ii}\lbl}
\\
\qquad\qquad \reason{}{\run\models\neventually{=\ii}\implies\costof\run\leq\kvar\cdot\ii}
\\
\leq\kvar\cdot\sum\limits_{i=n}^\infty
\ii\cdot
\probof\prog{\initconf\models\neventually{=\ii}\lbl}
\textreason{}{Algebra}
\\
\leq\kvar\cdot\sum\limits_{i=n}^\infty
\ii\cdot\eagernessof\prog^\ii
\textreason{}{Lemma~\ref{eagerness:lemma} and $\nn \geq \eagernessthresholdof\prog$}
\\
\leq\kvar\cdot
\frac{\eagernessof\prog^\nn}
{(1-\eagernessof\prog)^2}
\reason{}{\eagernessof\prog<1}
\\
=\costerrorvarof\nn
\textreason{}{Invariant~\ref{cerror:invariant}}
\end{array}
\]

Invariant (\ref{prob:probapprx:invariant}):
\[
\def\arraystretch{1.5}
\begin{array}{l}
\probof\prog\pth - \probapprxvarof\nn
\\
=\sum_{i=0}^\infty
\sum\limits_{\run\in\{\run \in\runsetof{\initconf}|\run\models\neventually{=\ii}\lbl\}}
\probof\prog\run
- 
\probapprxvarof\nn
\textreason{}{By definition}
\\
=\sum_{i=0}^\infty
\probof\prog{\initconf\models\neventually{=\ii}\lbl}
- 
\sum_{i=0}^\nn
\probof\prog{\initconf\models\neventually{=\ii}\lbl}
\textreason{}{Invariant~\ref{probapprx:invariant}}
\\
=\sum_{i=\nn}^\infty
\probof\prog{\initconf\models\neventually{=\ii}\lbl}
\textreason{}{Algebra}
\\
\leq\sum_{\ii=\nn}^\infty
\eagernessof\prog^\ii
\textreason{}{Lemma~\ref{eagerness:lemma}}
\\
\leq\frac{\eagernessof\prog^\nn}{1-\eagernessof\prog}
\reason{}{\eagernessof\prog<1}
\\
=\proberrorvarof\nn
\textreason{}{Invariant~\ref{perror:invariant}}
\end{array}
\]




\end{document}